\documentclass[12pt,preprint]{aastex} 
\usepackage{natbib}

\newcommand{\samis}{stars$\cdot$arcmin$^{-2}$} 
\newcommand{\sapcs}{stars$\cdot$pc$^{-2}$}
\newcommand{\Msun}{~$M_\odot$}
\newcommand{\kks}{~K$\cdot$km$\cdot$s$^{-1}$}

\slugcomment{}

\shorttitle{Embedded Population in the Rosette Complex}
\shortauthors{Rom\'an-Z\'u\~niga, Elston, Ferreira \& Lada}

\begin{document}


\title{A FLAMINGOS Deep Near Infrared Imaging Survey of the Rosette Complex I:
Identification and Distribution of the Embedded Population}

\author{Carlos G. Rom\'an-Z\'u\~niga\altaffilmark{1}, Richard
Elston\altaffilmark{1}, Bruno Ferreira\altaffilmark{1} and Elizabeth A.
Lada\altaffilmark{1}} 

\affil{Astronomy Department, University of Florida,
Gainesville, Fl 32611}

\altaffiltext{1}{Visiting Astronomer, Kitt Peak National Observatory. CTIO is
operated by AURA, Inc.\ under contract to the National Science Foundation.}

\begin{abstract}

We present the results of a deep near-infrared imaging survey of the Rosette Complex made with the Florida
Multi-Object Imaging Spectrograph (FLAMINGOS) at the 2.1m telescope at Kitt Peak National Observatory. We studied
the distribution of young embedded sources using a variation of the Nearest Neighbor Method applied to a
carefully selected sample of near-infrared excess (NIRX) stars which trace the latest episode of star formation in the
complex. Our analysis confirmed the existence of seven clusters previously detected in the molecular cloud, and
identified four more clusters across the complex. We determined that 60\% of the young stars in the complex
and 86\% of the stars within the molecular cloud are contained in clusters, implying that the majority of stars in
the Rosette formed in embedded clusters. Also, half of the young embedded population is contained in four clusters
that coincide with the central core of the cloud, where the main interaction between with the HII region is taking place. We compare the sizes, infrared excess fractions and average
extinction towards individual clusters to investigate their early evolution
and expansion. In particular, the size and degree of central condensation
within the clusters appear to be related to the degree of infrared
excess and mean extinction in a way that suggests that the clusters form
as compact entities and then quickly expand after formation.
We found that the average infrared excess fraction of clusters
increases as a function of distance from NGC 2244, implying a temporal
sequence of star formation across the complex. This sequence appears to
be primordial, possibly resulting from the formation and evolution of the
molecular cloud and not from the interaction with the HII region. Instead, 
the main influence of the HII region could be to enhance or inhibit the
underlying pattern of star formation in the cloud.
\end{abstract}


\section{Introduction} \label{section:introduction}

The study of the distribution of embedded stars in Giant Molecular Clouds
(GMCs) is important for our understanding of the problem of star
formation. At the embedded stage, these young stars represent the most
recent episodes of star forming activity and therefore their physical
properties should reflect the initial conditions of their parental cores. Near-infrared observations 
of the distribution of embedded stars in nearby clouds reveal
that the majority of stars form in embedded clusters \citep{la91,carp2MASS}
with {\it rich} clusters ($>100$ members) clearly dominating over small groups as
they contain more than 80\% of the embedded stellar population in GMCs 
\citep[][hereafter LL03]{porras03,ll03}. However, such surveys have been mostly restricted to nearby clouds ($d<500$ pc), mainly because the study of more distant complexes is limited by photometric depth and resolution of the existing observations. Data from available surveys like 2MASS are not sensitive to faint sources ($K>14.3$), limiting our ability to sample the young, low mass population in distant molecular clouds. This makes it difficult to make
meaningful comparisons between properties of star forming regions in different parts of the galaxy.

As part of the NOAO survey program {\it Toward a Complete Near-Infrared Spectroscopic and Imaging Survey of Giant
Molecular Clouds} (P.I.: Elizabeth A. Lada), we chose the Rosette Complex in the constellation of Monoceros for a new
systematic near-infrared study of a molecular cloud outside of the local neighborhood. A complete review of past and current studies on the Rosette Complex can be found in \citet{hbpaper}. The Rosette, located
in the constellation of Monoceros at a distance of $d=1.6\pm 0.2$ kpc \citep[]{parkandsung02,hensberge00}, stands out as a very
important astrophysical laboratory for the study of the early evolution of stars, because of its exquisite layout:
at the center of its best known feature, the Rosette Nebula, there is a giant OB association, NGC~2244, whose
powerful UV radiation has generated an expanding HII region. It has been suggested that the shock front
generated by the photo-dissociation bubble could have triggered the formation of a family of young clusters
\citep[][hereafter PL97]{pl97} located in the adjacent, highly structured Rosette Molecular Cloud \citep[RMC; e.g.][hereafter WBS95]{wbs95}, following a process known as Sequential Star Formation \citep[SSF;][]{elmelada77}. To test this hypothesis, it is crucial to investigate the influence of the local environment (particularly the expanding HII region) on the properties of the young clusters, as it would help us to understand the different initial conditions under which cluster formation occurs.

\begin{figure}
\plotone{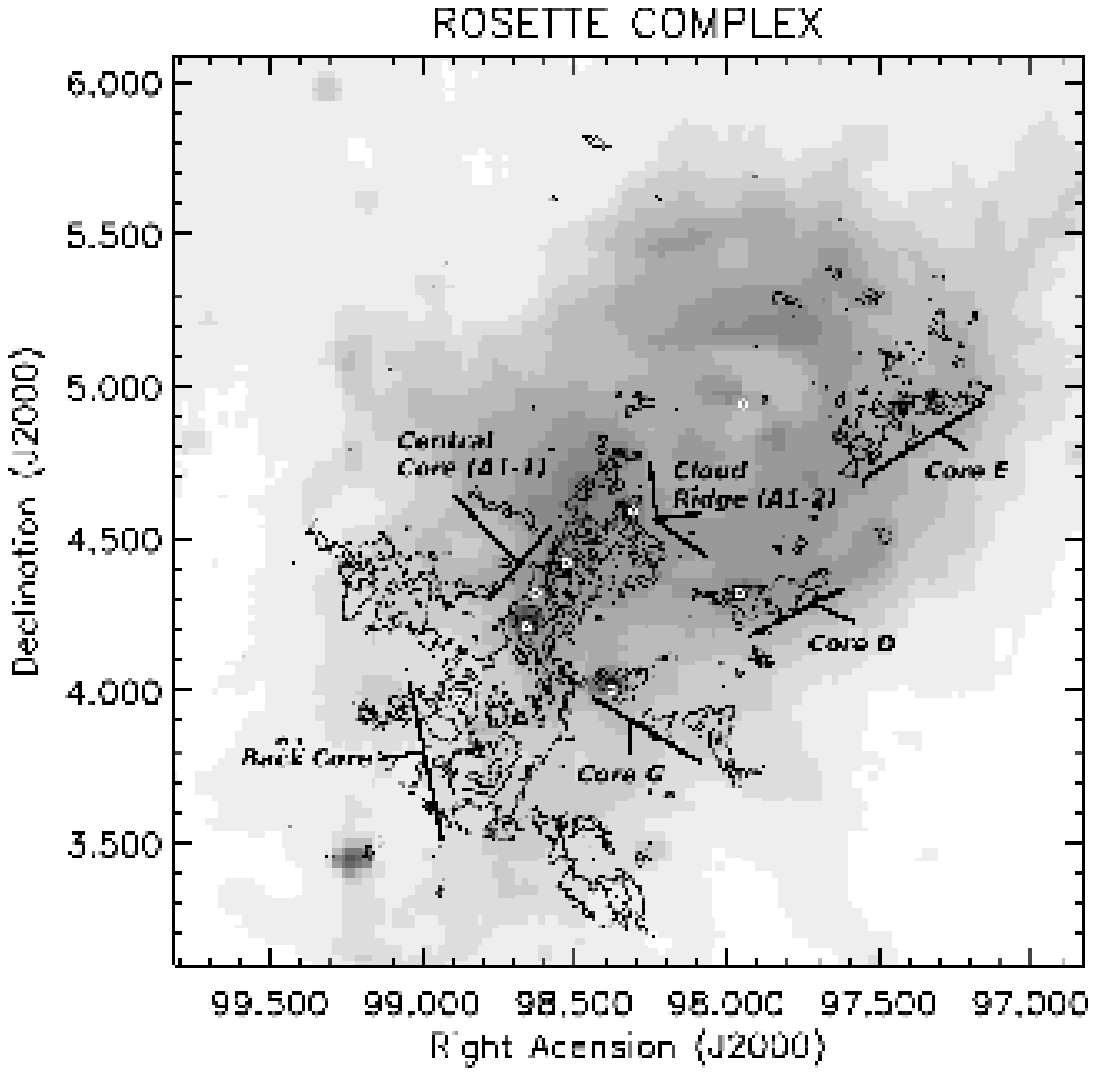} 
\caption{Integrated intensity $^{13}$CO map of the Rosette Molecular Cloud 
from the survey of Heyer et al. (2006). The locations of embedded clusters identified by Phelps \& Lada (1997) are indicated by solid dots. The principal regions of the Rosette Molecular Cloud, identified by Blitz \& Thaddeus (1980),
are labeled.} 
\label{fig:zones} 
\end{figure}

The Rosette has been previously surveyed in the near infrared. PL97 obtained $J$,$H$ and $K$ photometry with the Simultaneous Quad Infrared Imaging Device (SQIID) at the KPNO-2.1m telescope, and were able to identify seven embedded clusters by visual inspection of their images. In figure \ref{fig:zones} we show the location of the PL97
clusters\footnote{We use from now on the nomenclature ``PL01'' to ``PL07'' to refer to these clusters.} in the RMC,
in the context of the main regions of the cloud identified by \citet{bt80}: The regions A1-1 and A1-2, also known
as the ``Ridge'' and the ``Central Core'' of the cloud, host 4 of the PL97 clusters: PL02, PL04, PL05 and PL06. 
The cluster PL07 is located in region B or ``Back Core'' of the cloud. Finally two more clusters, PL01 and PL03 are located in two cores apparently separated from the main cloud, labeled as C and D. All of the clusters discovered by PL97 were
associated with a luminous IRAS source and a massive molecular clump from the study of WBS95, and until now, no
other study has been able to determine the existence of additional clusters.

Recently, a spatially uniform catalog
from the 2MASS survey was used by \citet{lismith3} to study the distribution of star forming activity in the
Rosette Molecular Cloud. They suggested the existence of three general areas of cluster formation, coincident with
the areas with highest extinction and temperature of the cloud. The extensions of the star formation regions
of \citeauthor{lismith3} were determined by eye, roughly tracing the division of BT80 and enclosing
the main clumps of WBS95 and the locations of the PL97 clusters. These general star forming
regions could not be compared with the clusters of PL97 because the extensions and membership of individual clusters were not determined in the later study. Also, due to the low 
sensitivity limits of 2MASS and SQIID ($K<14.2$ mag),
neither the study of PL97 nor that of \citeauthor{lismith3} could adequately sample the low mass population at the distance of the Rosette
Complex and therefore could not determine the fraction of stars forming in
clusters. Furthermore, the Rosette Complex is located at a low galactic latitude ($b\approx -1.8$), 
and the level of field star contamination in the direction of the cloud is very high, making the identification of cloud members even more difficult, specially for low sensitivity data.

In our study we try to overcome these
problems by a) using deeper observations of the region, capable of detecting sources below the limits previously
set by 2MASS and b) by using a selected sample of sources with near-infrared excess (NIRX) to trace young
populations with a minimum of non-member contamination \citep[e.g.][]{lievanslada97}, and c) by using the method of
Nearest Neighbors \citep[][hereafter CH85]{cahut85} to detect surface densities intrinsically larger than the
highly crowded field, thus allowing us to detect clusters of embedded stars.

The main goals of our survey are to investigate the distribution of star
formation in the Rosette, determine the fraction of stars that form in
embedded clusters and compare these results to those found in other local clouds
such as Orion. We also determine the distribution of the
physical properties (i.p. size and near-infrared excess fraction) of the
identified clusters in order to study the star forming history and
evolution of the clusters. Finally, we aim to investigate the distribution of embedded sources across the cloud to investigate the influence of the HII region on its star forming properties, and also to test the hypothesis of sequential formation. 

The paper is organized as follows: in section \ref{section:observations} we describe our
observations and the quality of our dataset; in section \ref{section:analysis} we explain the
methods used to determine the distribution of young stars in the Rosette
complex, specifically the selection of near-infrared excess sources and
the use of the nearest neighbor method; in section \ref{section:results}, we present the
results from our analysis, namely the identification and properties of
the embedded clusters; and in section \ref{section:discussion} we discuss the implications of
our results in the the context of the global properties of the Rosette
complex along with a summary of our work.

\section{Observations and Data Quality Assessment} \label{section:observations}

\begin{figure}
\plotone{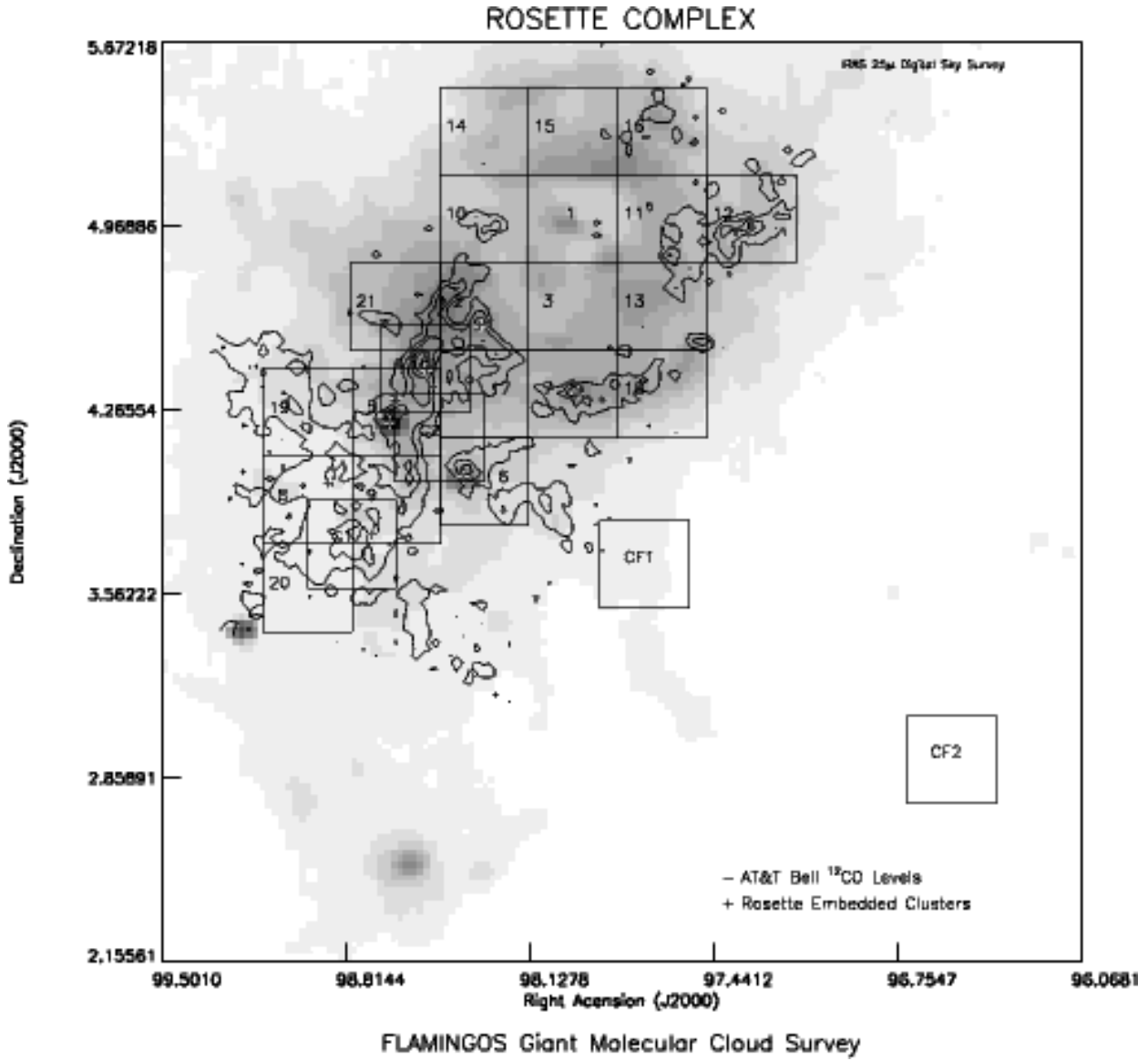} 
\caption{A map of the FLAMINGOS-NOAO-University
of Florida Rosette Molecular Cloud Survey. The boxes delimit individual
FLAMINGOS fields ($20\times20\arcmin$ after trimming) over a halftone image 
(in logaritmic scale) of the IRAS 25$\mu$m emission in the region. 
The labels at the left side of each box
refer hereafter to the fields detailed in table 1 and the text. Light solid
contours represent the extension of the Rosette Molecular Cloud in CO integrated emission
from the survey of Blitz \& Stark (1986) in steps of 20.0 \kks. Crosses mark
the centers of known embedded clusters from the previous study of Phelps \& Lada
(1997).} 
\label{fig:surveymap} 
\end{figure}

The data were obtained with the Florida Multi Object Imaging Near-Infrared Grism Observational Spectrometer
FLAMINGOS at the KPNO 2.1m telescope. This project is part of the NOAO survey program {\it Toward a Complete
Near-Infrared Spectroscopic and Imaging Survey of Giant Molecular Clouds}. At the 2.1m
telescope, FLAMINGOS has a FOV of $20\arcmin \times 20\arcmin$, and in the case of the Rosette Complex, a total of
23 FLAMINGOS fields were observed during winters of 2001 to 2004 using the $J$, $H$ and $K$ filters (1.25, 1.65 and 2.2 $\mu$m, respectively.) The Rosette observations
were designed to survey the regions of significant molecular gas density revealed by the $^{12}$CO and $^{13}$CO  maps of
\citet[][see also WBS95]{bs86} (integrated intensity levels above 20 \kks and 0.8 \kks, respectively) 
as well as the 25 $\mu$m emission map from the IRAS survey 
(we considered as significant those areas with emission above 5.0 MJy$\cdot$sr$^{-1}$). The result is a very complete coverage
of the Rosette Nebula and the Rosette Molecular Cloud regions. A map showing the positions of the observed fields
can be seen in Figure \ref{fig:surveymap}. Twenty of the fields are adjacent, while 3 of them (areas 4, G1 and G2 )
were added to enhance the quality of the observations in some particularly interesting regions. In addition, to
account for the field contamination, we observed 2 near control fields located away from the main molecular cloud
emission. These control fields have the same depth as any of our on-source fields, and their locations are also
indicated in Figure \ref{fig:surveymap}.

All the data of the FLAMINGOS star formation survey were processed with the aid of pipelines built in the standard
IRAF Command Language. One pipeline \citep[see][]{romanzunigathesis} processes raw data by applying a linearization correction, subtracting dark current, dividing by dome flat fields, applying a two pass sky
subtraction method, and combining reduced frames into image products which are free of geometric distortions, re-sampled to twice their size and trimmed into final frames of 4096$\times$4096 pixels. These products were then analyzed with a second pipeline \citep[see][]{levinethesis}, which extracts all possible sources from a given field, applies PSF photometry, and calculates an accurate astrometric solution and creates final catalogs after matching lists from individual filter observations. Both photometry and astrometry are calibrated with respect to 2MASS.

\begin{figure}

\plotone{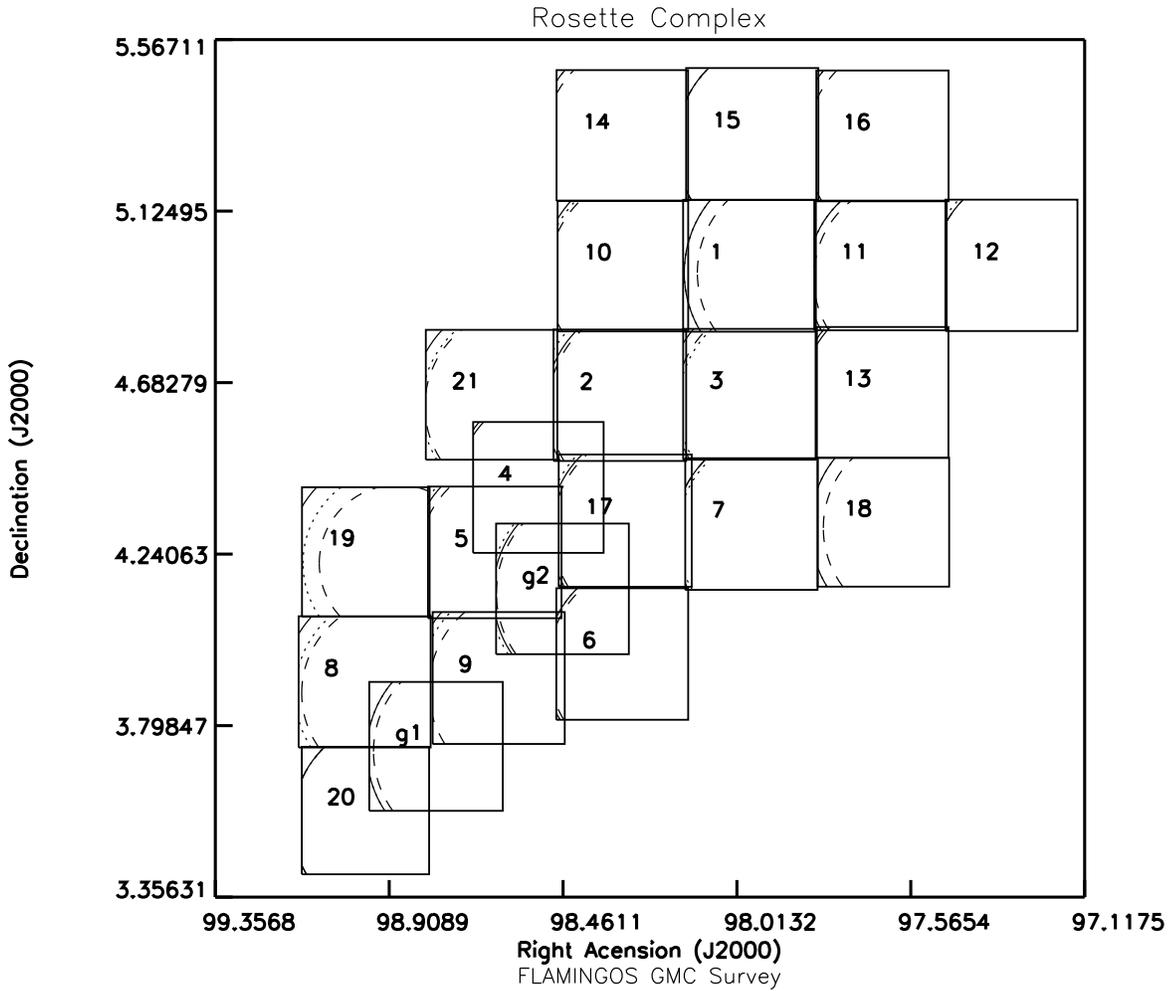} 
\caption{The extension of the areas of acceptable optical
distortion are marked for each field as circles with radii equal to the center
of the maximum bin at which the polynomial correction to the zero points (see
text and figure 5) can be applied within the detector.~This effect varies by
field (size of the acceptable area) and filter: the solid, dotted and dashed
linestyle circles represent the tolerance radii for J, H and K respectively.}
\label{fig:badareas} 
\end{figure}

For the Rosette survey, the completeness limits are $K=17.25$, $H=18.00$ and $J=18.50$ mag, and are slightly fainter in a few fields observed under optimal weather conditions. The completeness limits were determined with artificial star experiments, and apply well within 75 to 80\% of the area covered in each field, but are up to 0.25 mag brighter near the edges of the fields, where the photometric sensitivity of the FLAMINGOS detector degraded from the center due to a loss in optical quality \citep{romanzunigathesis}. This optical degradation also affected the shapes of the star images near the edges, and we required to apply a correction to the zero point of the photometric calibration. The correction is radial, with center on the point of best PSF adjustment (which varied, depending on field and epoch within a radius of approximately 100 pixels from position [3000,2400] in the 4096$\times$4096 final image grid), and for the Rosette Survey fields it was well defined with a sixth order polynomial form. Those stars located at radii where the correction was larger than 0.1 mag were not included in the final catalogs. In Figure \ref{fig:badareas} we show a map of the individual fields and the extension of the bad quality areas, which in all cases were smaller than 12\%. The zero point correction, however, helped to reduce photometric scatter in the accepted areas by an average of 0.02 mag in each field. Our final mean photometric uncertainties are $0.058\pm 0.012$, $0.064\pm 0.018$ and $0.056\pm 0.014$ for $J$, $H$ and $K$ respectively.

\begin{figure} 
\plotone{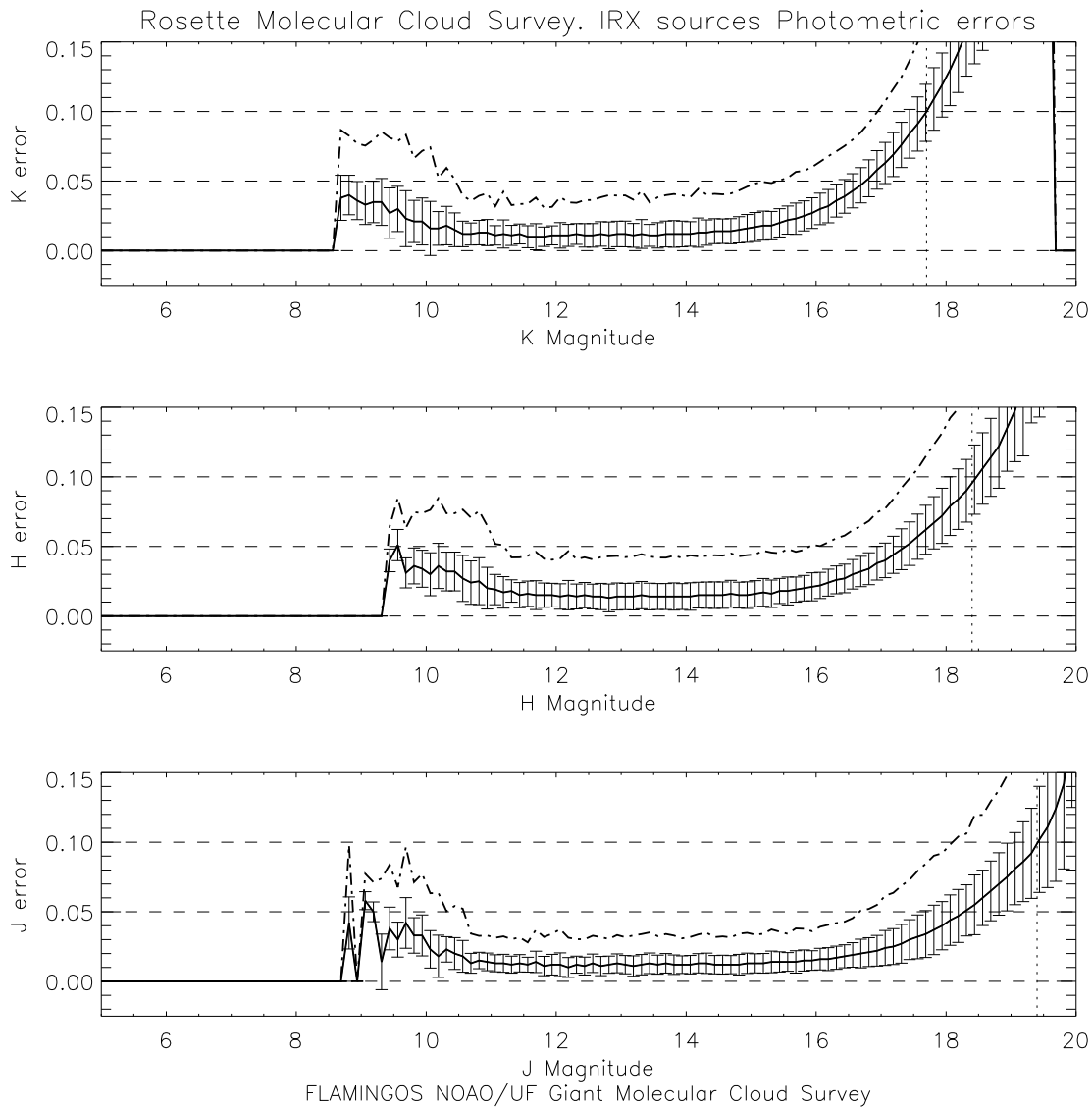} 
\caption{Intrinsic quality of
the Rosette Molecular Cloud survey. The points mark the mean value of the
photometric error in a given magnitude bin, with the size of one standard
deviation indicated by the error bars. The horizontal dashed lines marks the
zero level and the sensitivity level, estimated at 0.1 mag. Vertical dotted
lines indicate the sensitivity limits, marked as the bin at which the fiducial
curve generated by the mean values crosses the 0.1 line. The dash-dotted line
represents the 3-$\sigma$ level of the fiducial errors; any star in our catalog
with errors higher than this levels were rejected from the analysis.}
\label{fig:globalerrs} 
\end{figure}

Figure \ref{fig:globalerrs} shows the mean photometric uncertainties and their standard deviations per magnitude
bin as a function of brightness in each band. In this analysis we only used catalog stars with photometric
uncertainties deviating no more than 3.0$\sigma$ from the mean. Completeness limits for the analysis were not
significantly compromised with these cuts, because most of the sources rejected for having large photometric
uncertainties were fainter than $J=18.50$, $H=18.25$ and $K=17.50$ mag. Additionally, for those sources in which
all 3 bands were required for the analysis, we placed a new limit to the total color uncertainties, $\sigma
_{J-H}=\sqrt{{\sigma _J}^2+{\sigma _H}^2}$ and $\sigma _{H-K}=\sqrt{{\sigma _H}^2+{\sigma _K}^2}$, requiring them
to be smaller than 0.1 mag. This last constraint restricted the photometric errors in each band to be no larger
than 0.071 mag.

Sources brighter than approximately 11.0 mag suffered of saturation. To account for this, we rejected all sources
above $H=11.0$ from our catalogs (this cut also removed saturated stars in bands $J$ and $K$, which had slightly brighter saturation limits), and replaced them with the corresponding 2MASS point sources obtained from the All Sky Release database. The total number of sources replaced was 798. After applying the photometric quality cuts to the entire database we retained 74,285 sources for analysis.

\section{Analysis} 
\label{section:analysis}

The Rosette Complex is located at 1.6 kpc from the Sun, more than four times further away than regions such as
Orion and Perseus, where systematic searches for embedded clusters have been previously performed. Furthermore, the
Rosette is located at a very low galactic latitude ($b=-2$) and towards the anti-center of the Galaxy ($l=210$),
which results in a very high density of field sources located in the foreground and background of the cloud. These
characteristics make it more difficult to detect clusters using traditional star count methods. For that reason, our
analysis focuses on optimizing the detection of young, embedded sources in the cloud. This is done by a) selecting
them from a sample of infrared excess stars (NIRX) and b) by analyzing their distribution of surface densities using
a method of nearest neighbors.

\subsection{Infrared Excess Stars} 
\label{section:analysis:subsection:NIRX}

In the $J-H$ vs.~$H-K$ color-color diagram, NIRX stars fall to the right of the band defined by the projection of
the giant and dwarf sequences along the direction of interstellar reddening. This vector of extinction is defined
by a standard reddening law \citep[e.g. ][]{coh81, rile85}. If de-reddened, low mass NIRX stars would fall back to
the Classic T Tauri star (CTTS) locus \citep{mecahi97}, which is defined by large $H-K$ colors indicative of
dramatic intrinsic extinction due to the presence of circumstellar material.~The fraction of stars falling into
this NIRX region has been observed to be fairly large in deeply embedded clusters \citep[e.g
][]{1996AJ....111.1964L, 1997AJ....114..198C}. 

Studies have shown that the total NIRX fraction in young clusters decreases with age, as a result of inner disk evolution \citep{halala01}. For example, the $JHK$ fraction in the Trapezium and NGC 2024 (ages $\sim$1 Myr) is close to 50\%, while in slightly older clusters like IC 348 (age $\sim$2 Myr) the $JHK$ fraction is around 20\%. The deeply embedded population of the RMC can be expected to be younger than the exposed OB association NGC~2244, which is estimated to have an age of 2 Myr from spectroscopic observations 
\citep{perez2,parkandsung02}. Thus, the $JHK$ excess fraction in the embedded clusters of the Rosette can be expected to be between 20 and 50\%, significant enough to trace well the most recent episode of formation. 

For our study, we define an NIRX star as one with colors that place it 0.1 mag (5 times the standard deviation of
the $H-K$ uncertainty) to the right of the dwarf zero age main sequence reddening band (ZAMSRB), and above ${J-H} =
{0.47(H-K)+0.46}$, which defines the lower limit for the Classic T Tauri Star (CTTS) locus given the listed
uncertainties of \citeauthor{mecahi97}. The first of these two constraints avoids contamination of the sample from
non-NIRX stars with large $H-K$ uncertainties that could place the stars slightly outside the right edge of the
ZAMSRB.~The second one impedes the inclusion of objects that lie in the region located below the CTTS line. This
region, just as the one to the left of the ZAMSRB, contains detections that are mainly due to high color scatter.
Also, in some cases unresolved galaxies (which can pass as stars) with large photometric spreads \citep{labbe03} or
highly inclined intrinsic reddening vectors --due to dominant emission from HII regions and AO stars
\citep{hera96}-- are located in this region of the diagram.

\begin{figure}

\plottwo{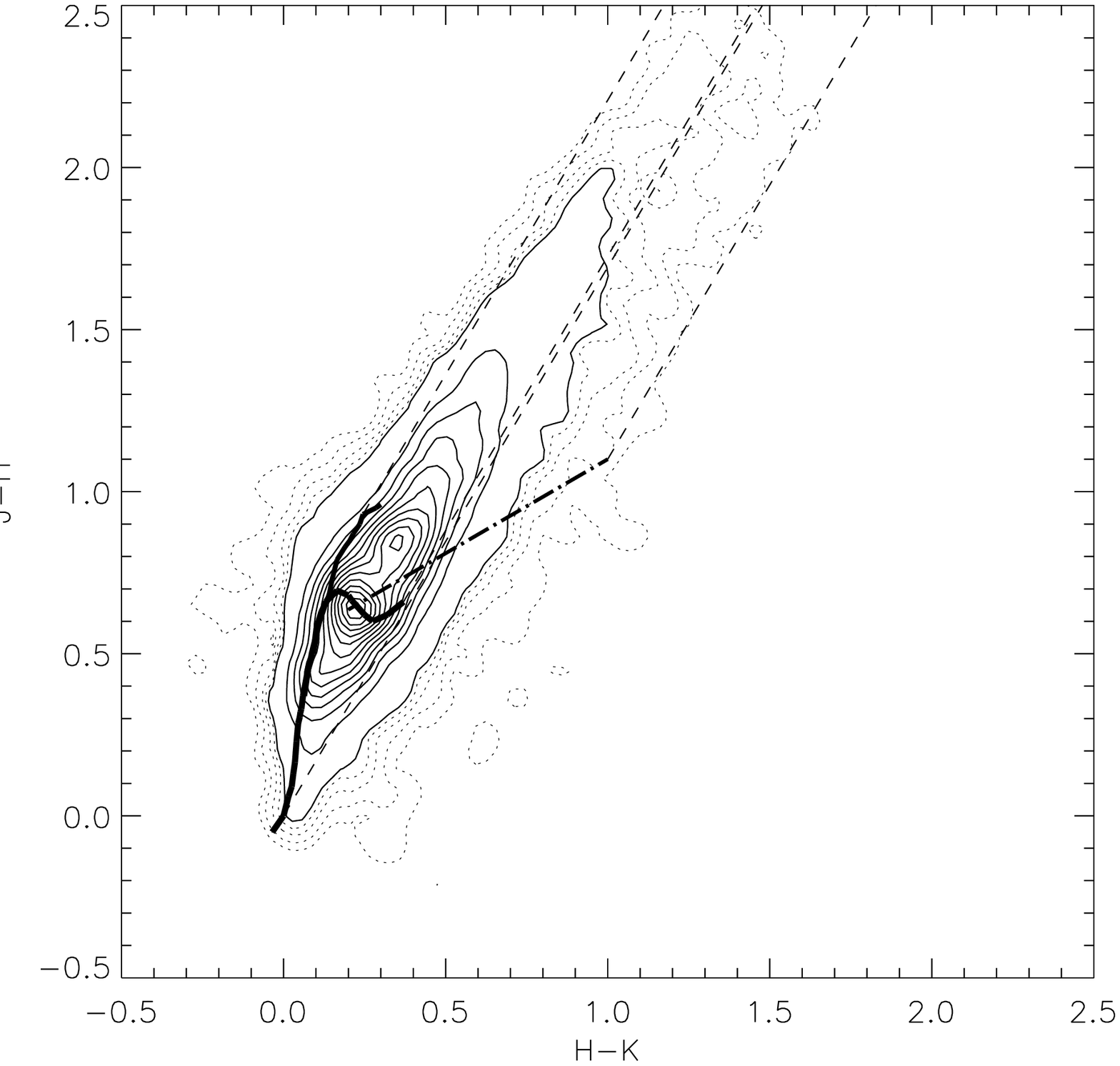}{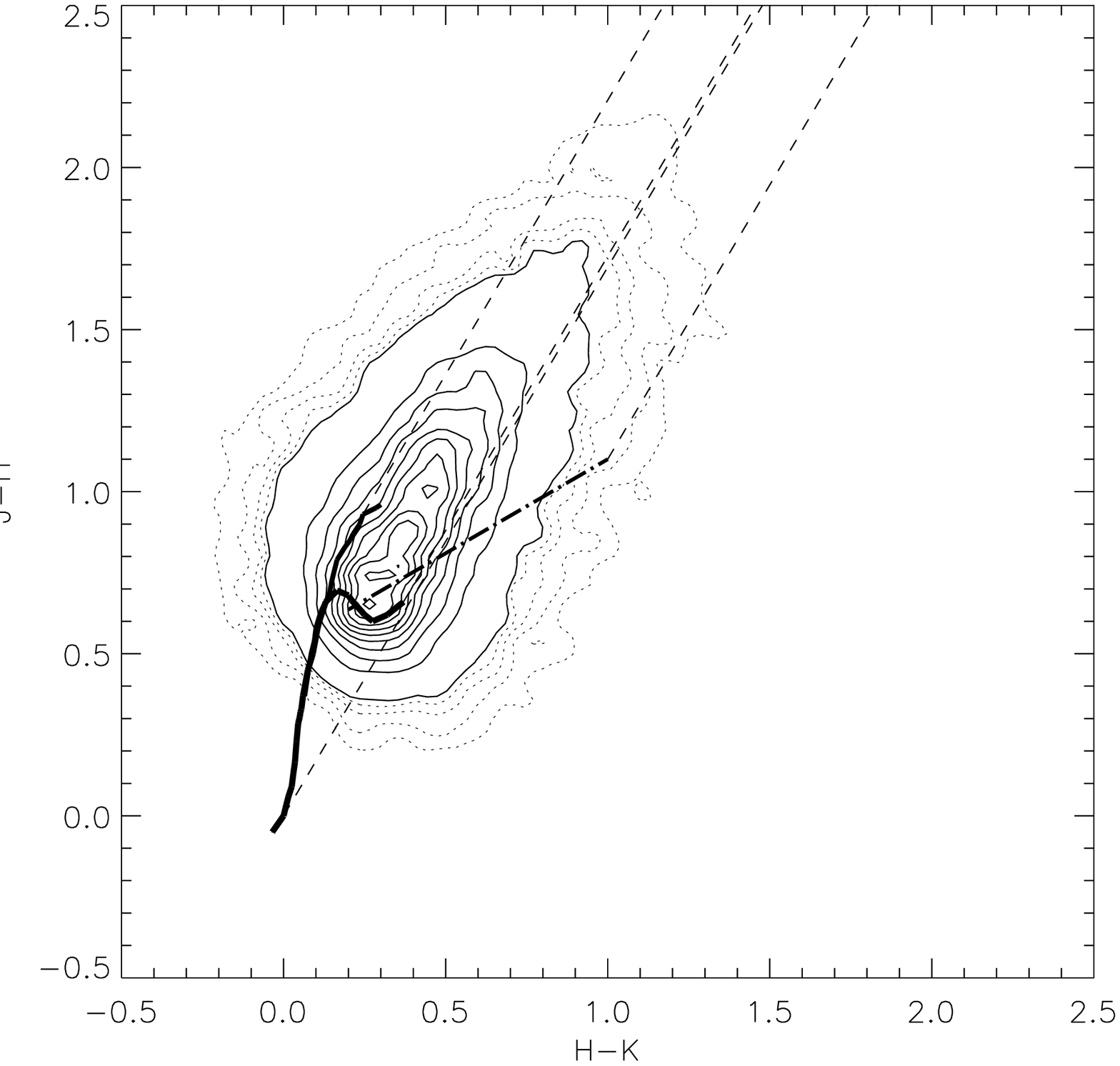} 
\caption{Contour level color-color diagrams for stars
in the FLAMINGOS RMC survey divided in two ample groups of brightness. Both
diagrams were constructed with a Nyquist box size of 0.1 mag. {\it Left:}
distribution of colors for `bright' stars within $5.0<K<15.75$~mag, {\it Right:}
distribution of colors for `faint' stars within $15.75<K<17.25$~mag. The solid line 
contours start at the mean level (360 and 466 dex$^{-2}$ for the left and right
panel, respectively) with subsequent steps of 1 sigma (1770 and 1910 dex$^{-2}$
for the left and right panel, respectively).} \label{fig:CCCs} 
\end{figure}

Combined effects of variable seeing quality over the seasons and variability of the focus quality across the wide
detector field of FLAMINGOS, resulted in a high dispersion of color values at the faint end of our sample, and it
cannot be reduced with the zero point corrections nor the uncertainty restrictions. The scatter in the $H-K$ colors
increases for sources fainter than $K=15.75$ mag. At this limit, the field object density also
increases significantly, and the larger the uncertainty in $H-K$, the more difficult it is to distinguish cloud
members from background sources. In Figure \ref{fig:CCCs} we show the color-color diagrams for stars with $K <
15.75$~mag and for stars with $15.75 < K < 17.25$~mag respectively. In these diagrams, the lowest contour level
shown represents the mean surface density of objects in the color-color space, and each subsequent level represents
a step of 1 standard deviation. The diagrams show how the scatter for the bright end of the sample is much smaller
than for the faint end, where we can see a quasi symmetric increase in the dispersion of colors at both sides of
the ZAMSRB near the most populated areas of the diagram. Our calculations indicate that the mean standard deviation
of the $H-K$ color dispersion is twice as large for the faint end bins (0.11 vs. 0.21 mag, calculated for the
entire catalog). The effect is also slightly worse for those fields in the survey which were observed under the
less favorable weather and airmass conditions.

Large scatter in photometric colors directly affects the calculation of the number of NIRX stars in our survey. We
performed a series of Monte Carlo experiments in which we simulated the colors of stars drawn from a model
population with an age of 1 Myr \citep{dantom97} located at the distance of the Rosette. We then added color errors and extinction to
these sources, similar to those observed in the survey fields. We found that the number of
stars with colors similar to those of NIRX stars due to photometric scatter can be up to 5 times larger for stars
fainter than $K=15.75$.

Because we are basing our analysis on the detection of infrared excess sources, we limited the primary aspect of
our analysis, the identification of embedded clusters, to those stars in the bright end of the sample ($K<15.75$
mag). This sample restriction is rather conservative, but assures a more robust detection across the entire survey
area. It is important to remark that these bright NIRX sources are only helping us to {\it trace} the location and
rough extension of clusters, and although our color uncertainties are high, individual K band magnitudes are still
good within 0.05 mag (averaged), and thus the sample is still robust enough to construct accurate luminosity
distributions with a bin resolution of 0.25 mag. Also, the photometric depth limit of $K=15.75$ implies a sample
1.5 magnitudes deeper than 2MASS and is equivalent (for dwarf type stars) to a stellar mass range of 0.09 to 0.18 \Msun~ --- for a population of 1 Myr embedded in a cloud with a typical extinction of 0 to 10 visual magnitudes
\citep{dantom97}

\subsection{The Nearest Neighbor Method}
\label{section:analysis:subsection:nnmethod}

In order to optimize cluster identification, we used a variation of the Nearest Neighbors Method \citep[][hereafter
NNM]{cahut85} to determine which objects in our fields have surface densities above the crowded background levels.
The NNM has been shown to have success in detecting and outlining the extensions of embedded clusters
\citep[][hereafter FL07]{nakajima,guter05,fl07}. The generalized form of the $j$th nearest neighbor surface density
estimator for a star in a field is simply:

\begin{equation} \mu_{j} = \frac{j-1}{\pi D^2_{j}}, \end{equation}

where $D_j$ is the distance from any given star to its $j$th neighbor. This estimator has only one degree of
freedom, the fixed number of neighbors, $j$, used to calculate the local density. For a large enough value of $j$,
the fluctuations in the local density estimations due to local irregularities will be small, which assures an
accurate determination of the extension and structure of large systems. However, for the same reason, $j$ also
limits the minimum number of grouped stars that can be detected without confusion: if $j$ is much larger than the
minimum number of stars that define a group or cluster, then small structures could be overlooked, as their surface
densities will be systematically lower than a defined average value. CH85 showed that $j=6$ is the minimum number
that can assure unbiased detection of groups with up to 1000 stars, but is 15\% more susceptible to local
statistical fluctuations than larger values of $j$. The selection of $j$ for each application will simply depend on
the goal: for example, in their study of young clusters using 2MASS data, FL07 chose a $j=20$ estimator, which
allowed them to identify clusters with radii as small as 0.3 pc and a minimum of 20$\pm$5 members. \citet{guter05}
used $j=5$ to determine the membership in cluster fields at high resolution in combination with
carefully corrected source counts.

At first, we tried to emulate FL07 by applying a $j=20$ estimator to our final RMC catalog and our control fields,
expecting to be able to detect and delimit the already known structures: NGC~2244 and the seven embedded clusters
reported by PL97. The result was that NGC~2244 could be easily identified as a high density region, along with the
clusters PL04 and PL05, located in the core of the cloud. The three mid size clusters PL01, PL03, and PL07 could be
identified too, but their apparent structures were not much more significant than some density fluctuations in our
fields due to patchy extinction. Finally, stars in the smaller clusters PL02 and PL06 had 20th NN densities below
the 3$\sigma$ level, and thus these clusters could not stand out above the noise levels of the distribution.

Cluster PL06, which is associated with the B-type proto binary AFGL961, contains large quantities of obscuring
material near its center, complicating the detection of its faintest and most
embedded members even in carefully
constructed near-infrared maps \citep[e.g][]{afgl961}. Cluster PL02 is deeply embedded near the front part of the
cloud, where material appears to be in contact with the expanding HII region. Likely, for small embedded clusters
like PL02 and PL06, the combination of small number of members, high background contamination and non-uniform
extinction, hampers the effectiveness of the NNM. Furthermore, in regions of high extinction, some embedded
clusters could have surface densities intrinsically lower than the field so that their members could be
undetectable with a large $j$. Or, random patches of low extinction in the cloud could have larger than average
field densities and may pass as ``false'' clusters.

In order to overcome these complications, we applied the NNM only to a JHK color selected sample of sources that
exhibited a near infrared excess. This allowed us to trace preferentially the youngest and most probable members of
the cloud. For this reason, the number of NIRX sources is expected to greatly diminish outside the molecular cloud,
and therefore background contamination is reduced, along with false cluster occurrences. Also, the number of young
sources with NIRX is expected to be larger in deeply embedded populations, and consequently high extinction regions
could actually present larger NIRX densities, opposite from the ``all-star count'' case, in which these regions would be
overlooked.

Since it is expected that only a fraction of the members in a given young cluster exhibit infrared excess, then our
sample will diminish in size, and our $j$ estimator value has to be reduced accordingly. We experimented with
different values of $j$, from 6 to 15, and we ultimately adopted a value of $j=10$ for our analysis. This choice
allows us to detect NIRX groups of 10 members, which for a typical NIRX fraction of 30 percent (see section 
\ref{section:analysis:subsection:NIRX}) would trace a cluster with $\sim$ 30 members. A preliminary visual
inspection of the images of the less populated clusters, PL01, PL02 and PL06, indicated that the typical number of
stars within visible areas of bright nebulosity --usually coincident with embedded cluster cores-- could be indeed,
around 30. This is close to the minimum number that defines a cluster dynamically \citep{am01,ll03}, and below
that, any groups could at most be considered associations or isolated populations, although they will be harder to
distinguish from local fluctuations. In principle, a lower value of $j$ would allow us to detect those smaller
stellar groups, but our tests indicate that analysis with lower values (i.e. $j$=6 to 10) do not differ
significantly from the case of $j=10$, and thus we opted for this, more robust choice.

\subsection{Nearest Neighbor Analysis for Infrared Excess Stars}
\label{section:analysis:subsection:NIRXnn}

\begin{figure} 
\includegraphics[scale=.70]{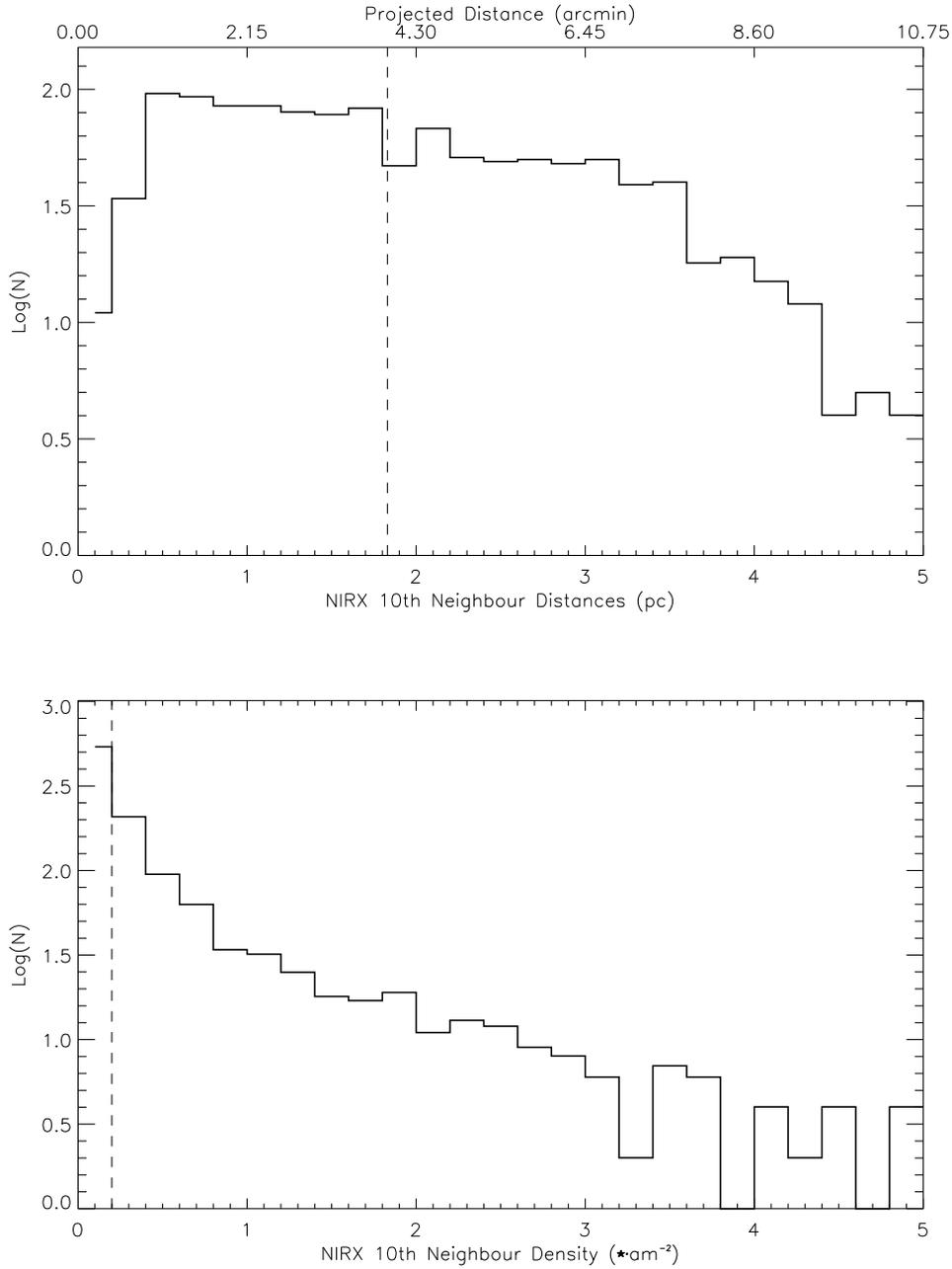}

\caption{Nearest Neighbor distributions for bright NIRX stars. The top panel
shows the distribution of 10th neighbor distances. The bottom panel is the
distribution of 10th neighbor densities. In the top panel line A indicate the
limit of distances shorter than 1.0 pc, while the dashed line B indicates the
midpoint value at 1.83 pc. At the bottom panel the equivalent limits in density
space are also indicated.} \label{fig:10NN_NIRX_1575} 
\end{figure}

We detected a total of 1169$\pm$34 NIRX sources with $K<15.75$. In Figure \ref{fig:10NN_NIRX_1575} we show their
distributions of 10th Nearest Neighbor distances, $D_{10}$ and local surface densities $\mu _{10}$. The mean value
of $D_{10}$ is 3.79 arcmin (1.83~pc) which corresponds to a $\mu_{10}=0.2$ \samis ~(0.92 \sapcs) . This limit is also indicated in
the plot.

For the control fields we found 19 sources that had NIRX colors down to a maximum brightness of $K=15.75$. The mean
10th Nearest Neighbor surface density of the 19 sources together is 0.18 \samis ~(0.82 \sapcs), which compares well with the mean
value 0.2 \samis~ for the $\mu_{10}$ distribution in the cloud survey areas. Thus, we considered a round value of
0.2 \samis~ as a {\it background} level for the NIRX population, below which we cannot assure that an embedded NIRX
source has a density high enough to be distinguished from a field source with similar colors.

The minimum value of the $D_{10}$ distribution in the survey is 0.311 arcmin, which  represents a density of 29.5
\samis ~(136.3 \sapcs). This value is a good estimation of the typical local  surface density in the central regions of RMC
clusters. The midpoint between this minimum distance and the mean is 2.387 arcmin (1.1 pc) which corresponds approximately  to $\mu_{10}$=0.6 \samis ~(2.77 \sapcs). It turns out that this value is a good
estimate of the  average embedded cluster size in the RMC.

\section{Results} 
\label{section:results} 

\subsection{Identification of clusters} \label{section:results:subsection:clusteriden}

\begin{figure} 
\plotone{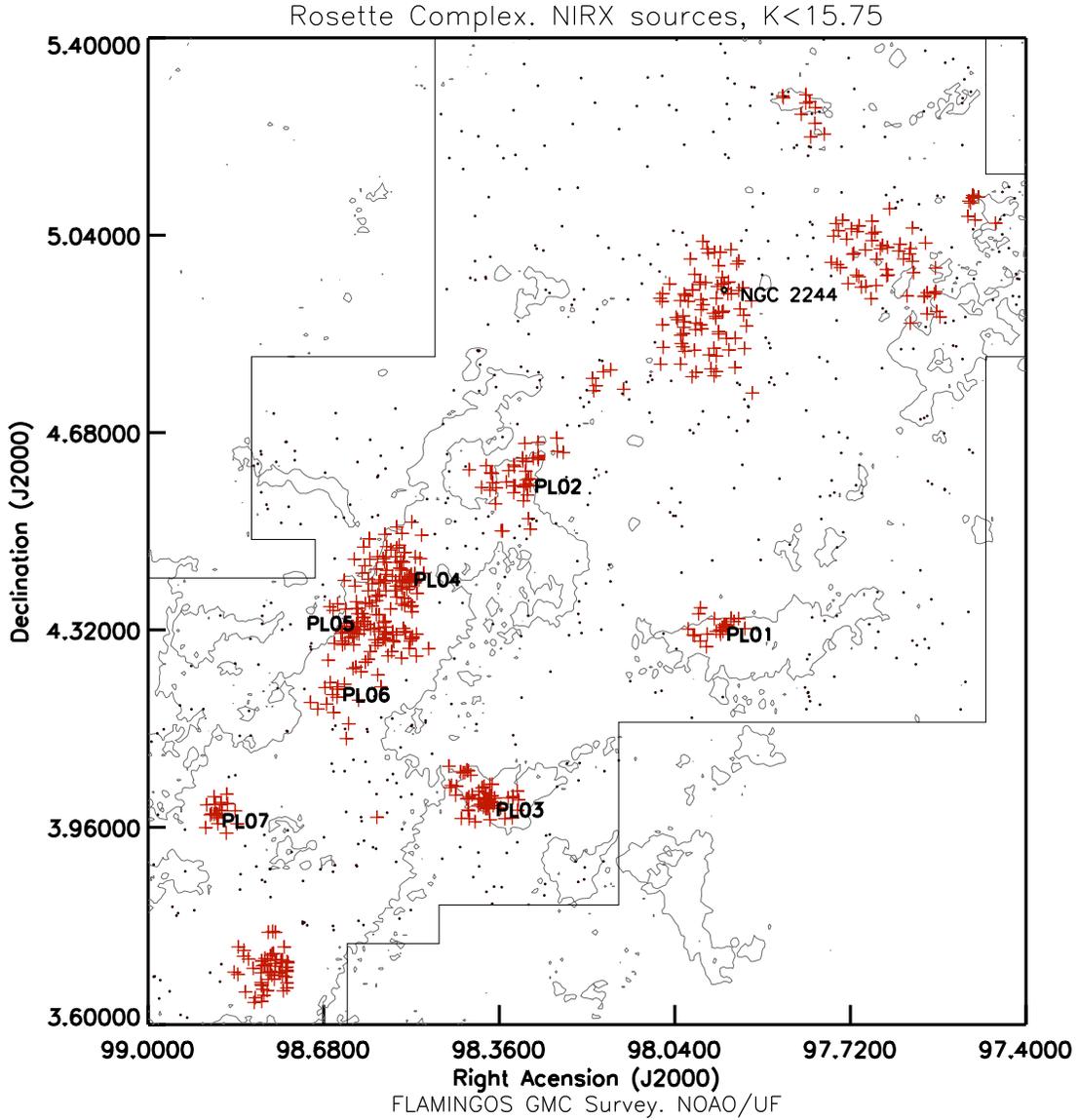} \vspace{0.4cm} 
\caption{The location of NIRX
stars in the Rosette survey with brightness $K<15.75$~mag. The "plus" symbols
are NIRX stars with 10th neighbor densities higher than 0.2 \samis, while black
dots are stars with densities below 0.2 \samis. We also indicate the expected
centers for the known clusters NGC~2244 and PL01-PL07. The contours indicate the baseline level of $^{13}$CO emission of 0.8 \kks that we used to define the extension of the main molecular cloud regions. The solid thin line indicates the limits of the survey coverage.} \label{fig:NIRXpos_1575} 
\end{figure}

In Figure \ref{fig:NIRXpos_1575} we show the location of NIRX stars with surface densities higher than the background
level at 0.2 \samis. All of the known clusters are traced well by these sources, which confirms that they
are among the main regions of star formation in the Rosette Complex. In this plot we also show the contours of CO
from the study of \citet{bs86}. From now on we will use the nomenclature of the main RMC features from
\citet{bt80}. As seen in the figure, most NIRX stars with high NN densities are located within the molecular cloud,
except for those located in the field of the Rosette Nebula.

\begin{figure} \plotone{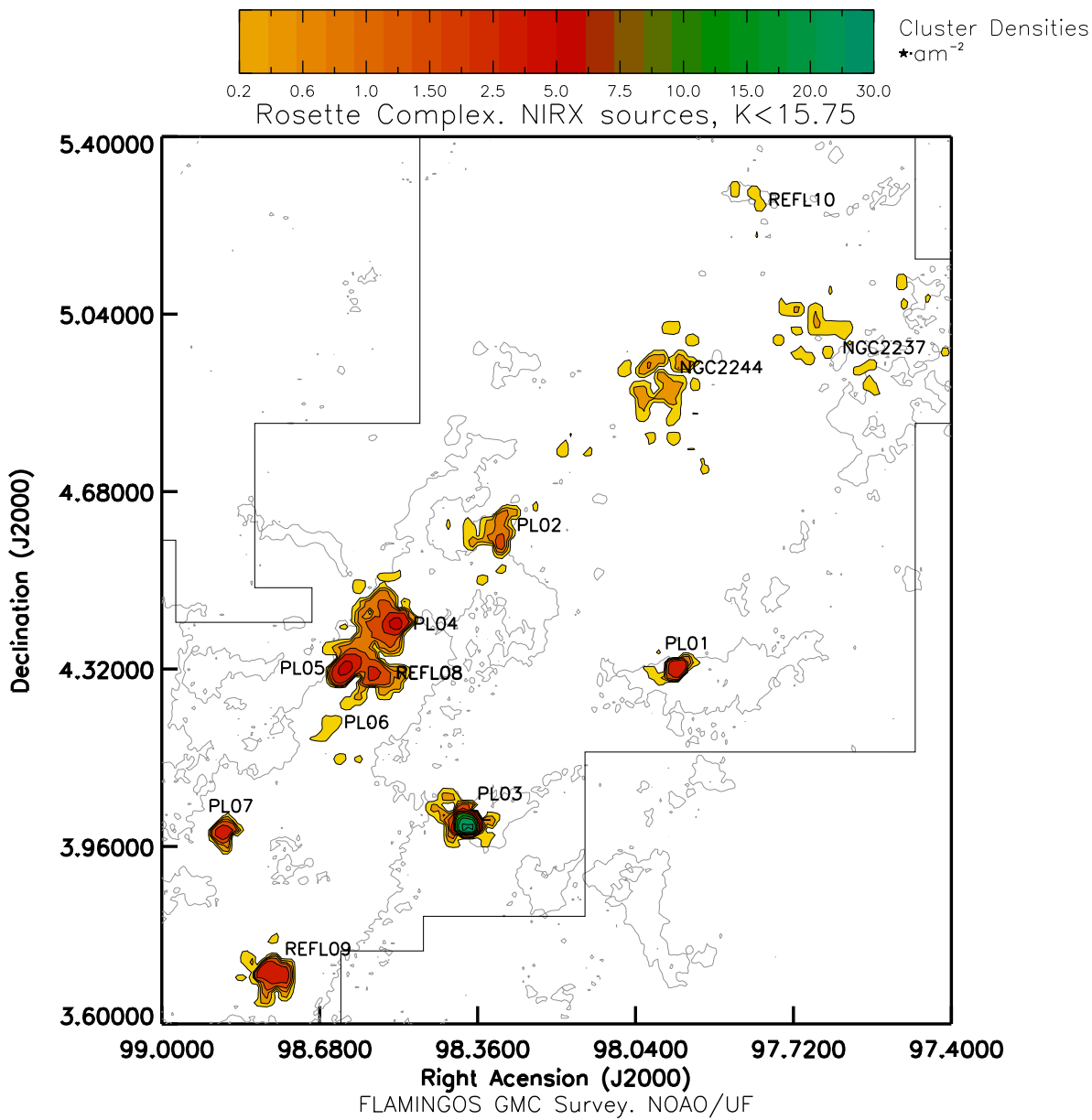} \vspace{0.4cm} 
\caption{Final identification of
clusters in the Rosette Complex. The colored scale contours indicate levels of
surface density from the 10th NN analysis. The contours were constructed using a
Nyquist box sampling of 1.5 arcmin. The dot line contours indicate the baseline level
of $^{13}$CO emission of 0.8 \kks that we used to define the extension of the
main molecular cloud regions. The solid thin line indicates the limits of the
survey coverage.} \label{fig:clusteriden} 
\end{figure}

In Figure \ref{fig:clusteriden} we show a contour level map of the local surface densities calculated with the NN
method ($j=10$). The contours were constructed using a Nyquist sampling box of 90$\arcsec$. Using this map, we
define an embedded cluster as a region for which we can observe a {\it closed} contour at a level of 0.2 \samis,
containing at least 10 NIRX sources. Using this definition we found, in addition to NGC~2244 and the seven clusters
from PL97, four additional areas that arise as significant but have not been previously studied:

The first new cluster, REFL08 is located in the ``Central Core'' of the cloud, East of cluster PL05 and South of
cluster PL04. These two large, neighbor clusters present patchy reddening and have a number of members already
visible in optical plates (e.g. DSS), thus they can be considered to be partially emerged from the cloud. In
contrast, REFL08, is not visible in any optical plates and has a large number of conspicuously reddened sources.
Therefore, REFL08 appears to be a third, more deeply embedded substructure in this active region of star formation
at the center of the cloud, in which clusters PL04 and PL05 are the largest clusters.

The second new cluster, REFL09 is a relatively extended, highly reddened cluster located in the southeastern edge
or ``Back Core'' of the cloud. Along with REFL08, these are clear examples of clusters which are located in regions
of very high extinction, resulting in stellar surface densities comparable or lower than the field if a color
selection is not applied. However, these clusters contain a large number of heavily reddened sources -easily
distinguishable in JHK color composite images- and a significant number of NIRX sources that let them stand
unequivocally as embedded clusters.

The third new cluster is located to the east of NGC~2244, in the region of the cloud historically identified as
NGC~2237. The existence of this cluster was also suggested by \citet{lismith3} in their study of 2MASS data.

The fourth cluster, REFL10, is much less prominent. It is located North of NGC~2244, and despite of its low NIRX
surface density compared other star forming regions in the complex, its 13 NIRX sources make it qualify as a
cluster. REFL10 is an example of cluster that would not be significant if the NN distribution of NIRX sources was
analyzed with a large value of $j$, or if a color selection was not applied.

\subsection{Properties of Clusters}
\label{section:results:subsection:clusterproperties}

Once the clusters were identified, we proceeded to investigate their basic properties and how are they related in the context of the evolution of the star forming complex. For each cluster we determined sizes, centers, number of members and infrared excess
fractions. These cluster properties were determined individually, isolating regions centered at each cluster density peak. The area of these individual cluster analysis regions varied from 25 to 120 arcmin$^2$, depending on the apparent extension of the cluster.~This way we
were able to observe in detail structures occupying areas between 6 and 60 arcmin$^2$.

\subsubsection{Determination of cluster properties}

Cluster boundaries are defined by the 0.2 \samis~contour level in our NIRX NN density plots. All sources with
$K<17.25$~mag within this contour level are considered potential cluster members. Cluster sizes are determined by
measuring the area, $A_{P}$, within the cluster boundary and calculating the equivalent radius,
$R_{eq}=\sqrt{A_{P}/\pi}$.

The density peak, $\vec{X}_d$, and the core radius, $R_{core}$, of each cluster were calculated with the formulation of CH85. The
density peak defines the natural center of a cluster, and is simply the density weighted average of the star
positions in a given region:

\begin{equation} \vec{X}_{d} = \frac{\sum _i \vec{X}(i)\mu _j(i)}{\sum _i \mu _j(i)}, \end{equation}

Similarly, the core radius, $R_{core}$ is defined as the density weighted average of the distance of each star to
$\vec{X}_{d,j}$:

\begin{equation} R_{core} = \frac{\sum _i|\vec{X}(i) - \vec{X}_{d,j}|\mu _j(i)}{\sum _i \mu _j(i)} \end{equation}

In Table~\ref{tab:clustertable} we present the center coordinates (density peaks), core radii, equivalent radii,and average
extinction values for each cluster.  The latter were calculated from the $H-K$ colors.  We also list the number of
NIRX sources to $K < 15.75$~mag and the corresponding excess fraction. The excess fraction was calculated by
comparing the number of IR excess sources to the number of cluster members. The number of cluster members was
determined by subtracting the expected number of background field stars from the number of stars having $K< 15.75$
within the cluster boundary. The number of field stars was determined for each cluster by scaling our off fields to
the cluster areas and applying appropriate extinction corrections.

\begin{figure}

\includegraphics[width=7cm]{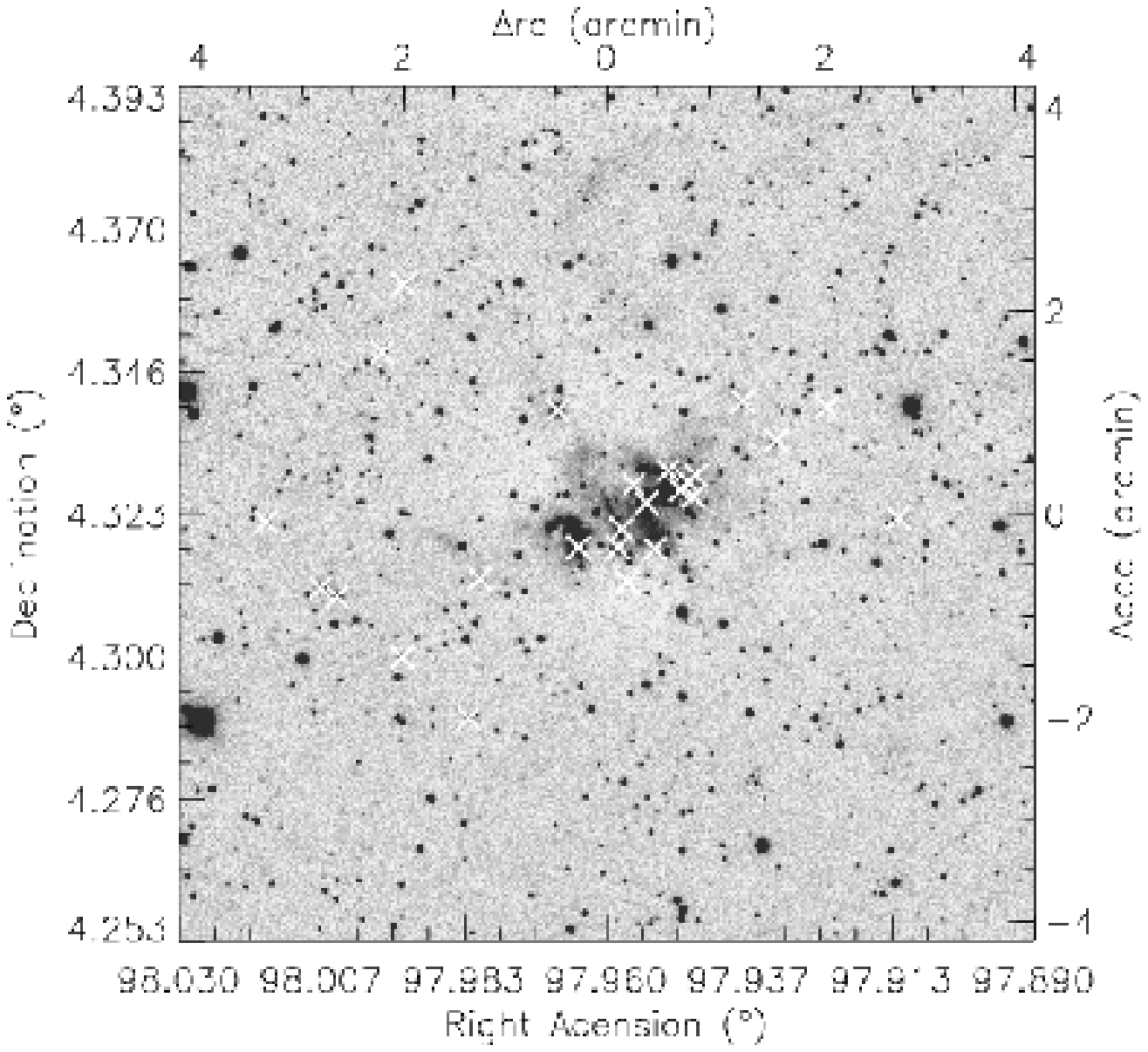}\hspace{1cm}\includegraphics[width=7cm]{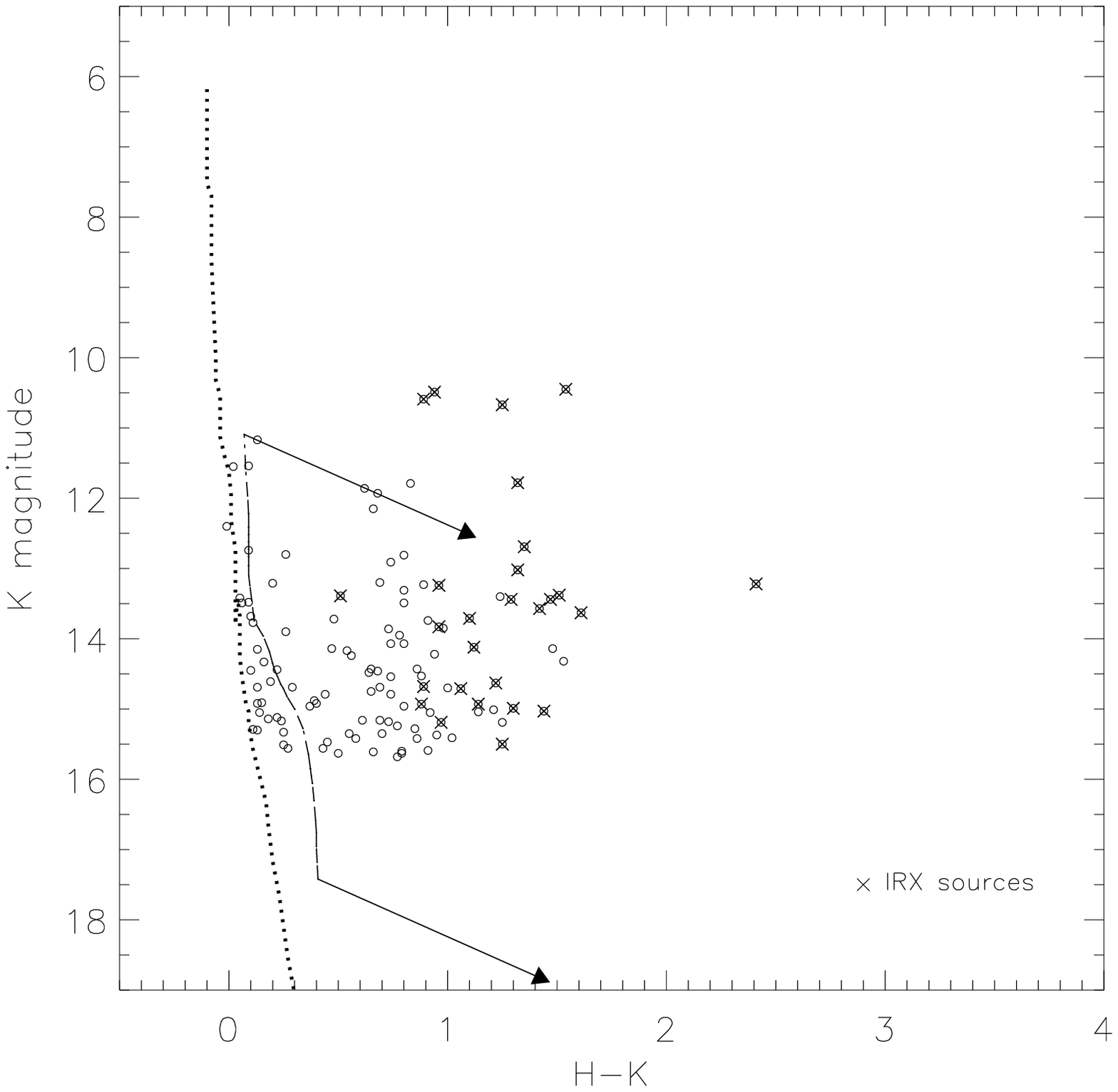}
\includegraphics[width=7cm]{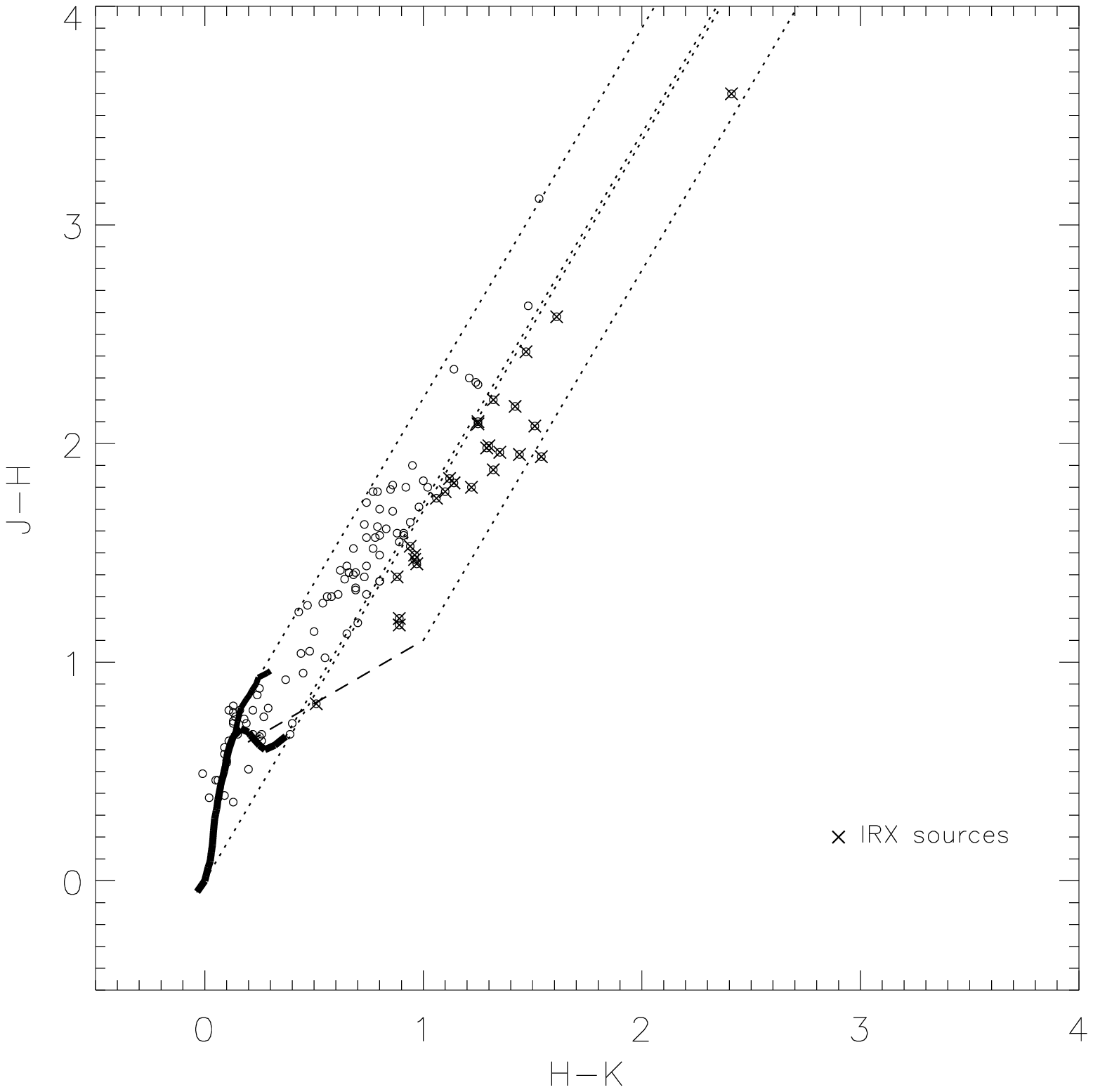}\hspace{1cm}\includegraphics[width=7cm]{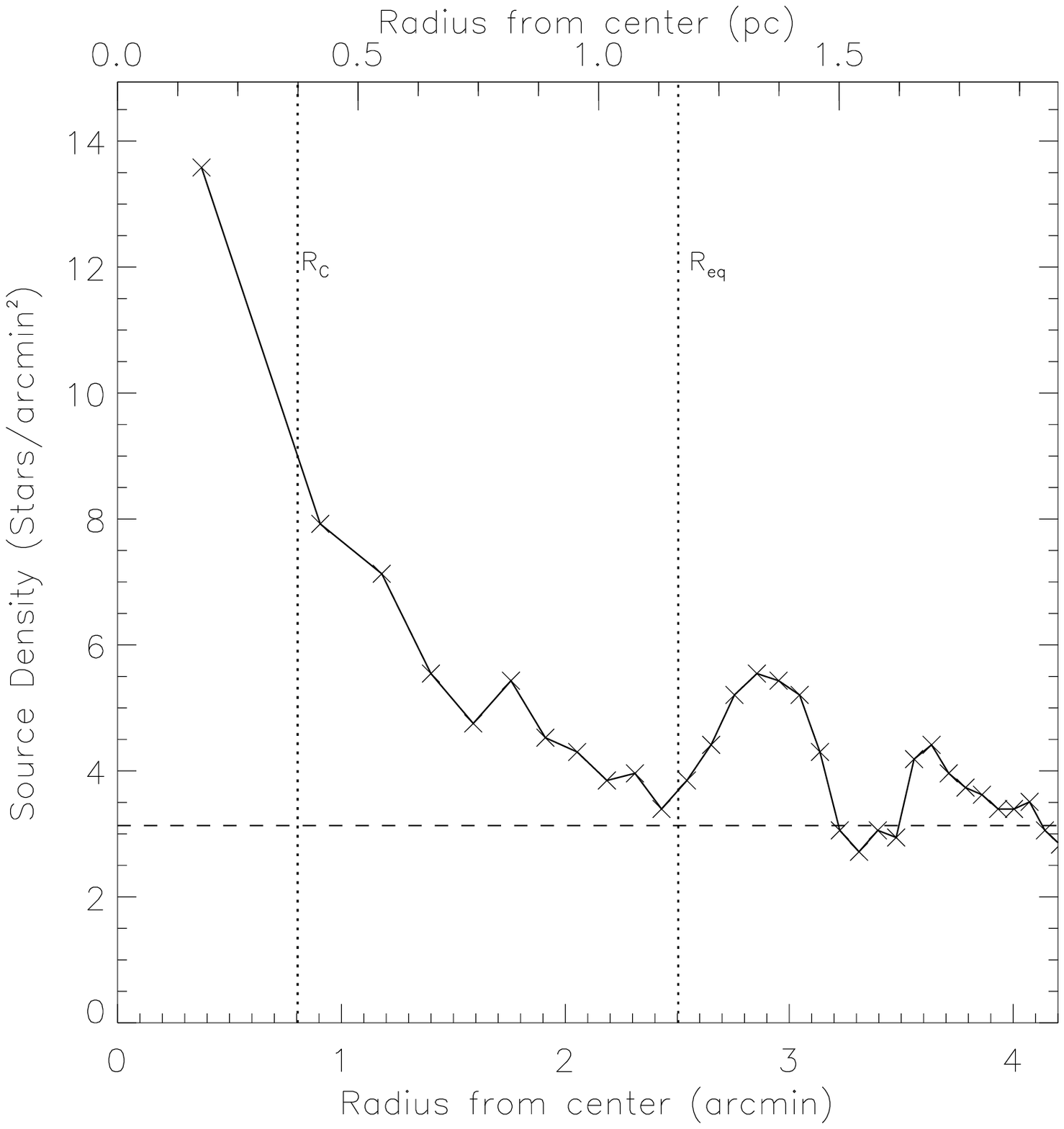}

\caption{{\it Top Left:} K band image (white symbols indicate
locations of NIRX sources), {\it Top Right:} control magnitude
diagram, {\it Bottom Left:} color color diagram and {\it Bottom Right:} Radial
Density Profile for the area corresponding to cluster PL01. See text for
explanation.} \label{fig:pl01panels} 
\end{figure}

\begin{figure}

\includegraphics[width=7cm]{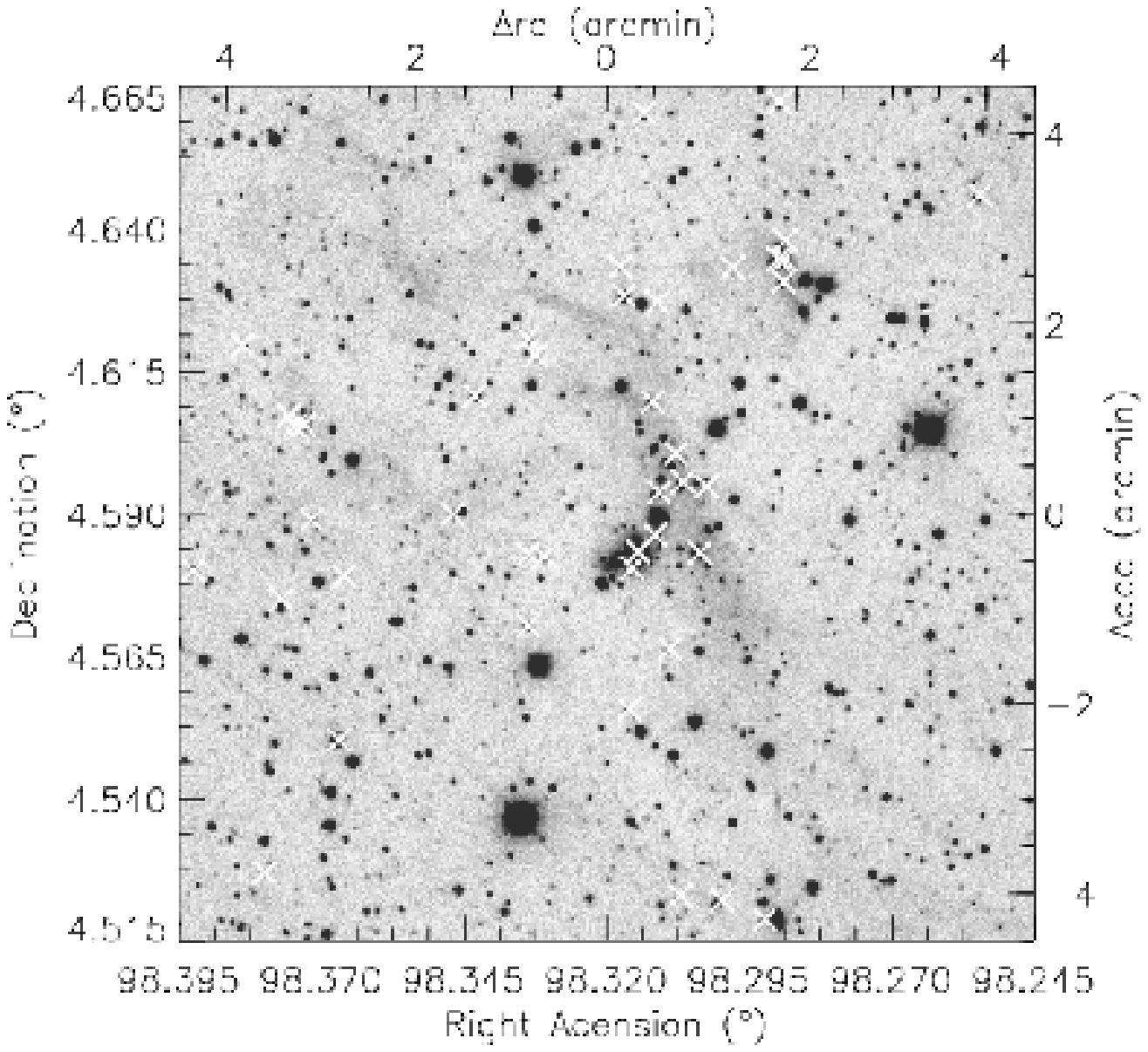}\hspace{1cm}\includegraphics[width=7cm]{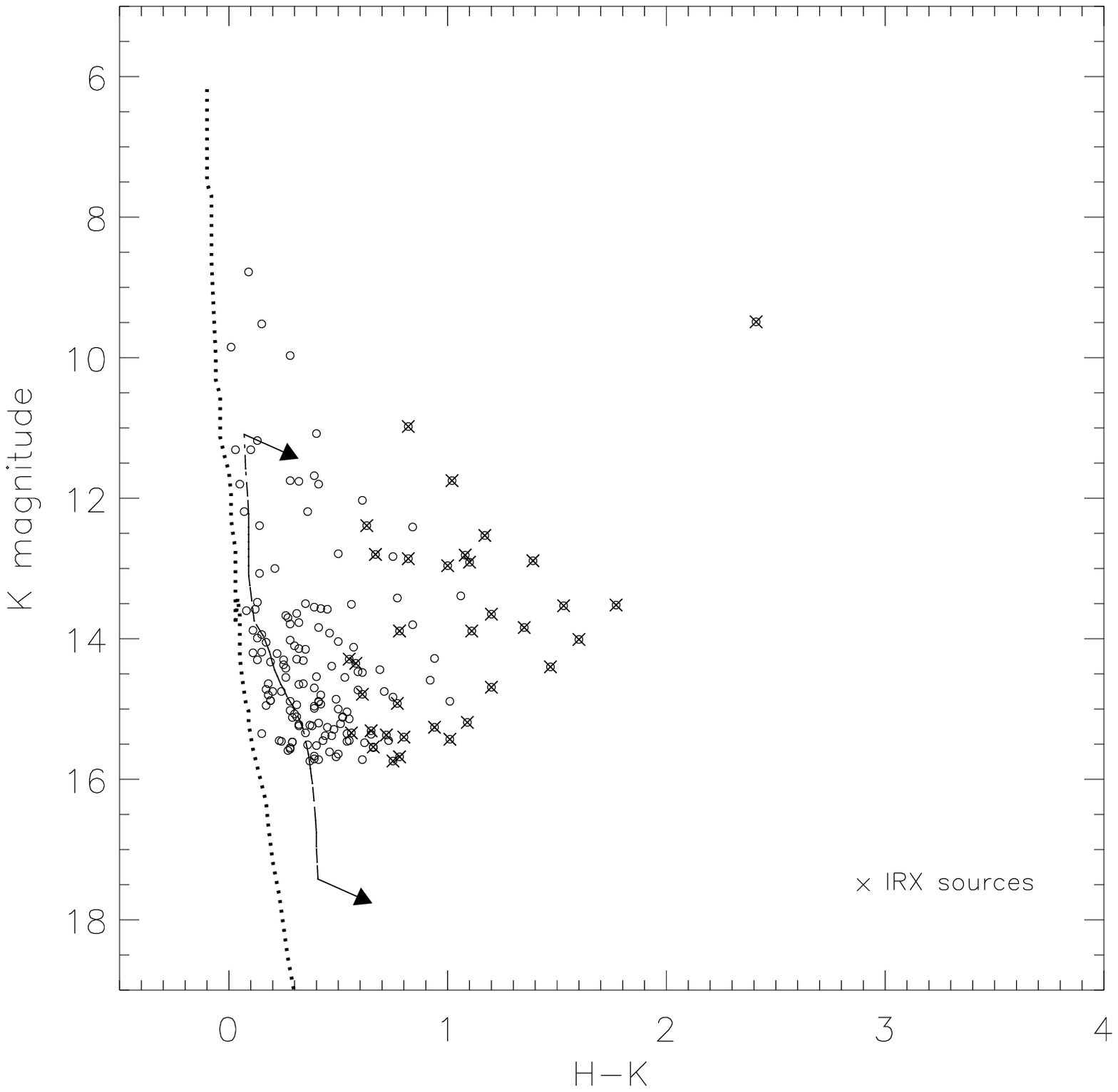}
\includegraphics[width=7cm]{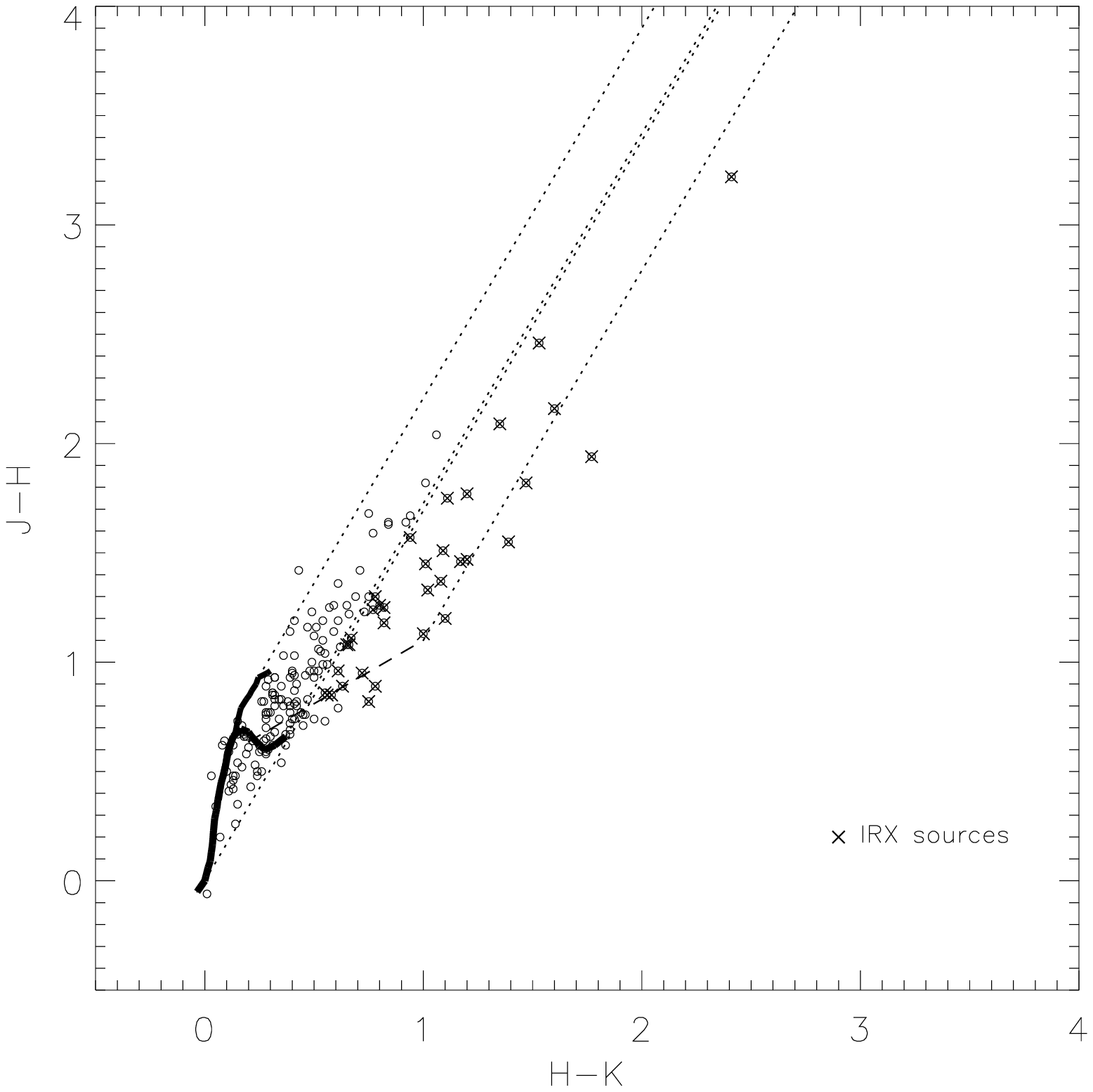}\hspace{1cm}\includegraphics[width=7cm]{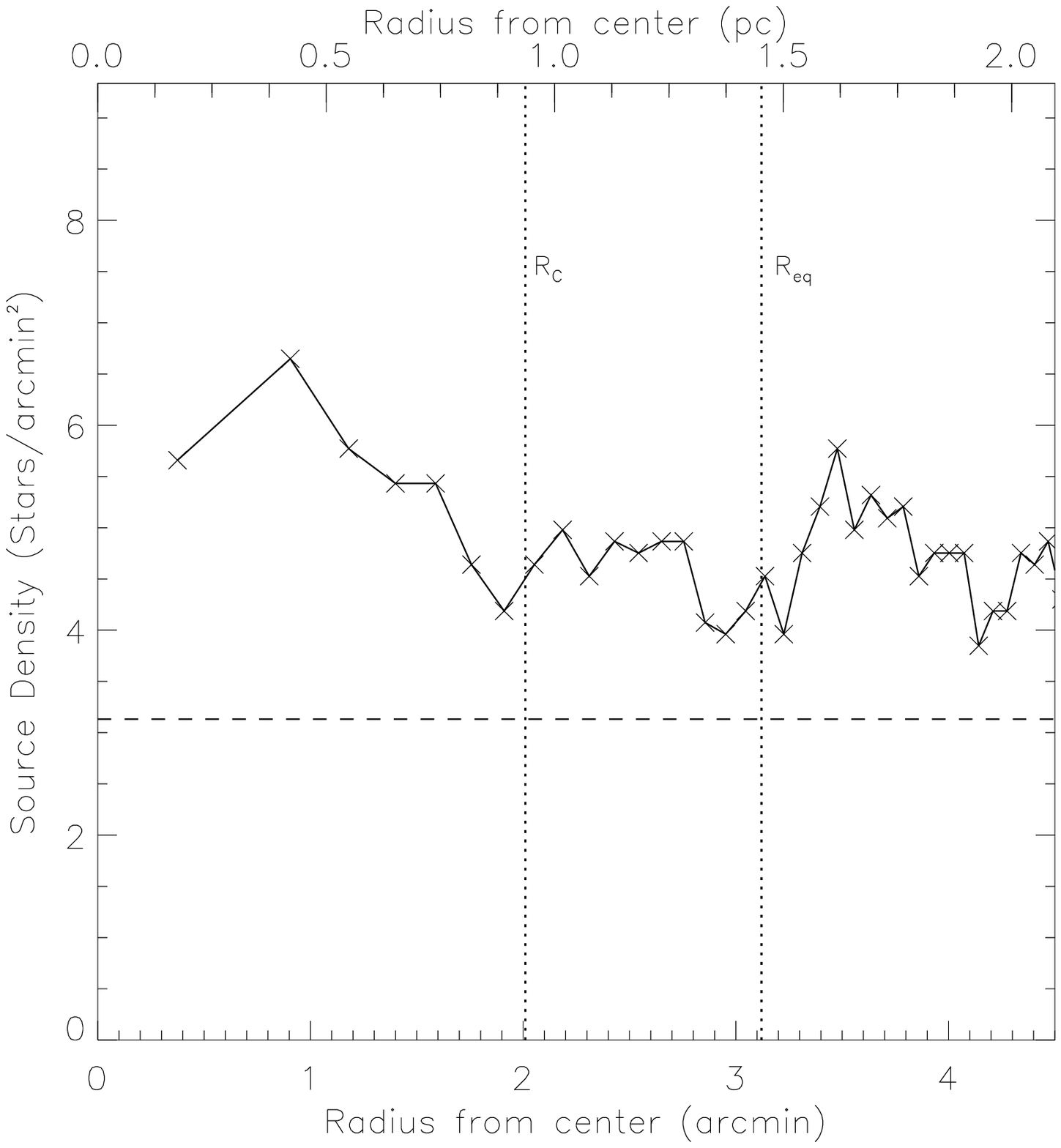}

\caption{Same as Figure \ref{fig:pl01panels}, for cluster PL02.}
\label{fig:pl02panels} 
\end{figure}

\begin{figure}

\includegraphics[width=7cm]{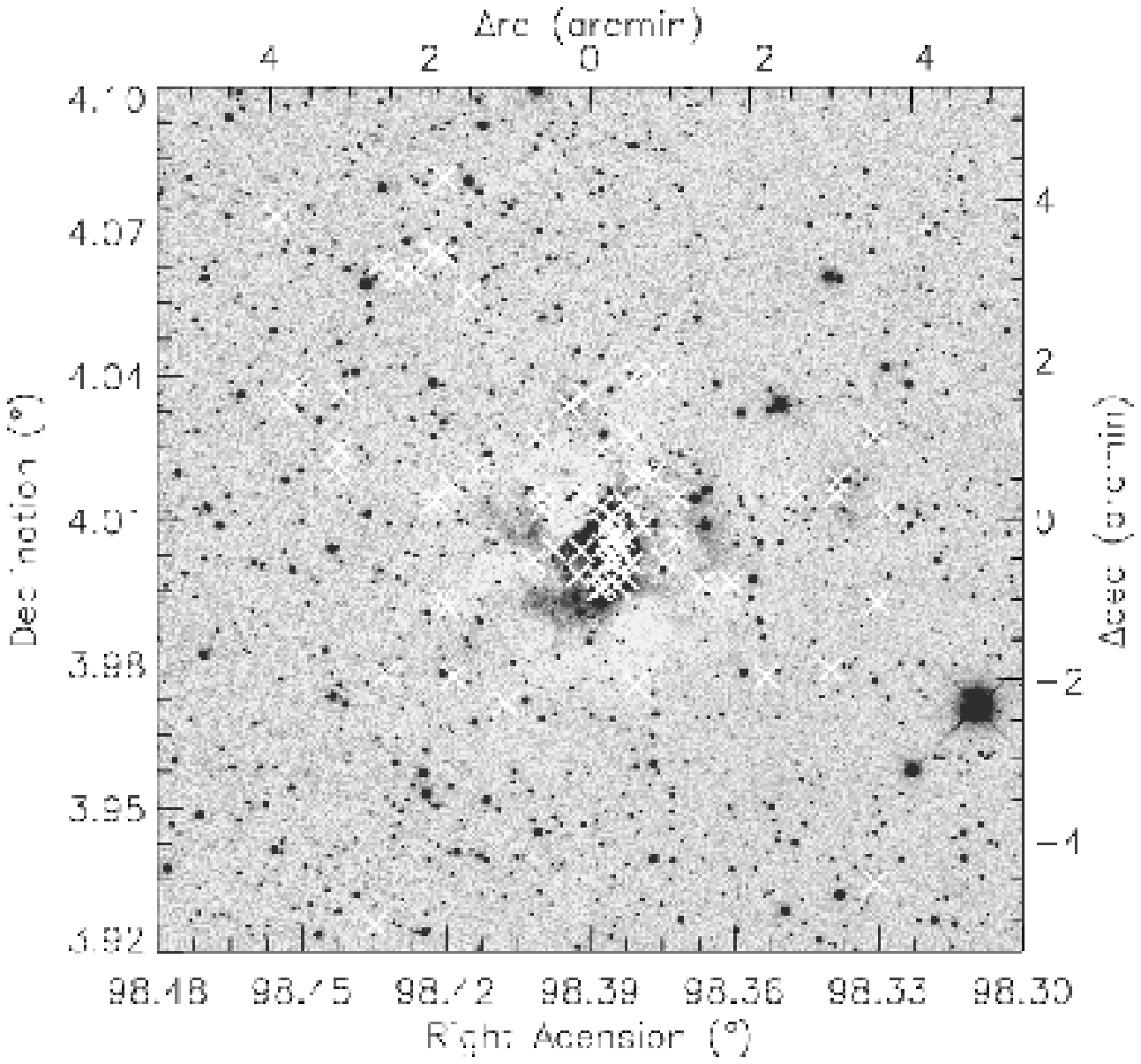}\hspace{1cm}\includegraphics[width=7cm]{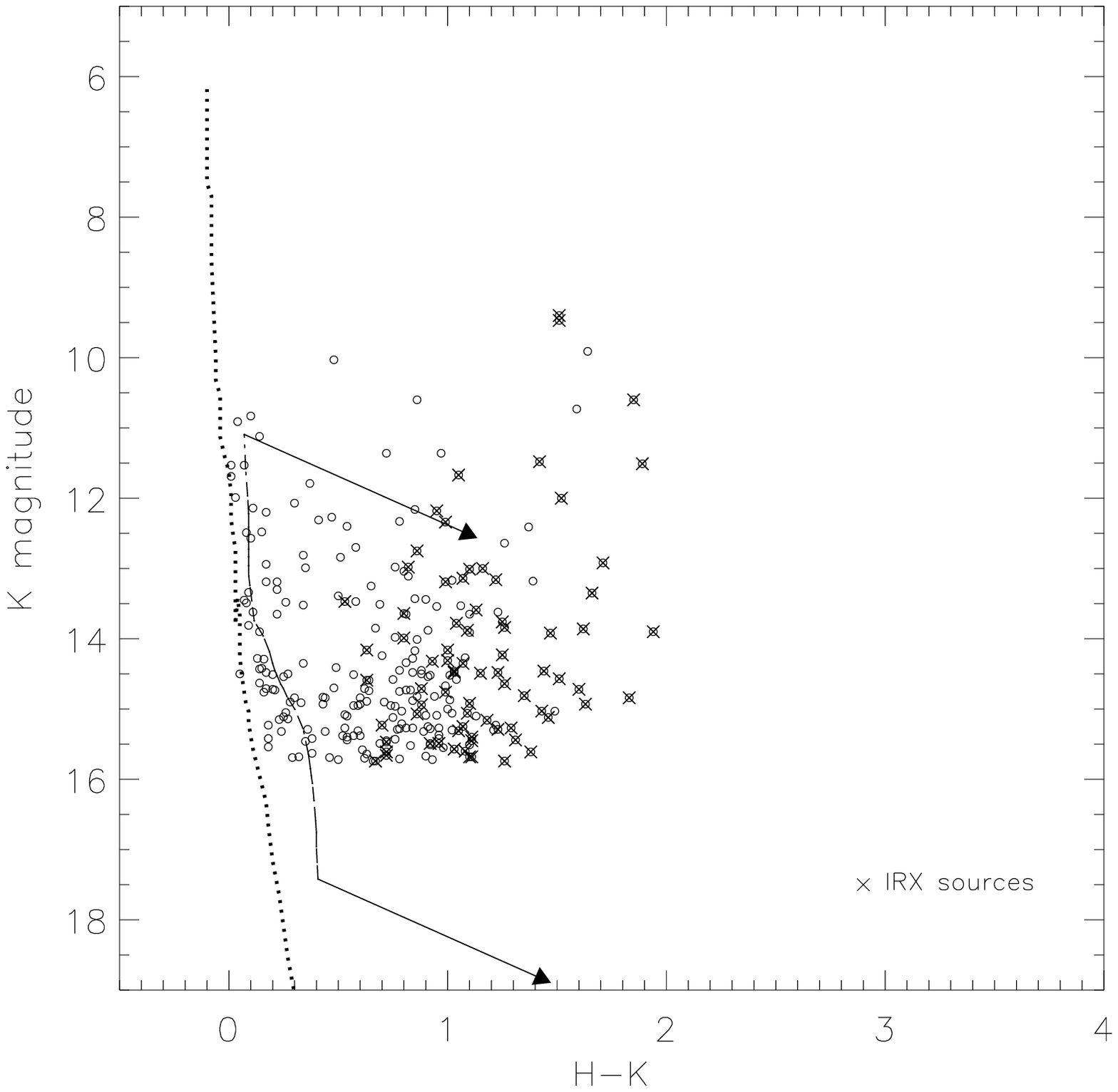}
\includegraphics[width=7cm]{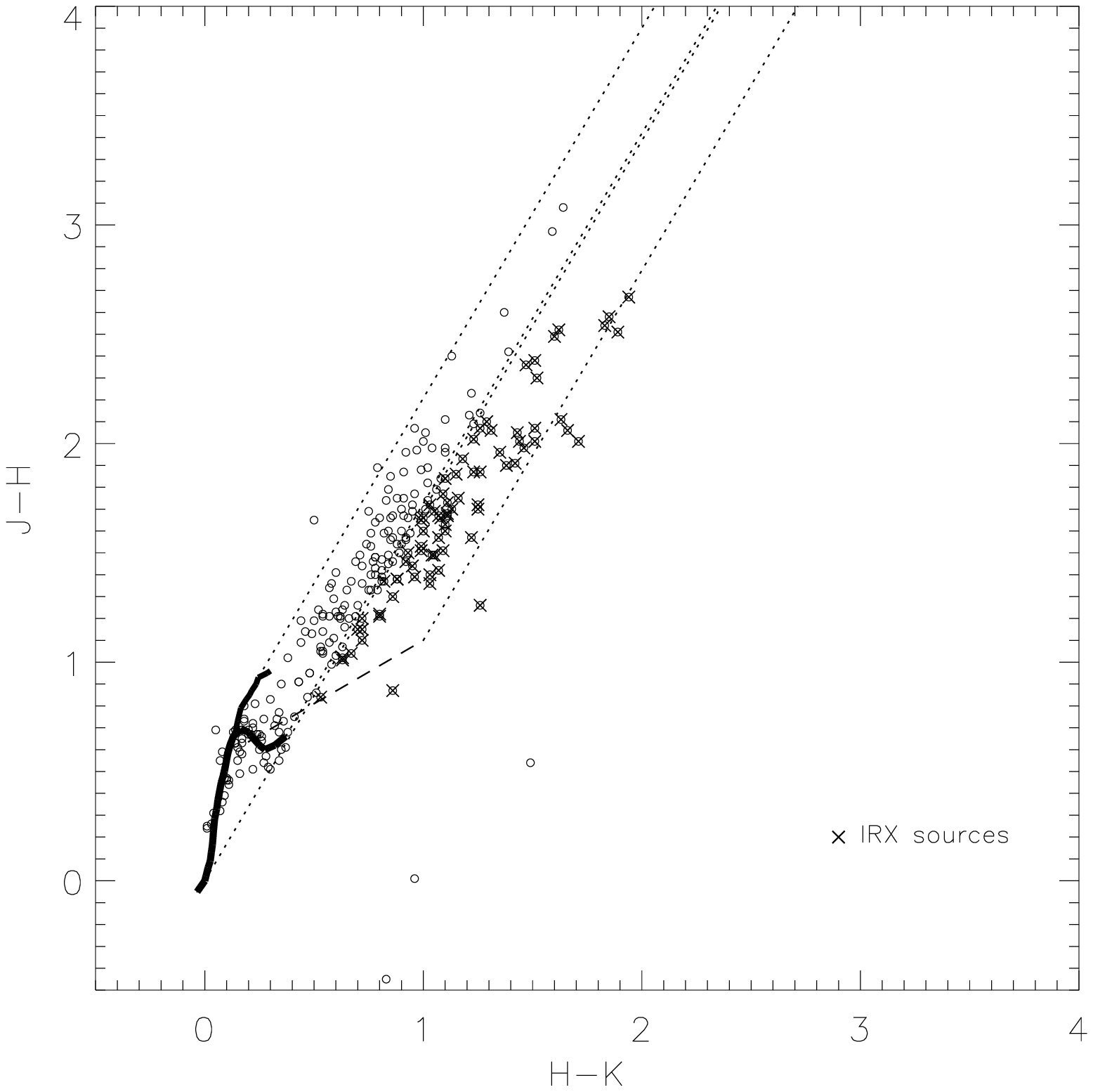}\hspace{1cm}\includegraphics[width=7cm]{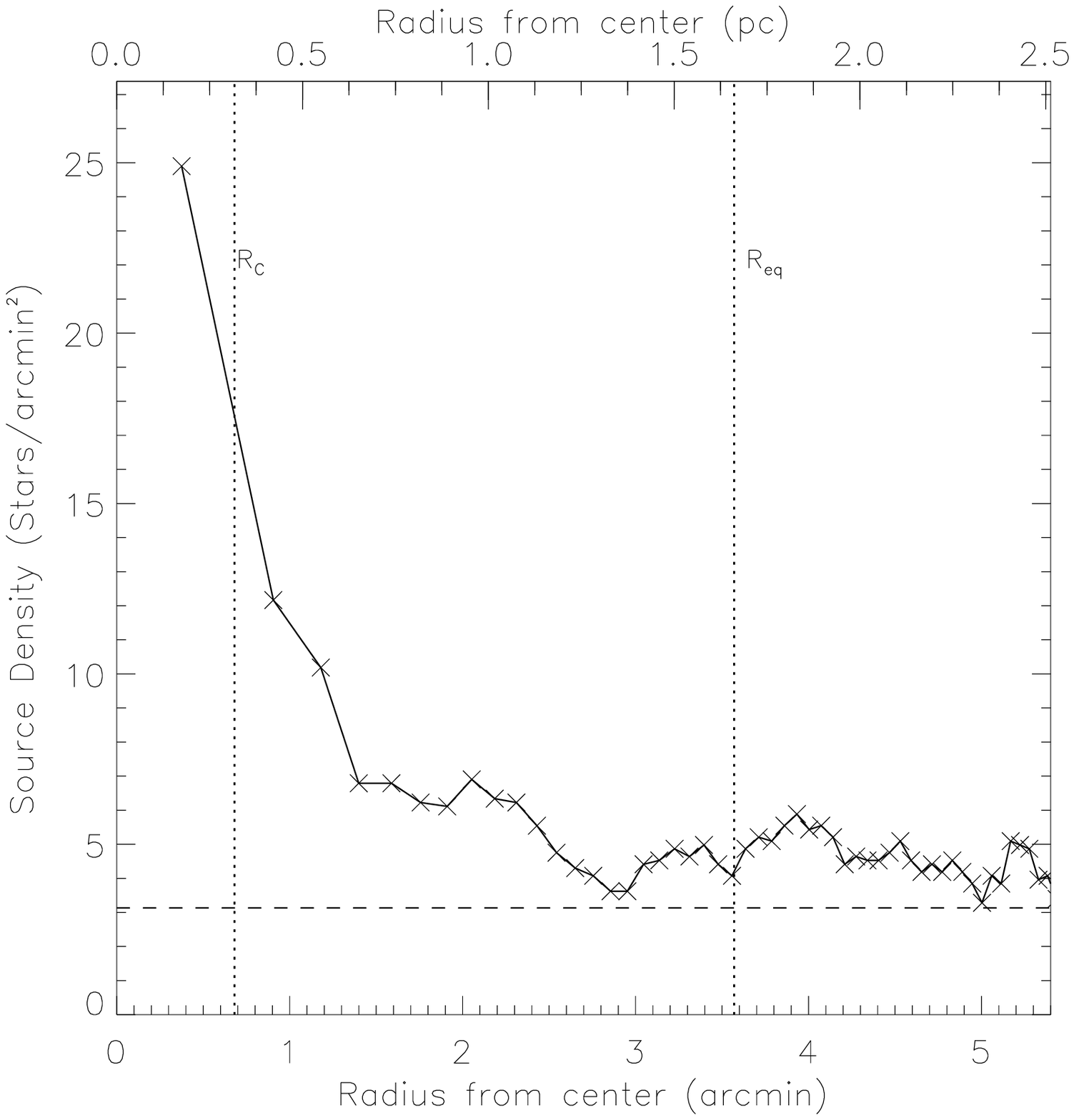}

\caption{Same as Figure \ref{fig:pl01panels}, for cluster PL03.}
\label{fig:pl03panels} 
\end{figure}

\begin{figure}

\includegraphics[width=7cm]{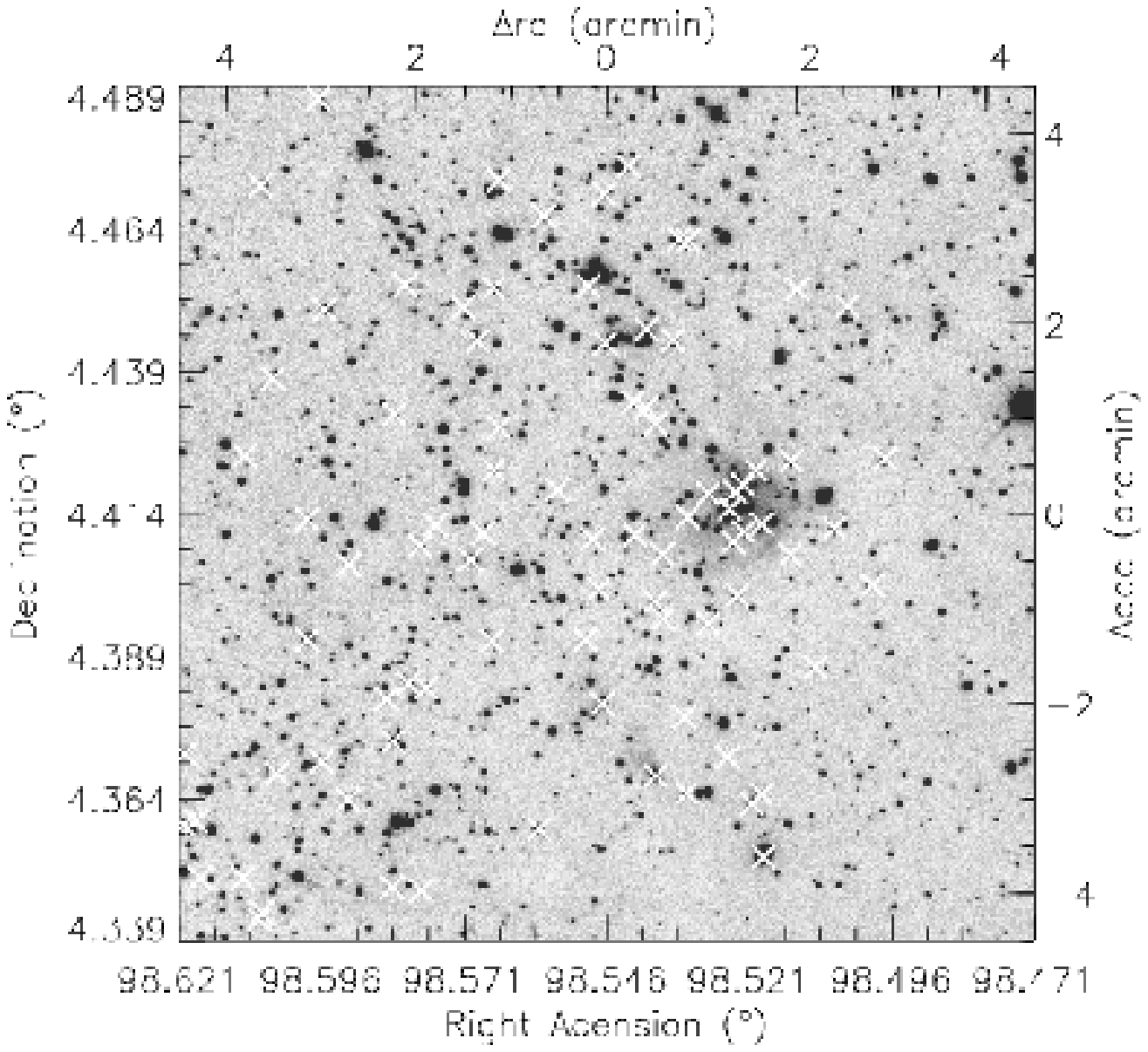}\hspace{1cm}\includegraphics[width=7cm]{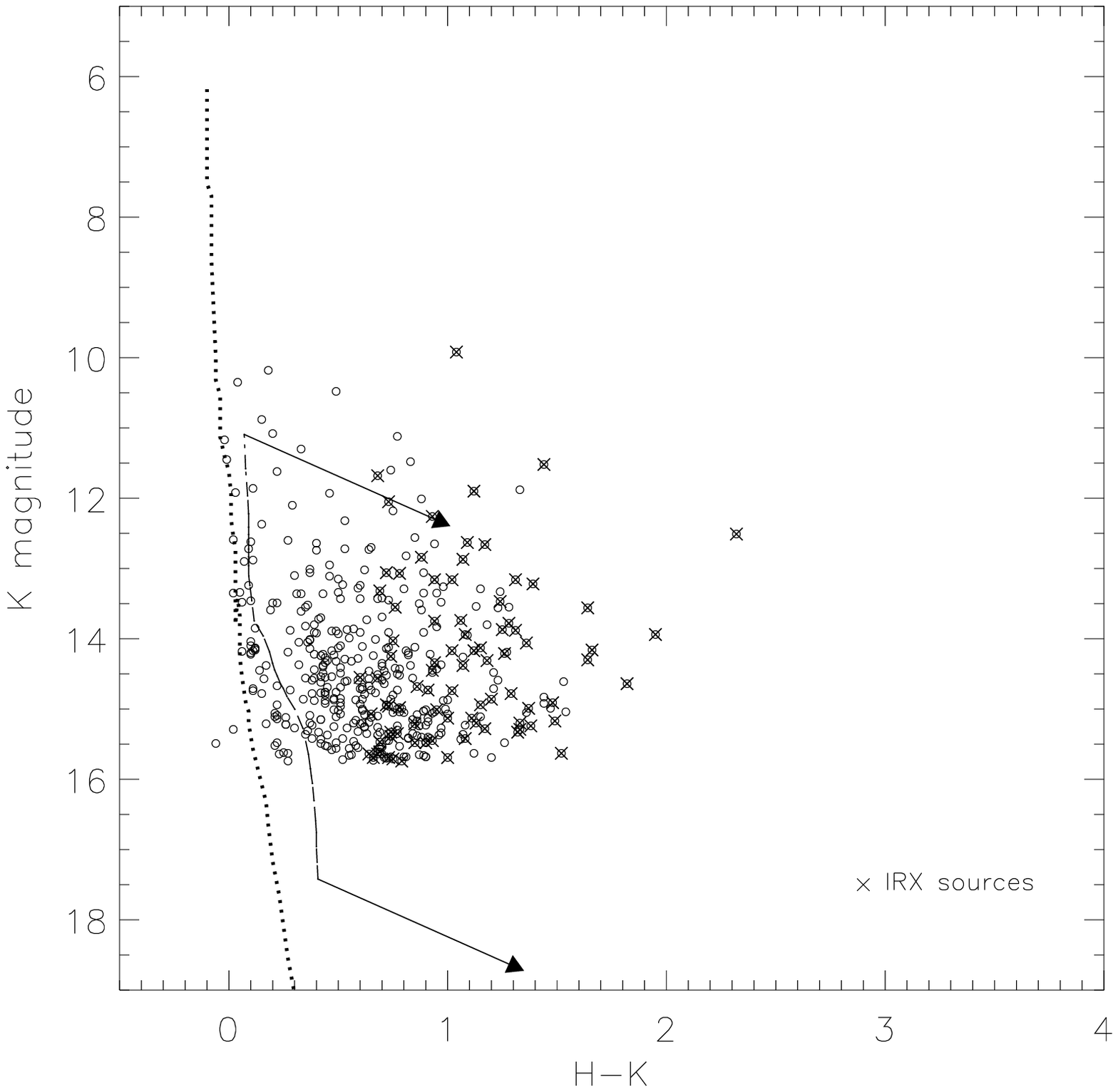}

\includegraphics[width=7cm]{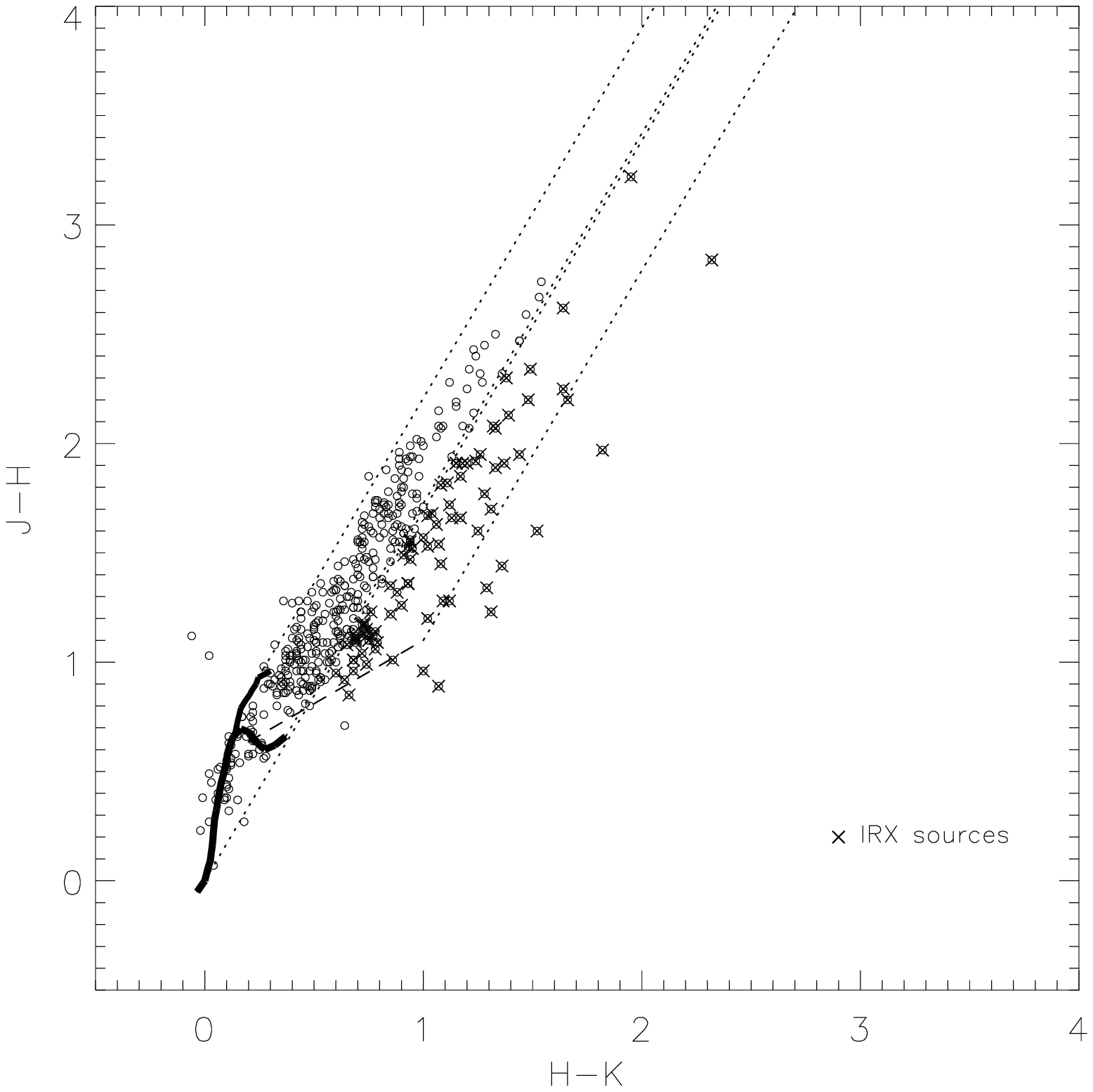}\hspace{1cm}\includegraphics[width=7cm]{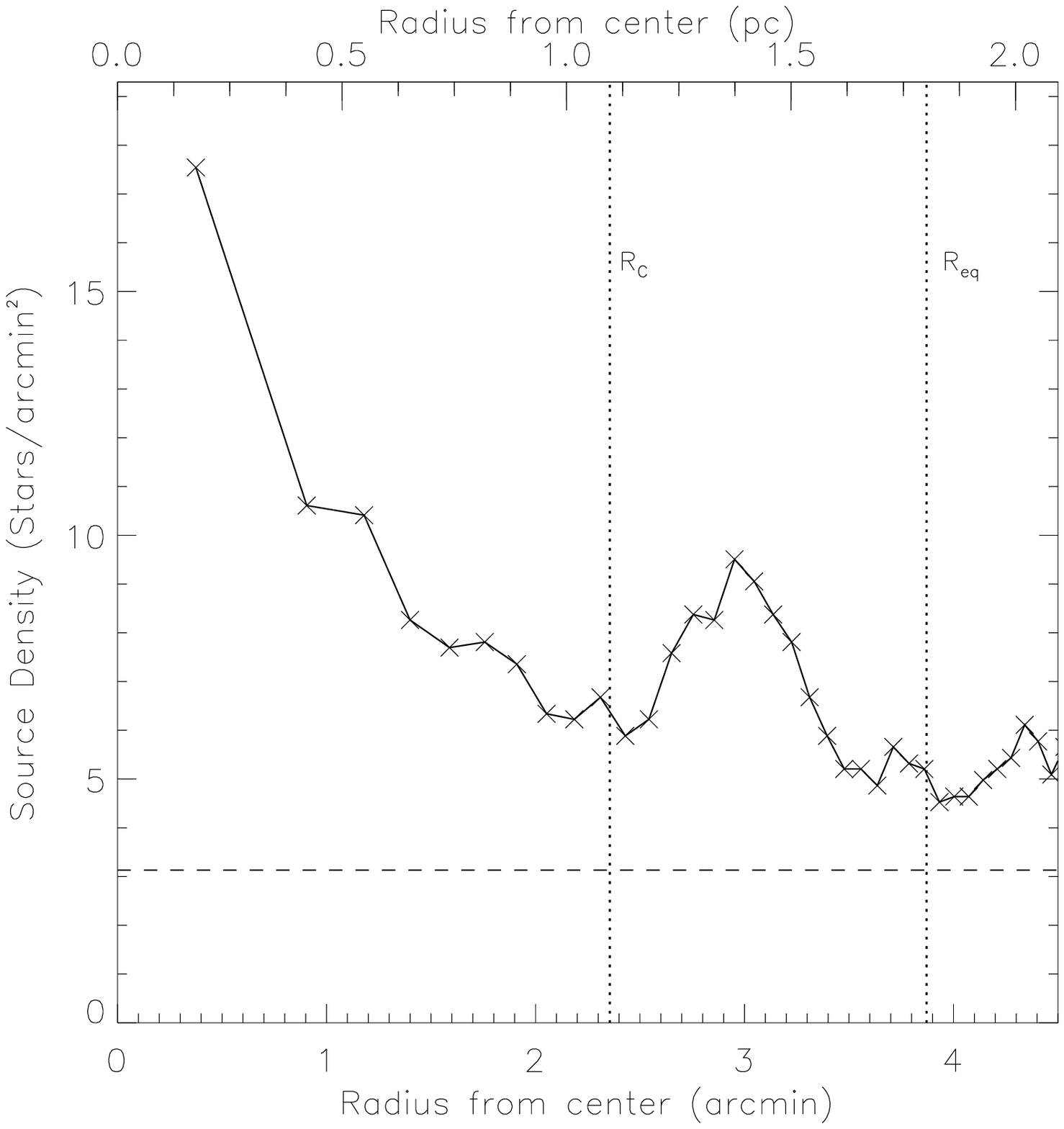}

\caption{Same as Figure \ref{fig:pl01panels}, for cluster PL04.}
\label{fig:pl04panels} 
\end{figure}

\begin{figure}

\includegraphics[width=7cm]{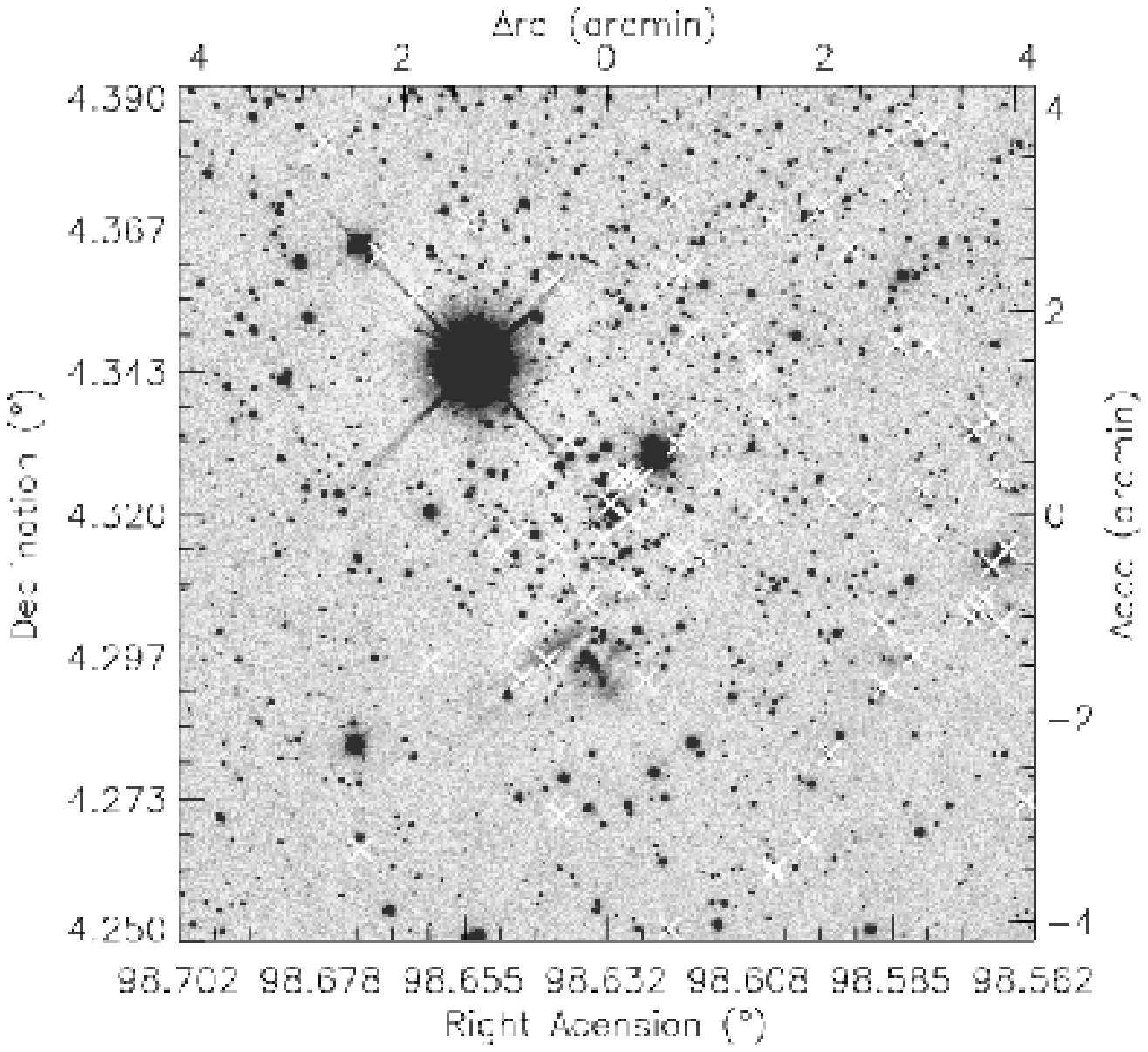}\hspace{1cm}\includegraphics[width=7cm]{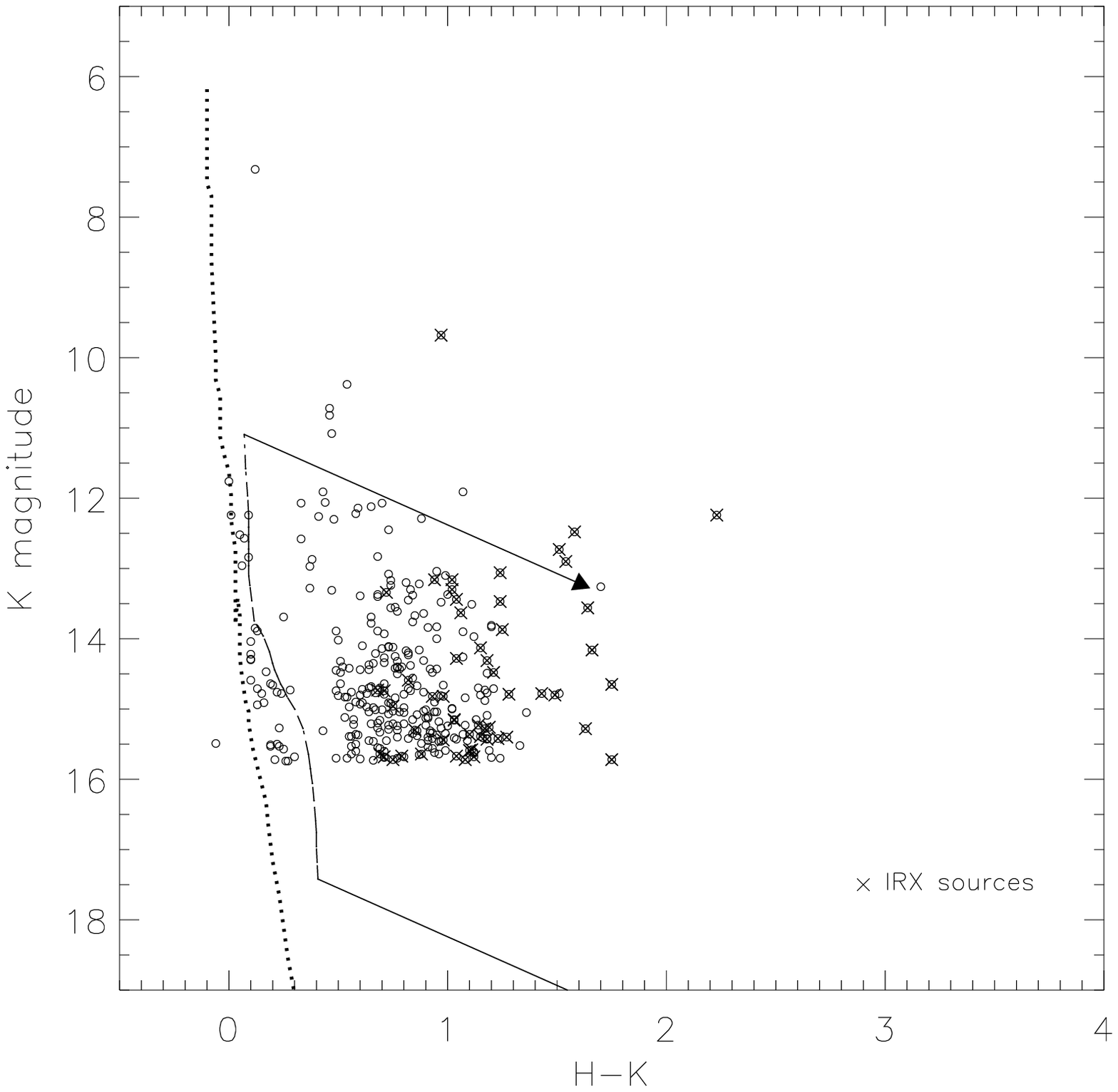}

\includegraphics[width=7cm]{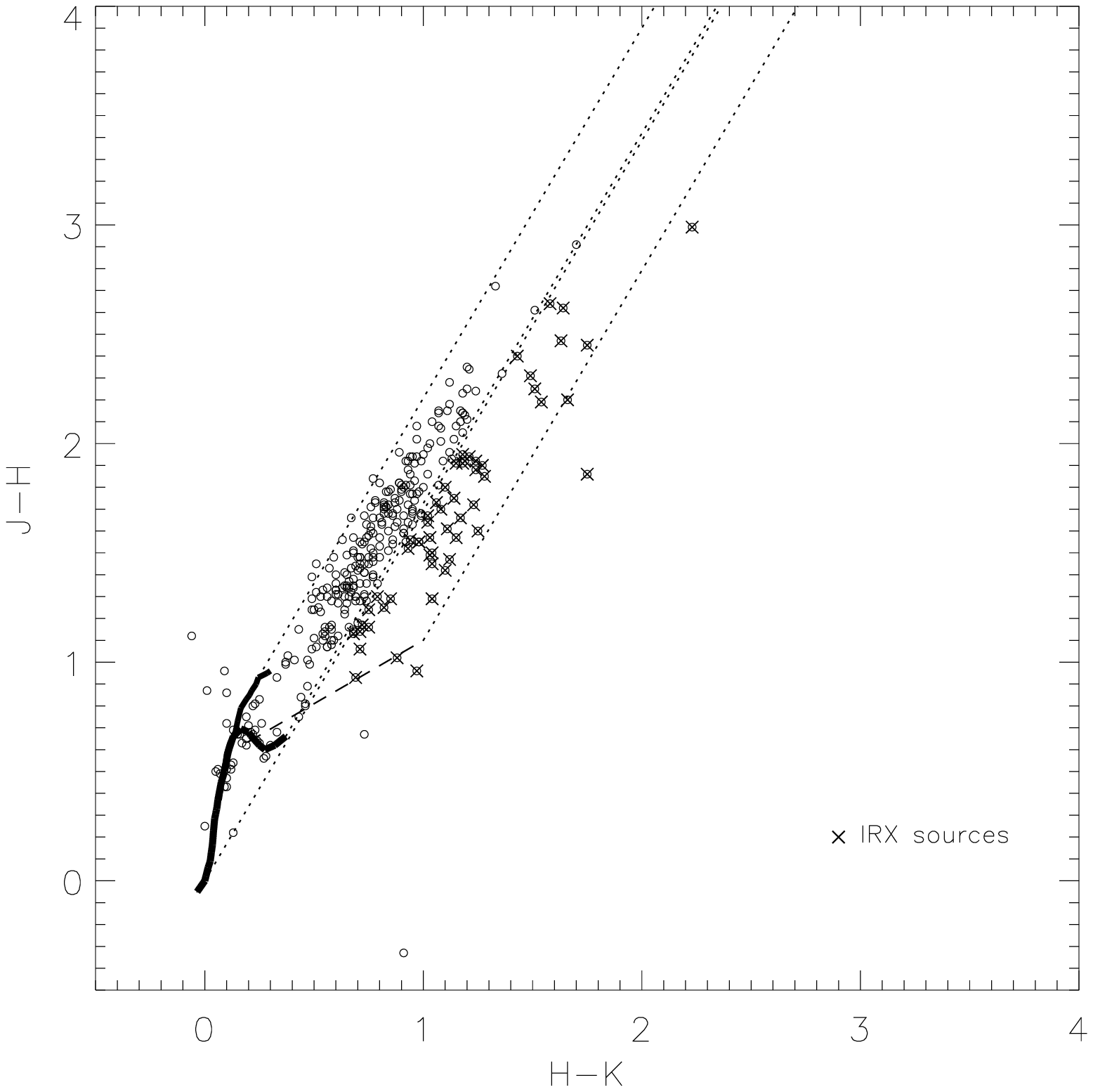}\hspace{1cm}\includegraphics[width=7cm]{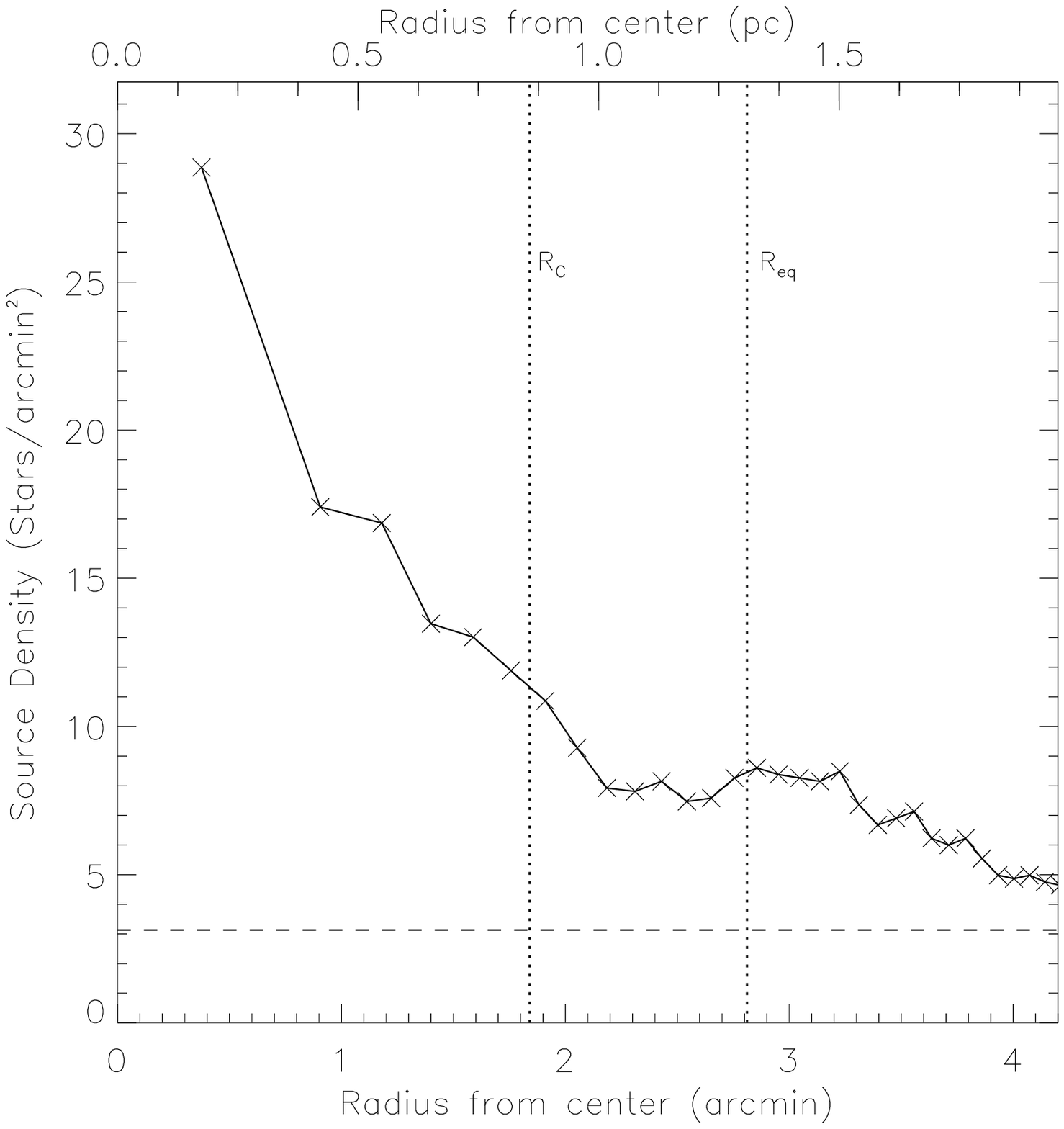}

\caption{Same as Figure \ref{fig:pl01panels}, for cluster PL05.}
\label{fig:pl05panels} 
\end{figure}

\begin{figure}

\includegraphics[width=7cm]{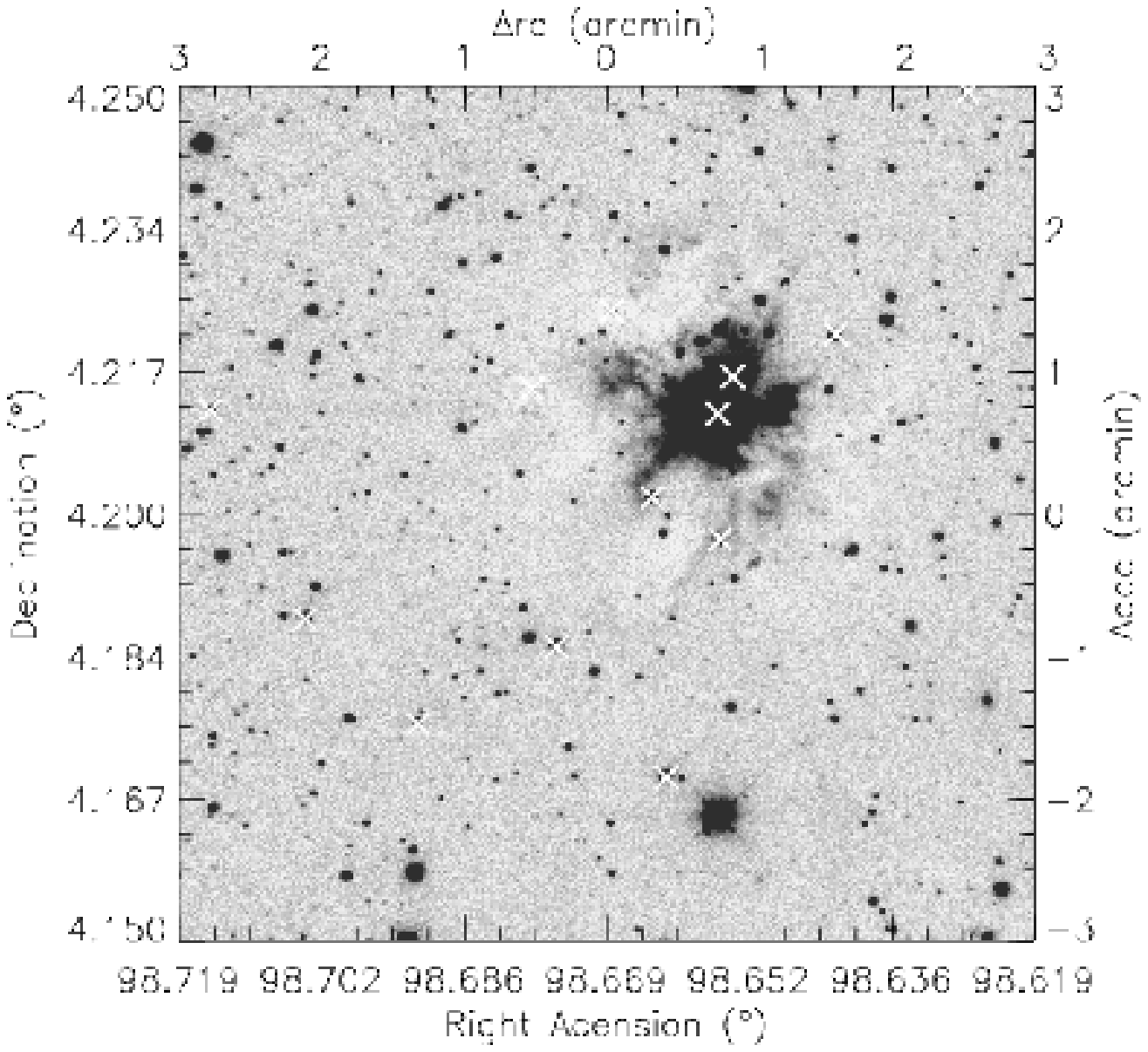}\hspace{1cm}\includegraphics[width=7cm]{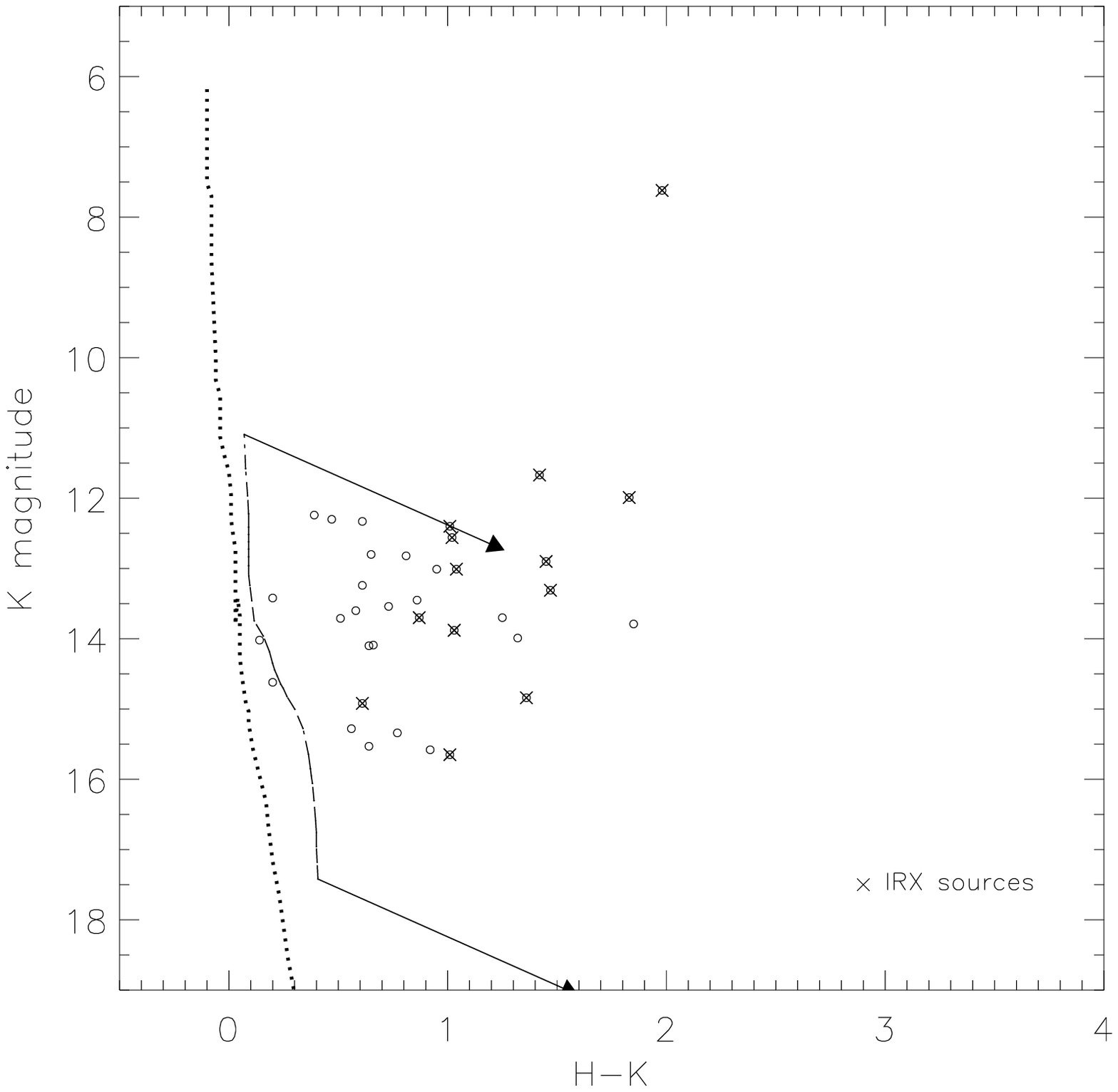}

\includegraphics[width=7cm]{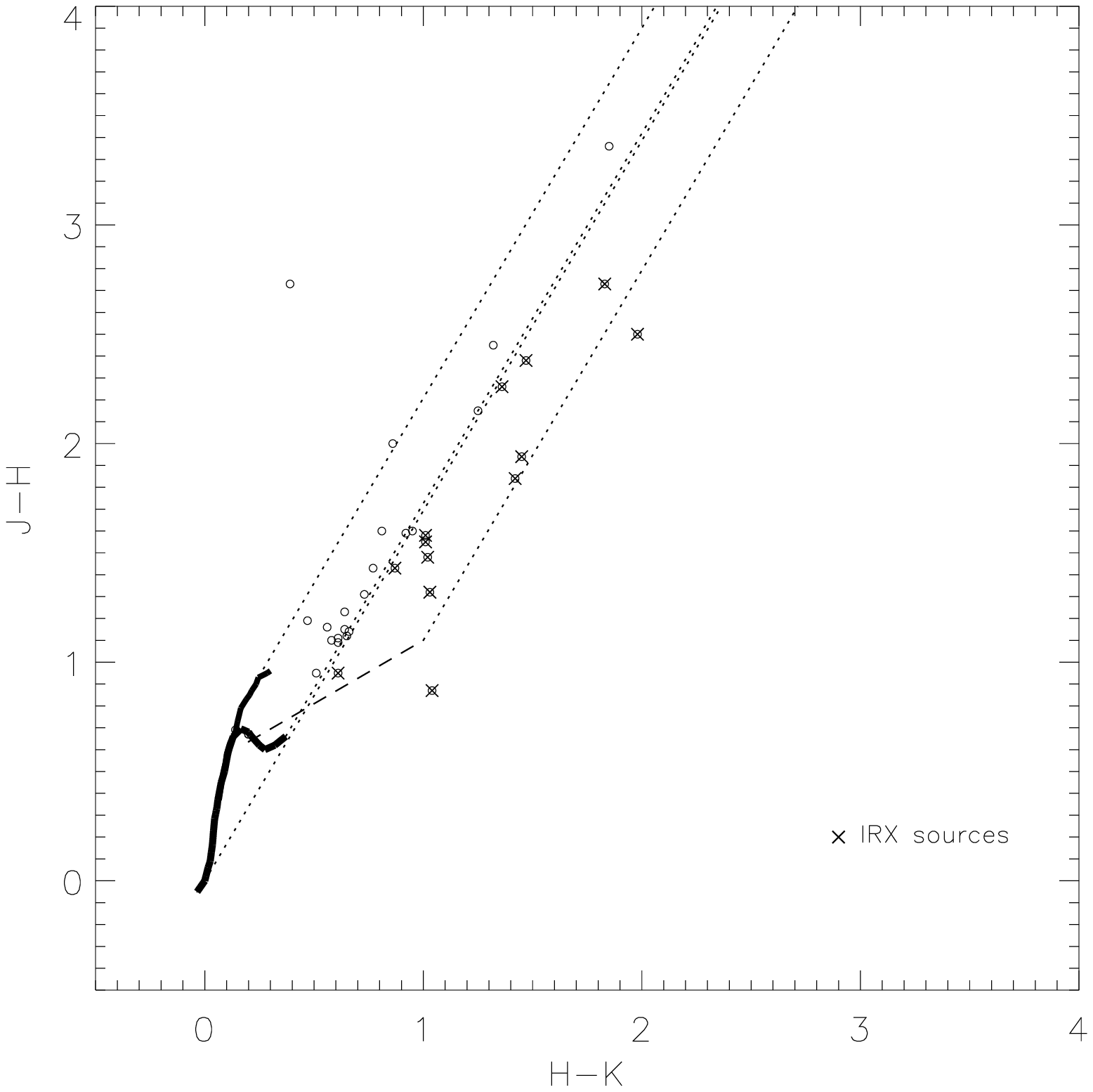}\hspace{1cm}\includegraphics[width=7cm]{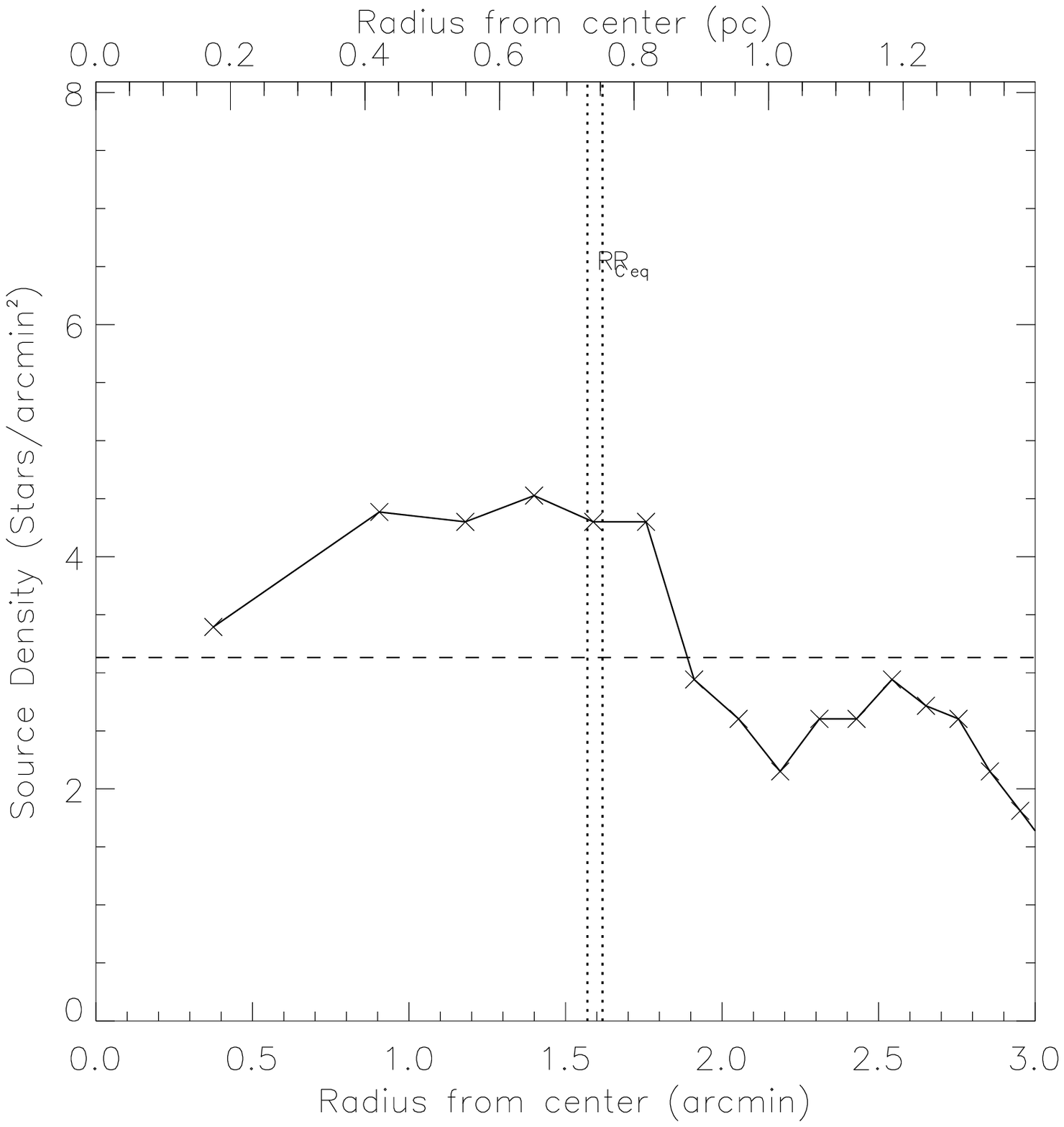}

\caption{Same as Figure \ref{fig:pl01panels}, for cluster PL06.}
\label{fig:pl06panels} 
\end{figure}

\begin{figure}

\includegraphics[width=7cm]{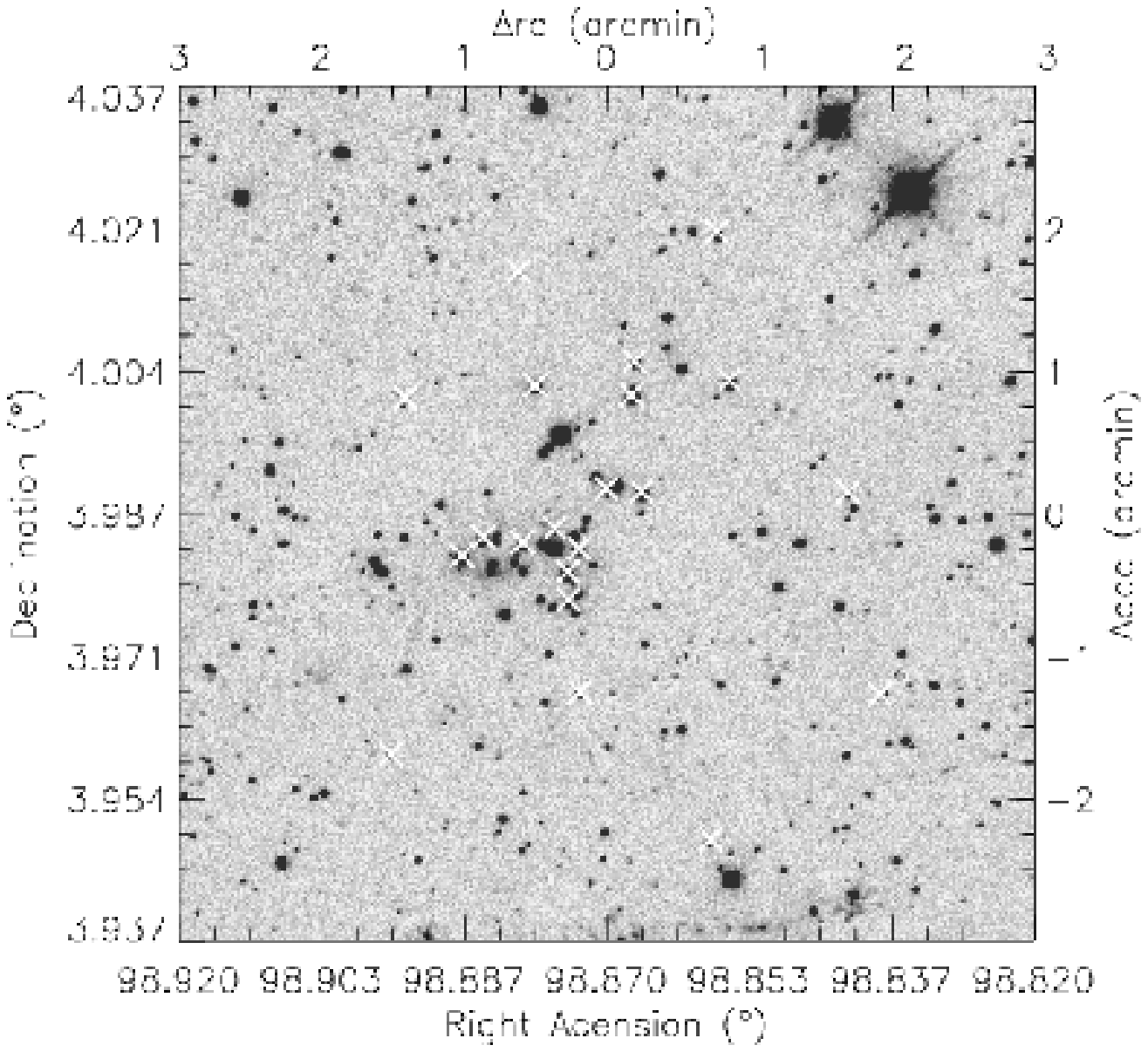}\hspace{1cm}\includegraphics[width=7cm]{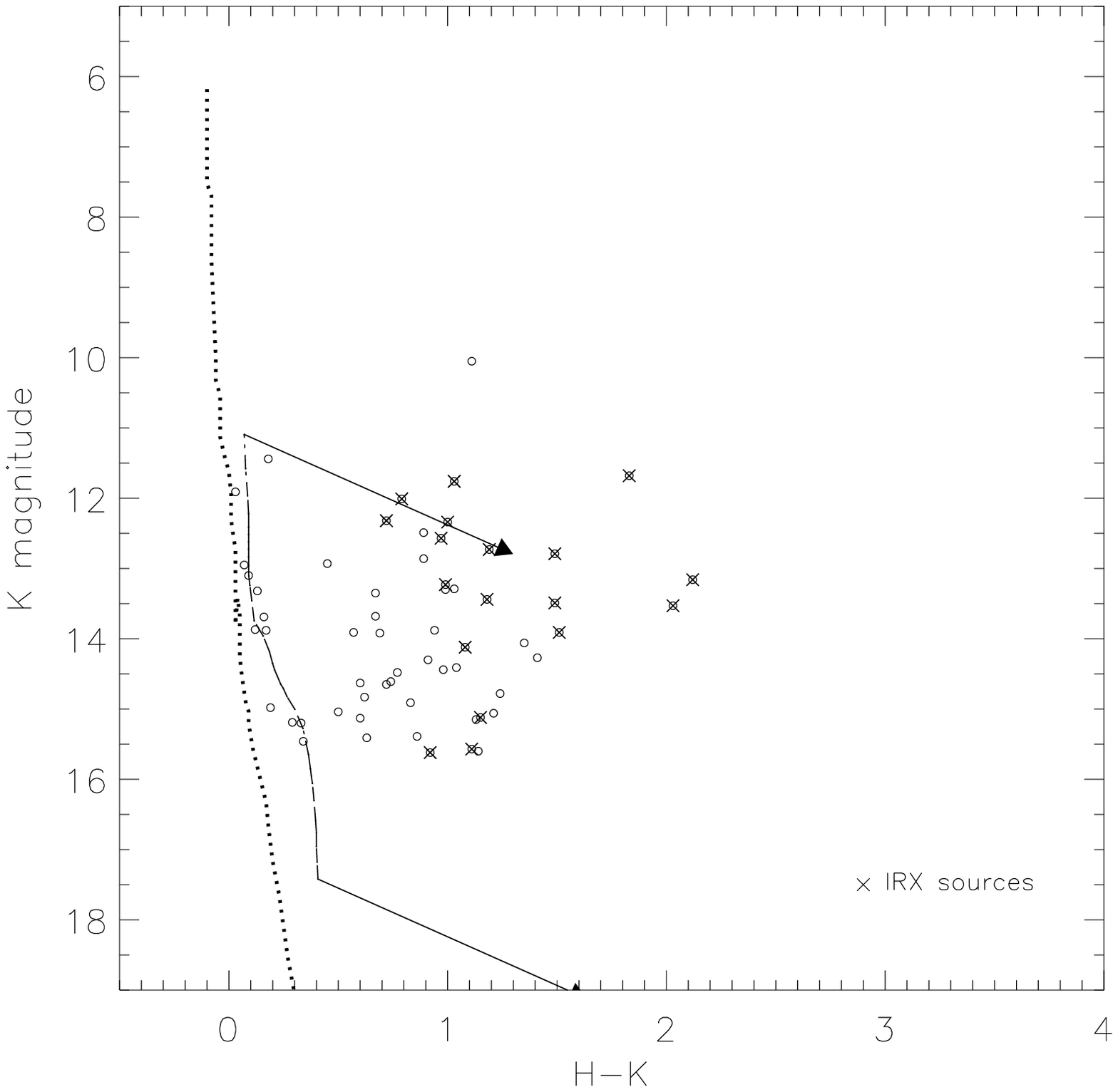}

\includegraphics[width=7cm]{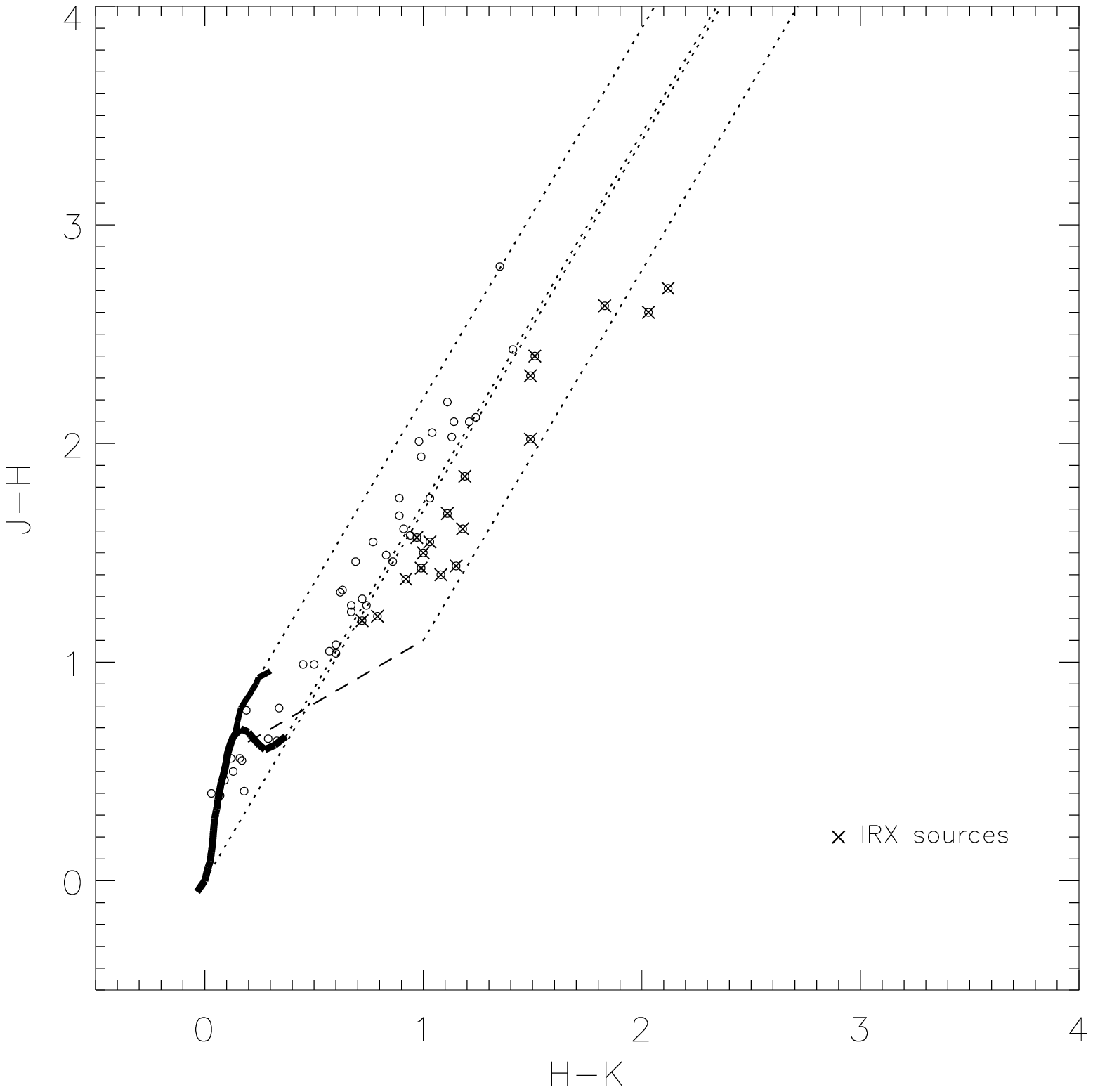}\hspace{1cm}\includegraphics[width=7cm]{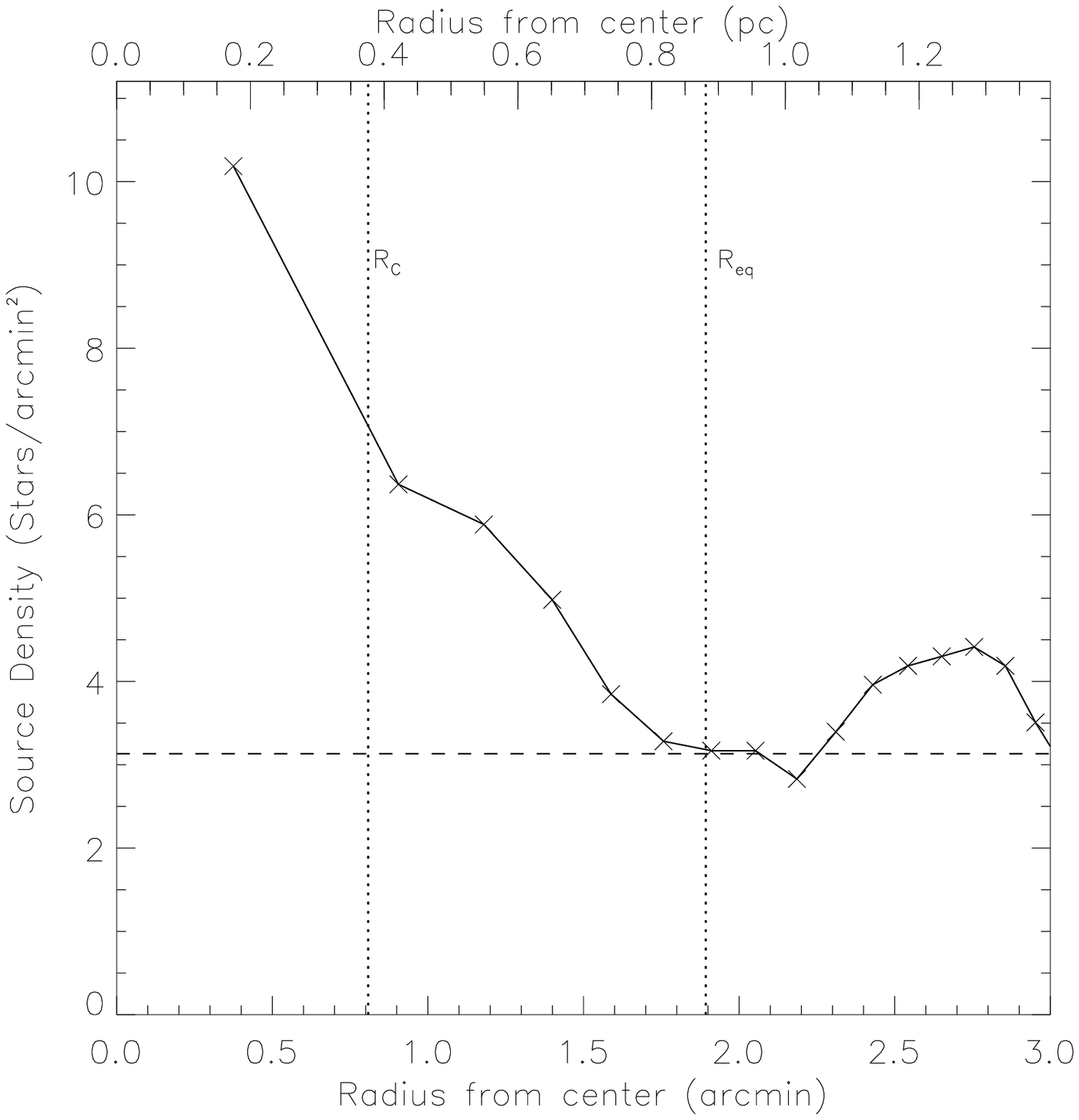}

\caption{Same as Figure \ref{fig:pl01panels}, for cluster PL07.}
\label{fig:pl07panels} 
\end{figure}

\begin{figure}

\includegraphics[width=7cm]{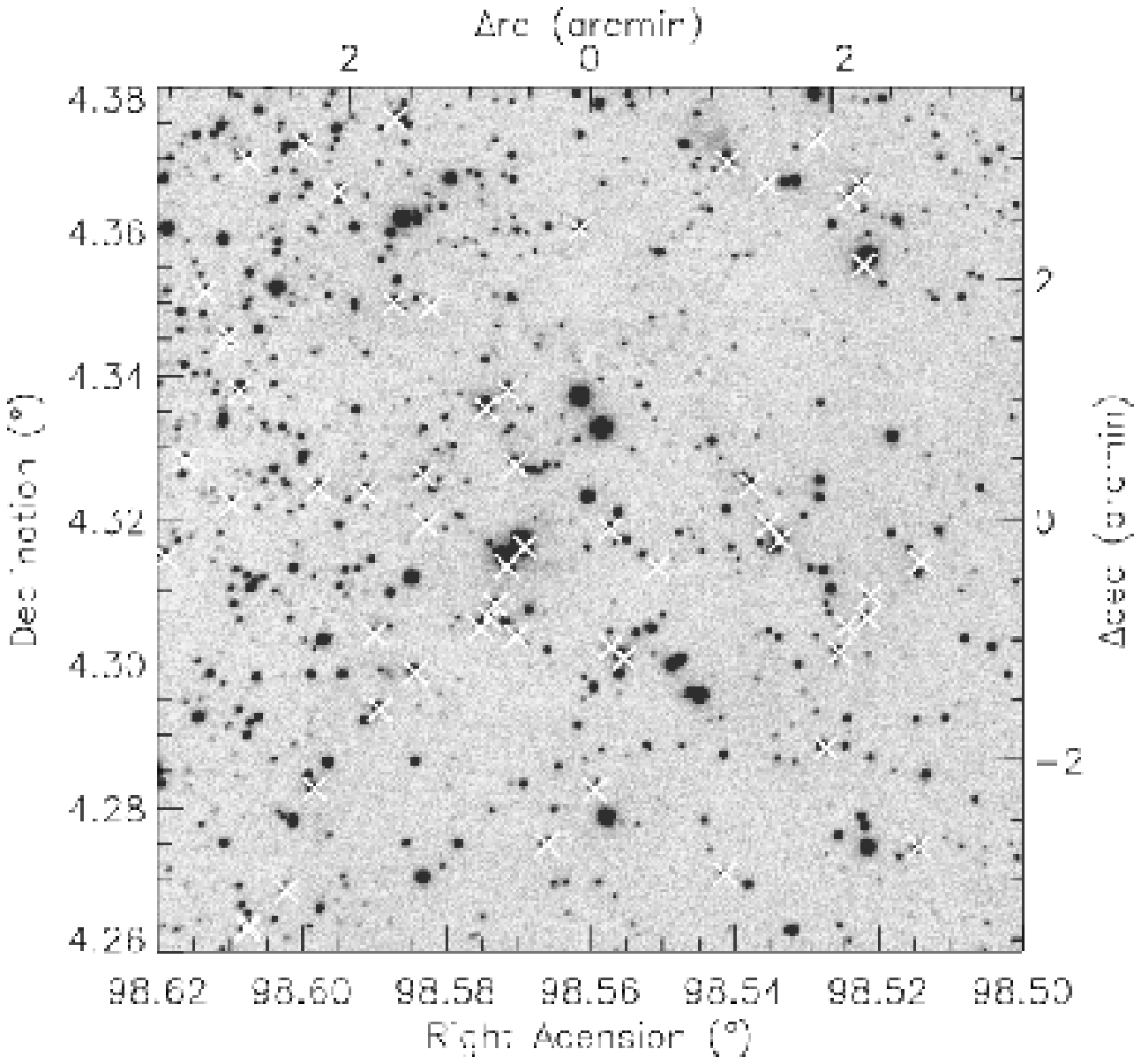}\hspace{1cm}\includegraphics[width=7cm]{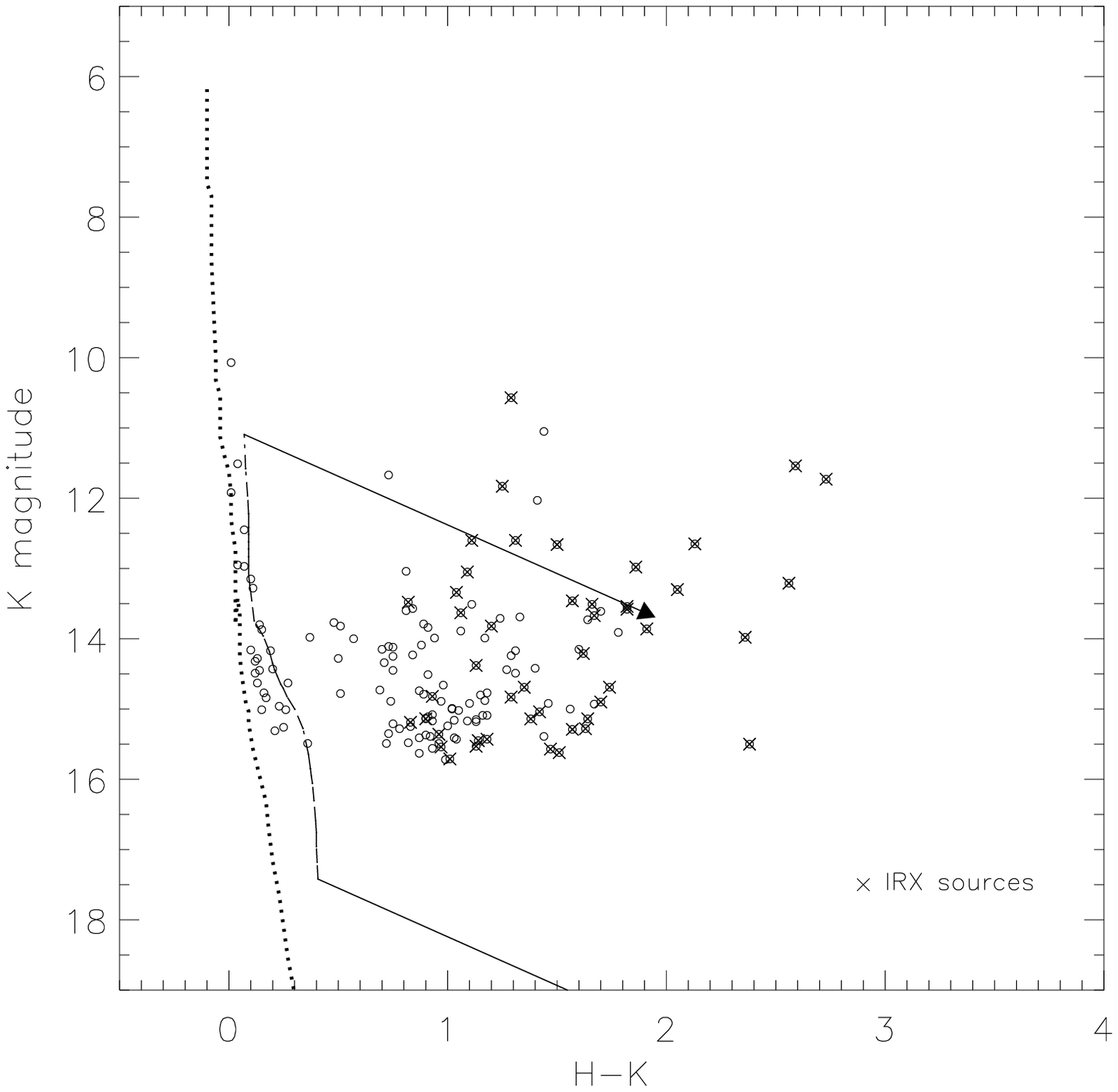}

\includegraphics[width=7cm]{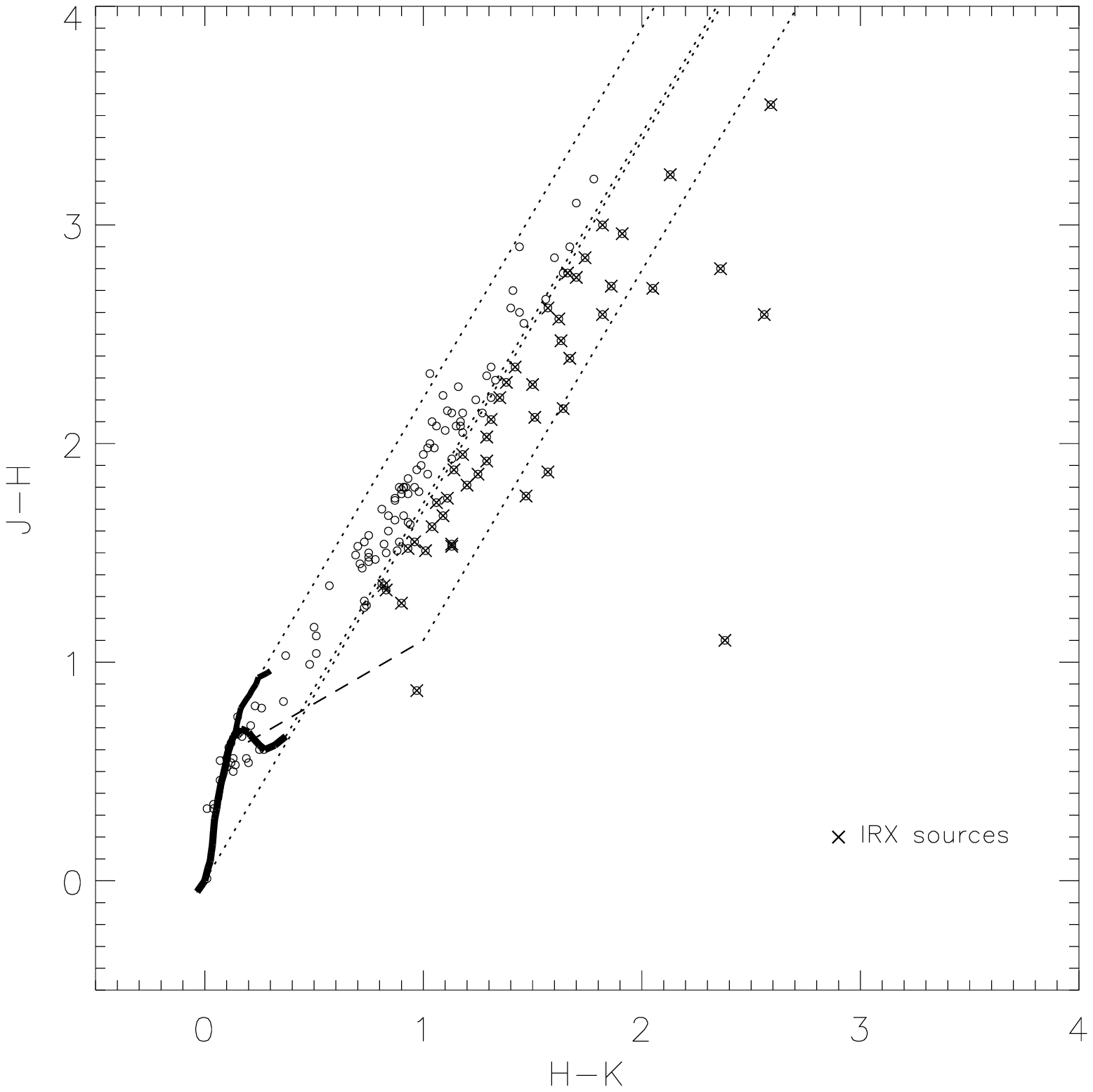}\hspace{1cm}\includegraphics[width=7cm]{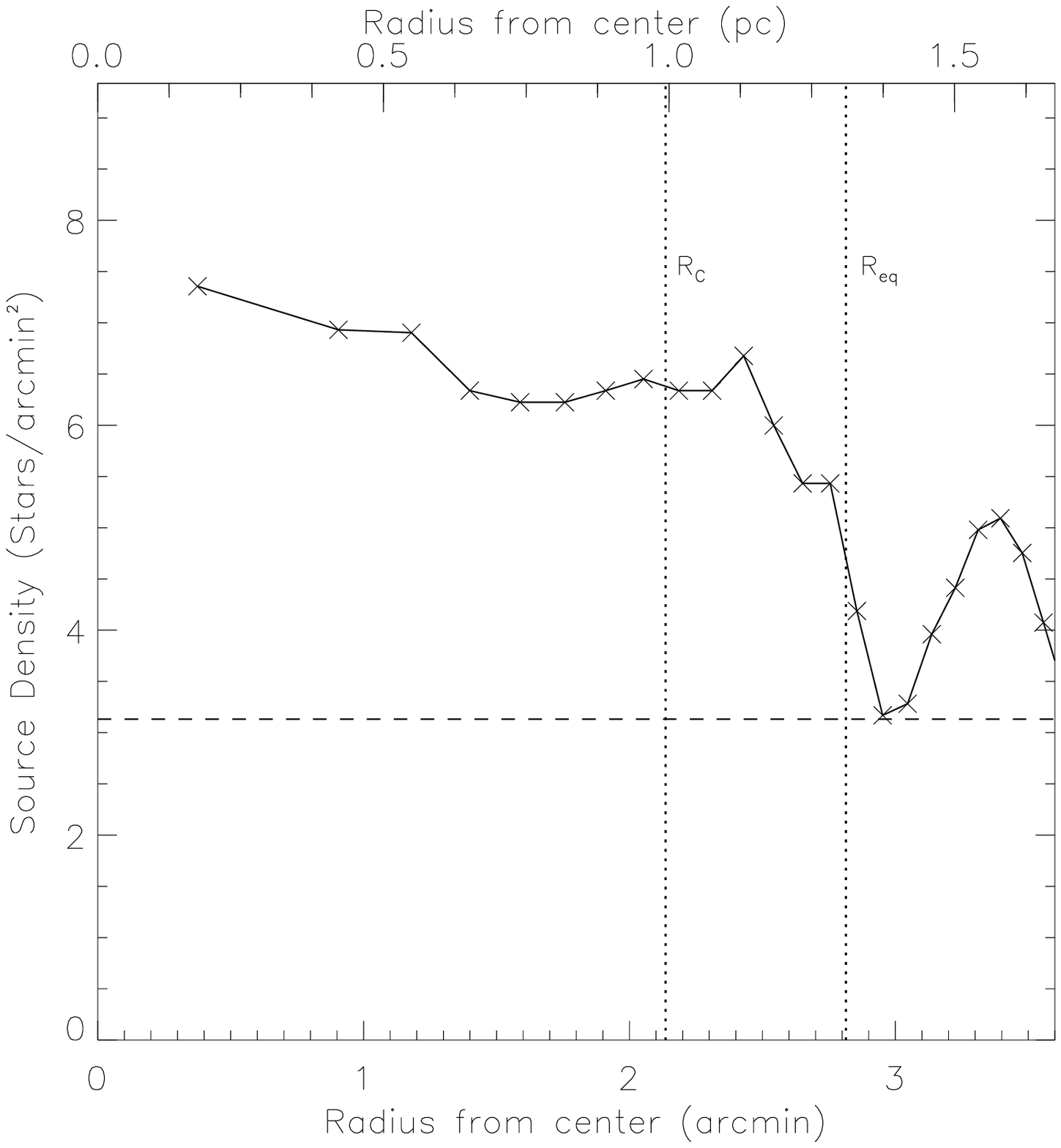}

\caption{Same as Figure \ref{fig:pl01panels}, for cluster REFL08.}
\label{fig:REFL08panels} 
\end{figure}

\begin{figure}

\includegraphics[width=7cm]{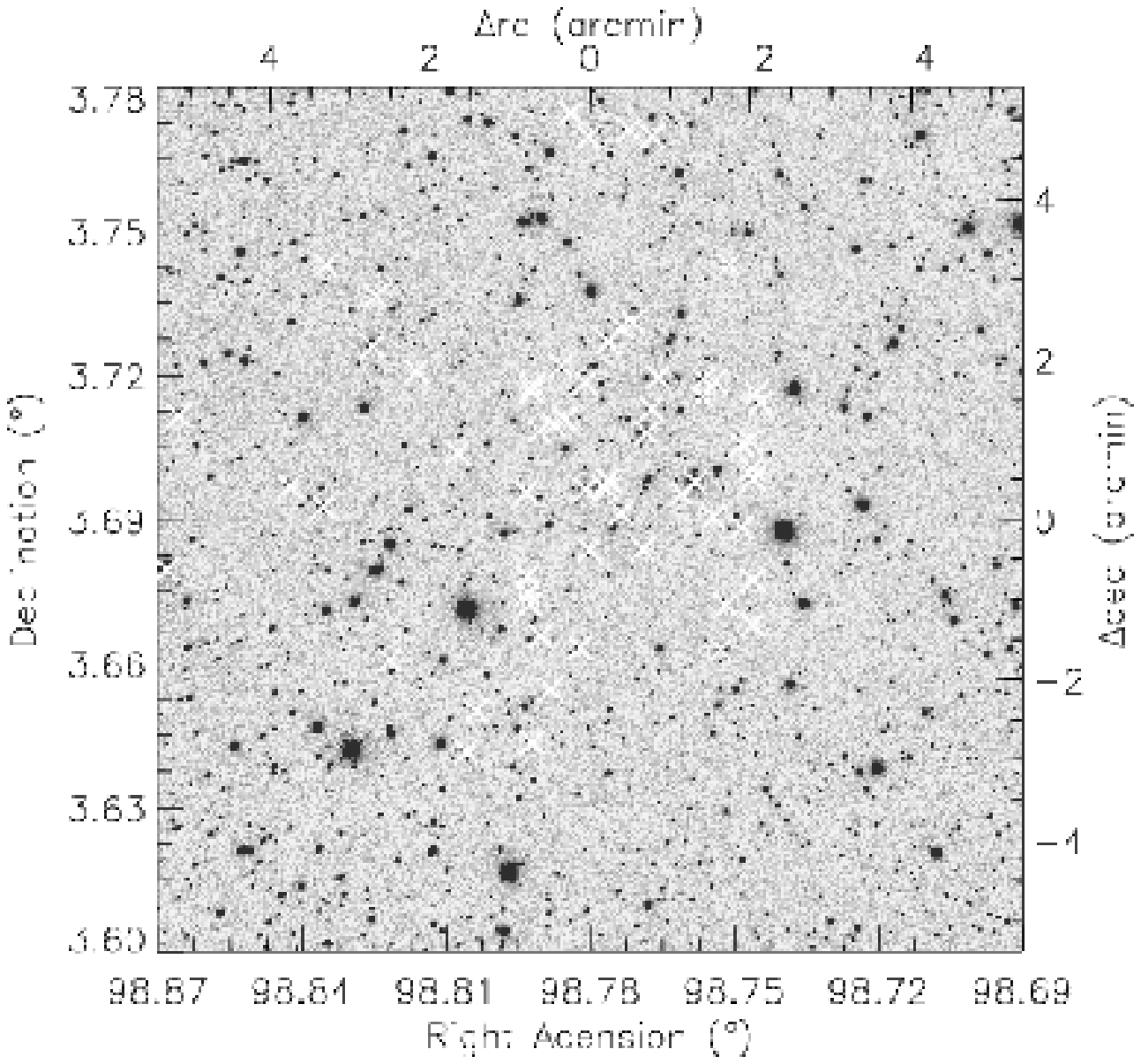}\hspace{1cm}\includegraphics[width=7cm]{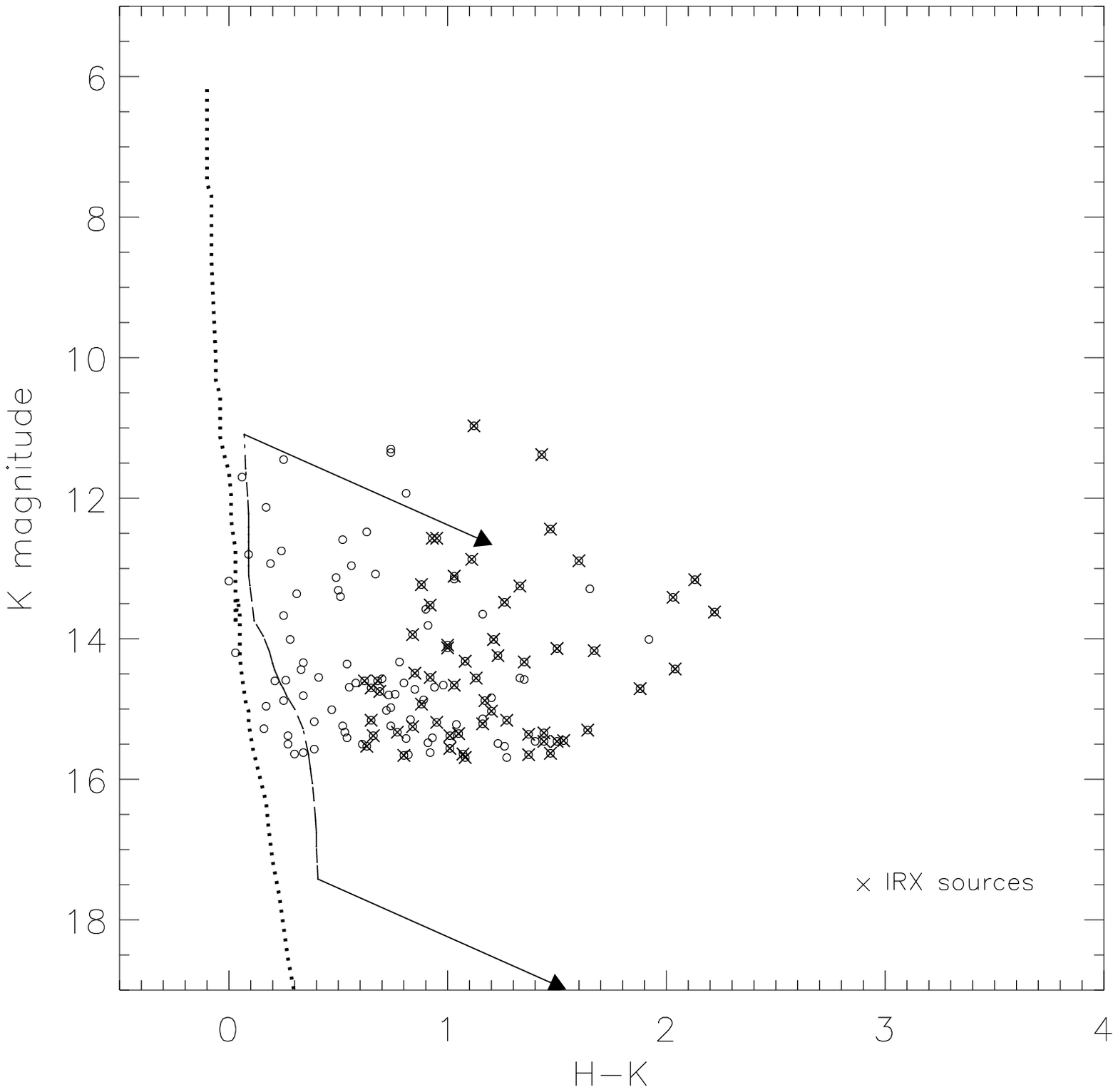}

\includegraphics[width=7cm]{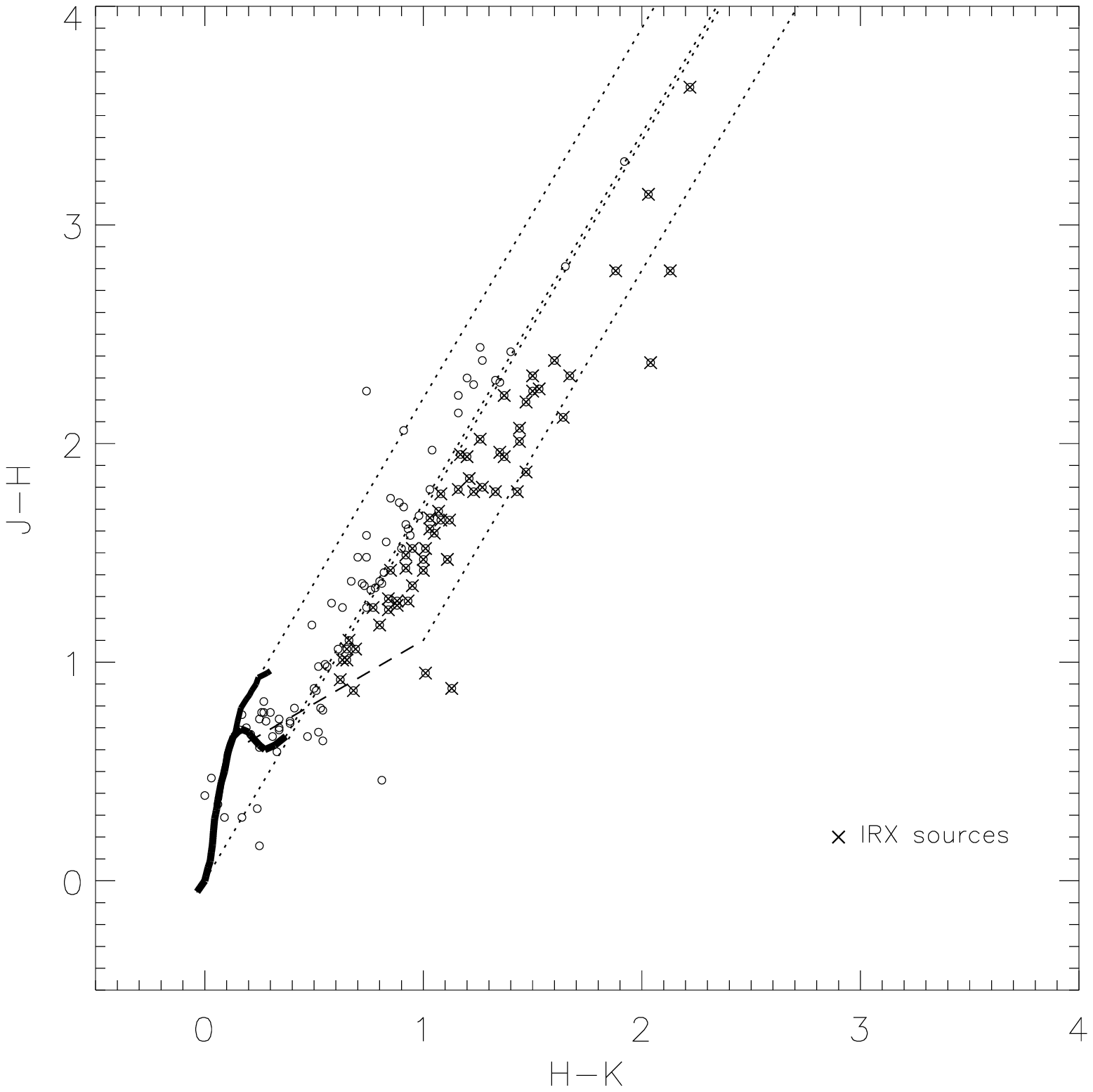}\hspace{1cm}\includegraphics[width=7cm]{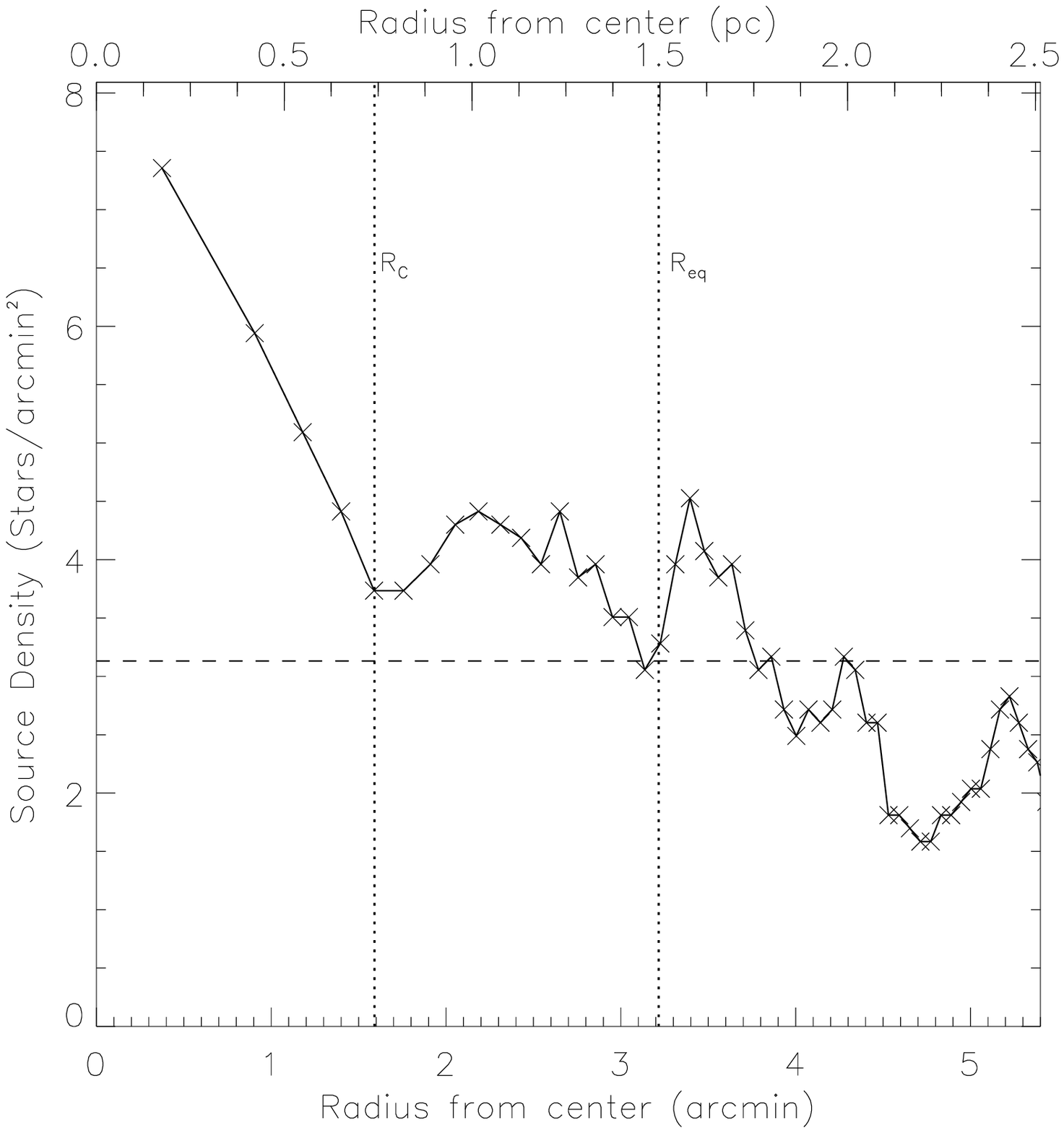}

\caption{ Same as fig \ref{fig:pl01panels}, for clusters REFL09.}
\label{fig:REFL09panels} 
\end{figure}

\begin{figure}

\includegraphics[width=7cm]{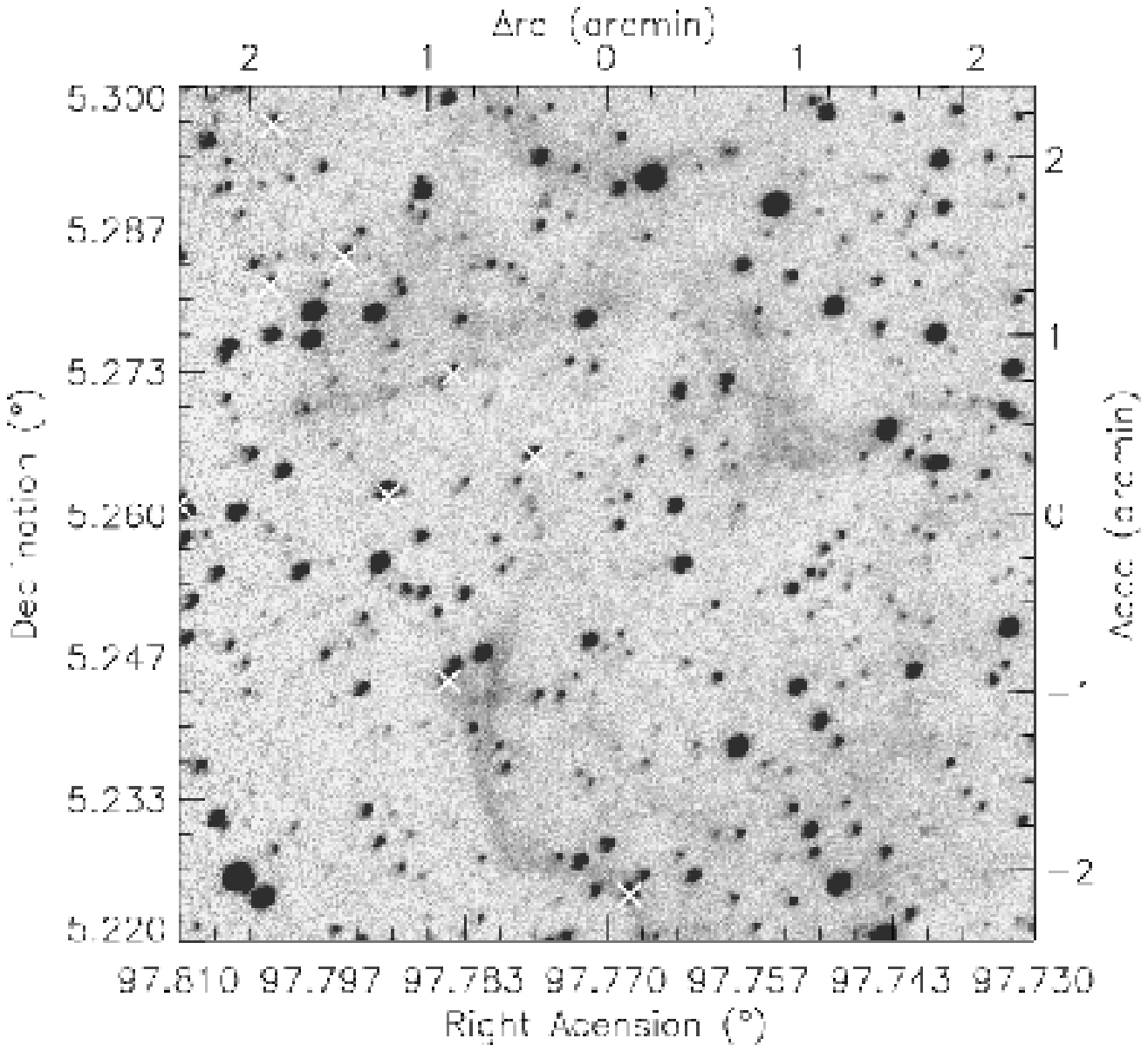}\hspace{1cm}\includegraphics[width=7cm]{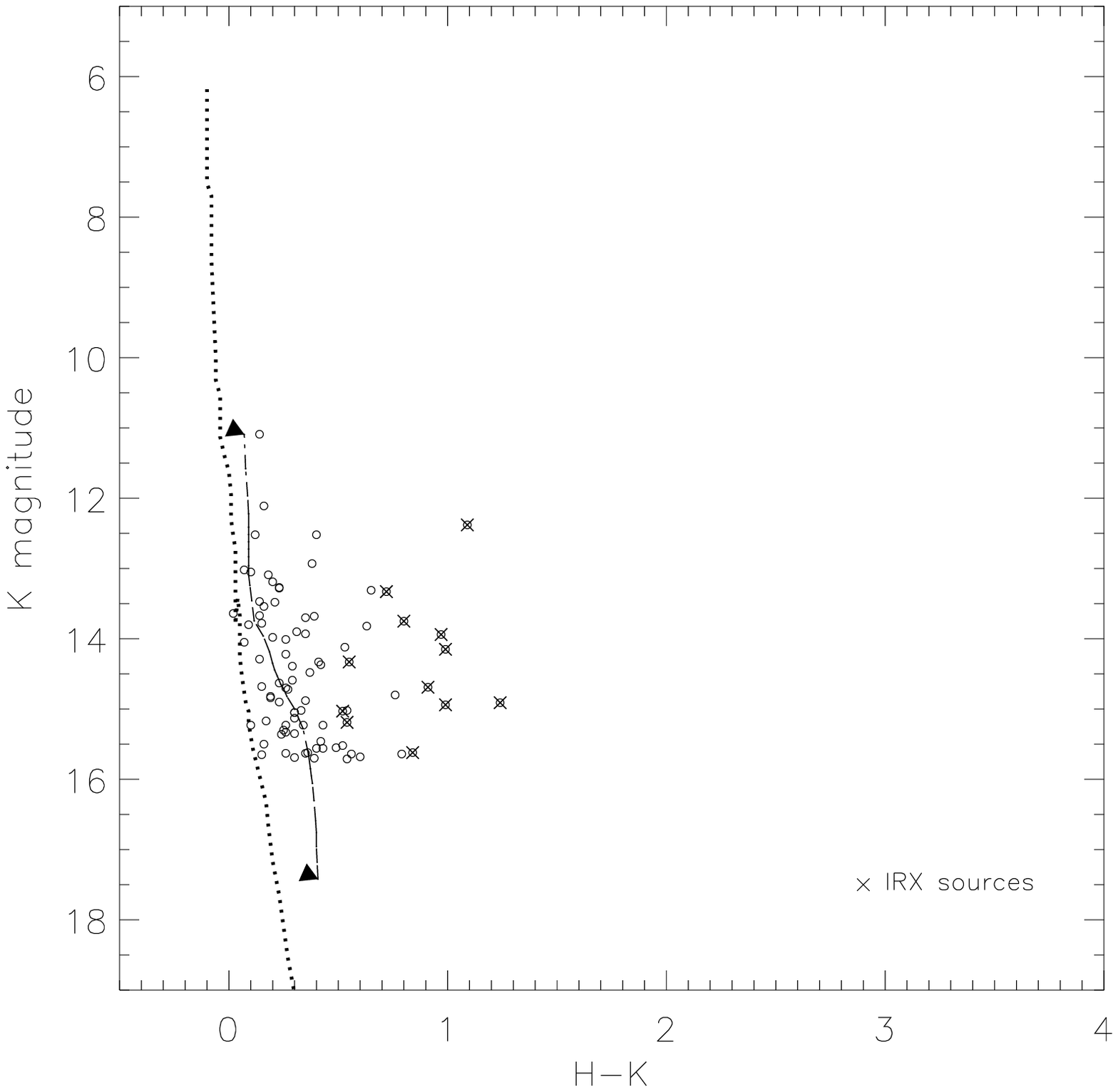}

\includegraphics[width=7cm]{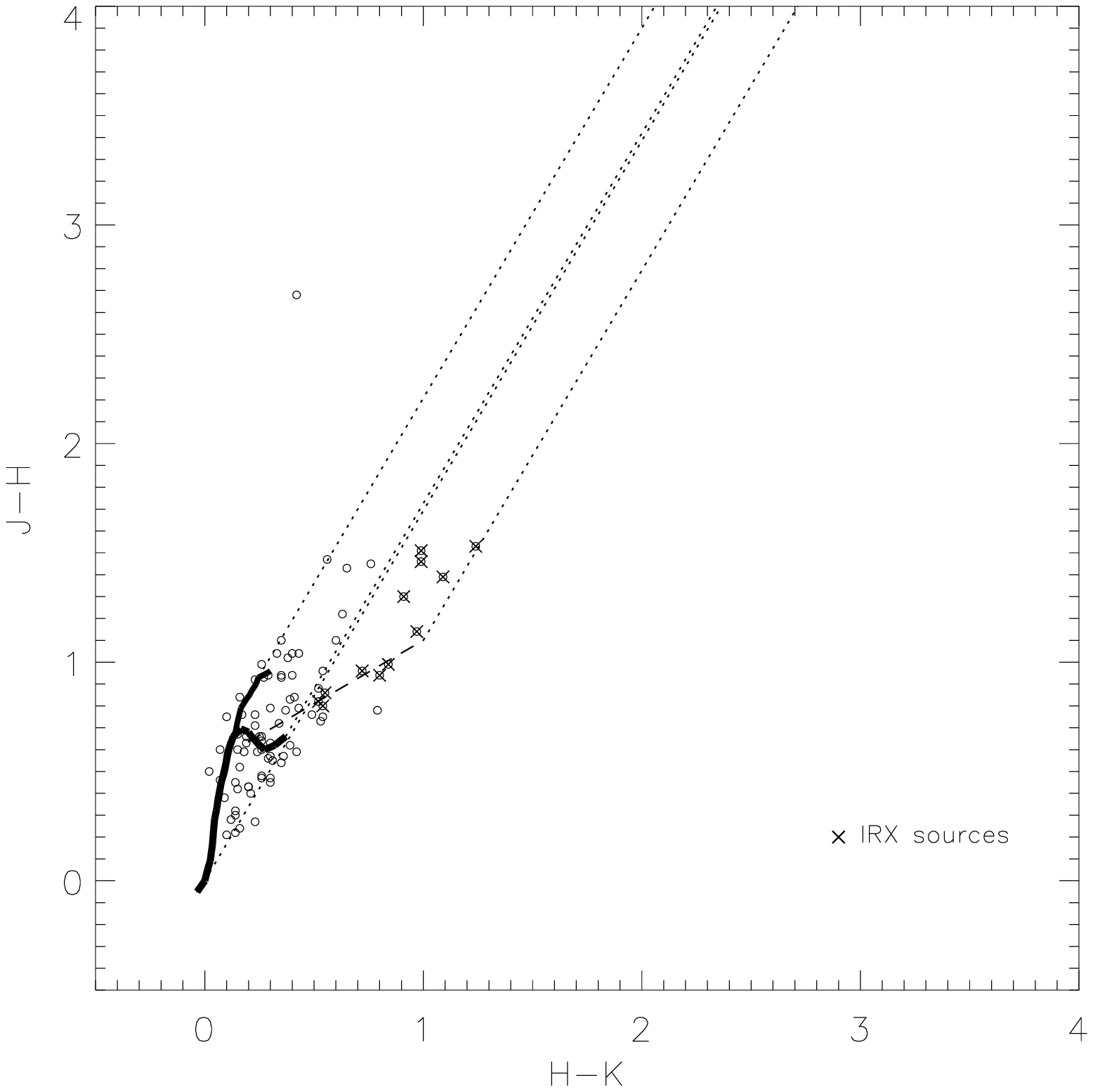}\hspace{1cm}\includegraphics[width=7cm]{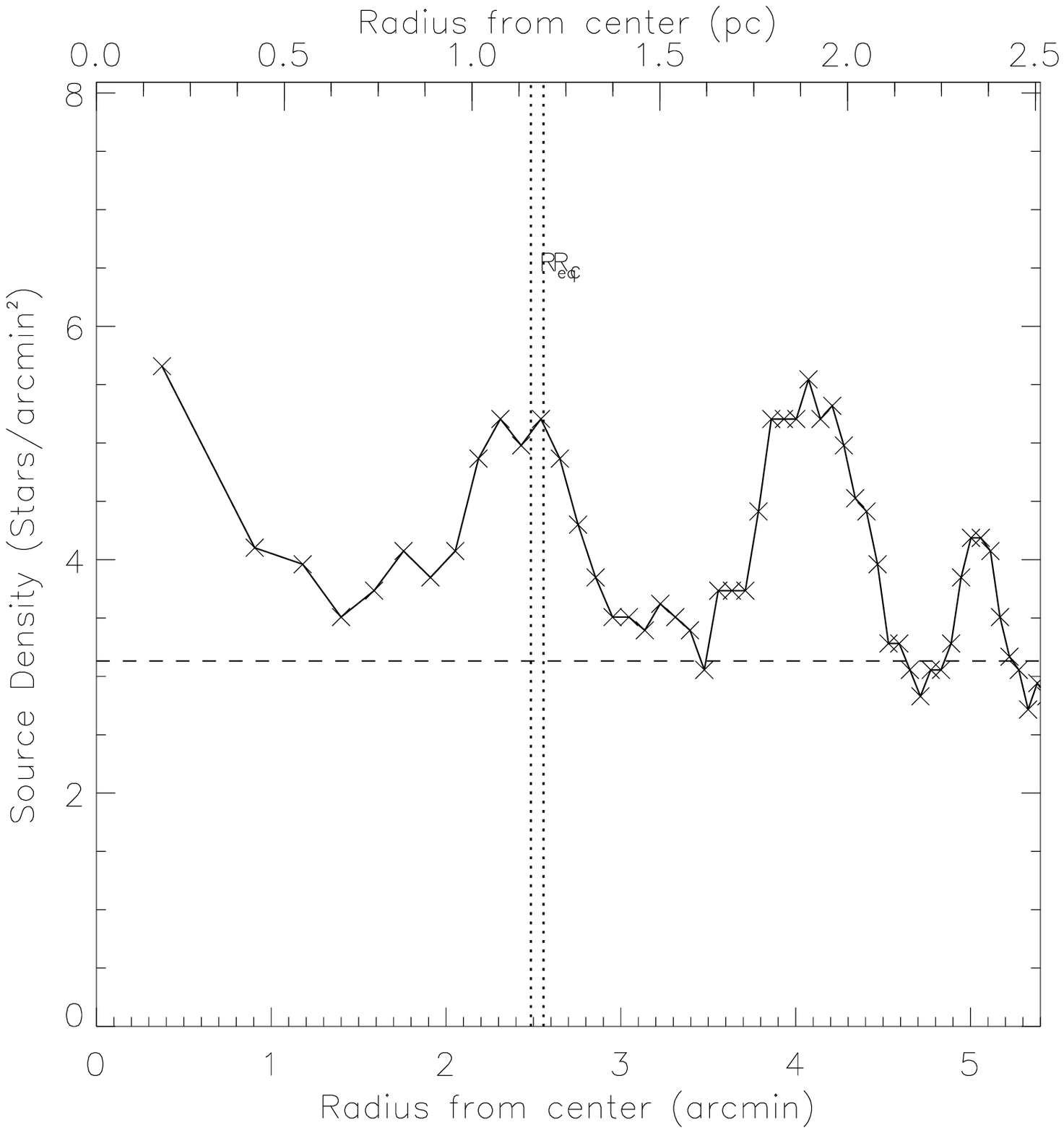}

\caption{Same as Figure \ref{fig:pl01panels}, for cluster REFL10.}
\label{fig:REFL10panels}
 \end{figure}

\begin{figure}

\includegraphics[width=7cm]{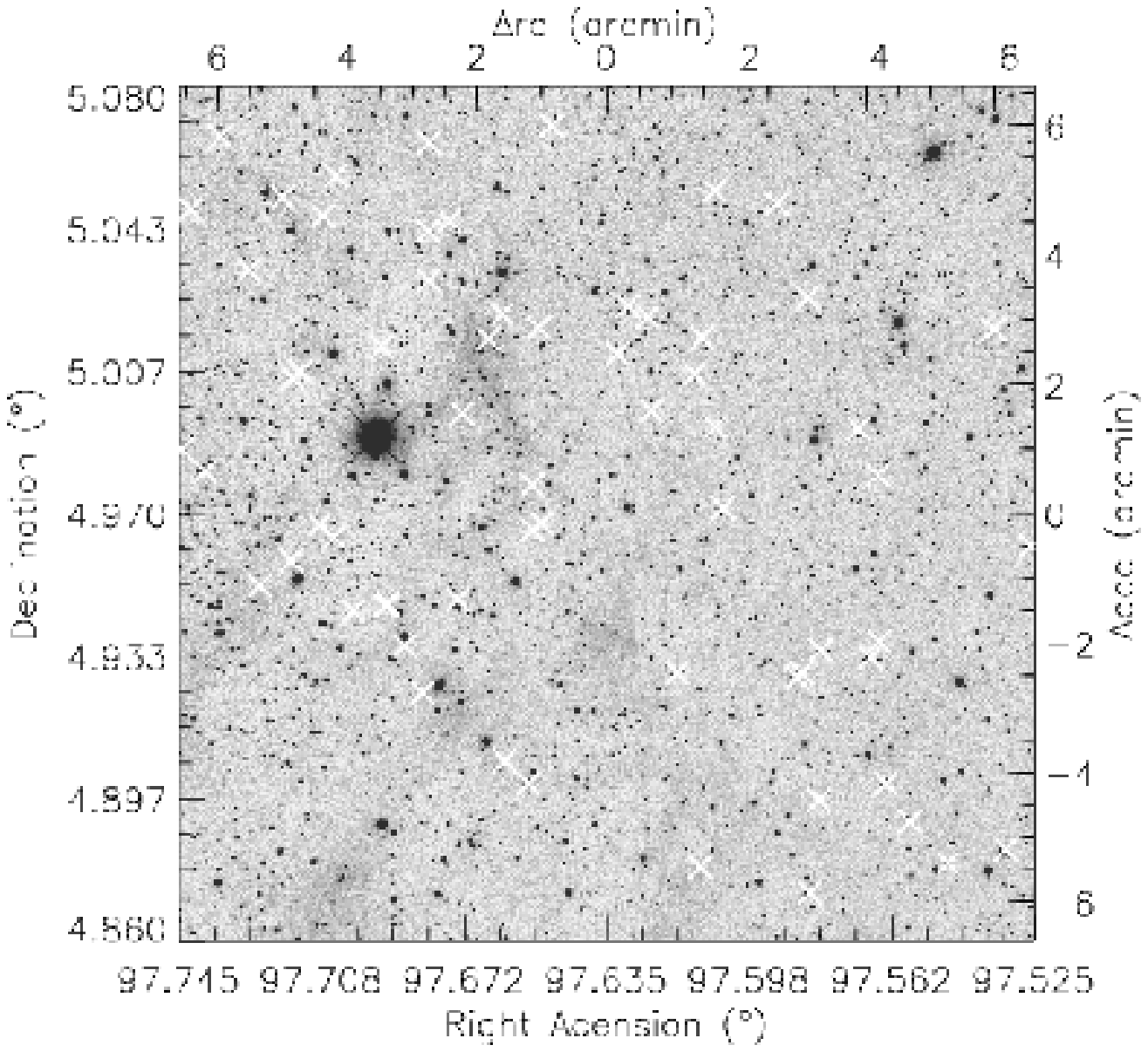}\hspace{1cm}\includegraphics[width=7cm]{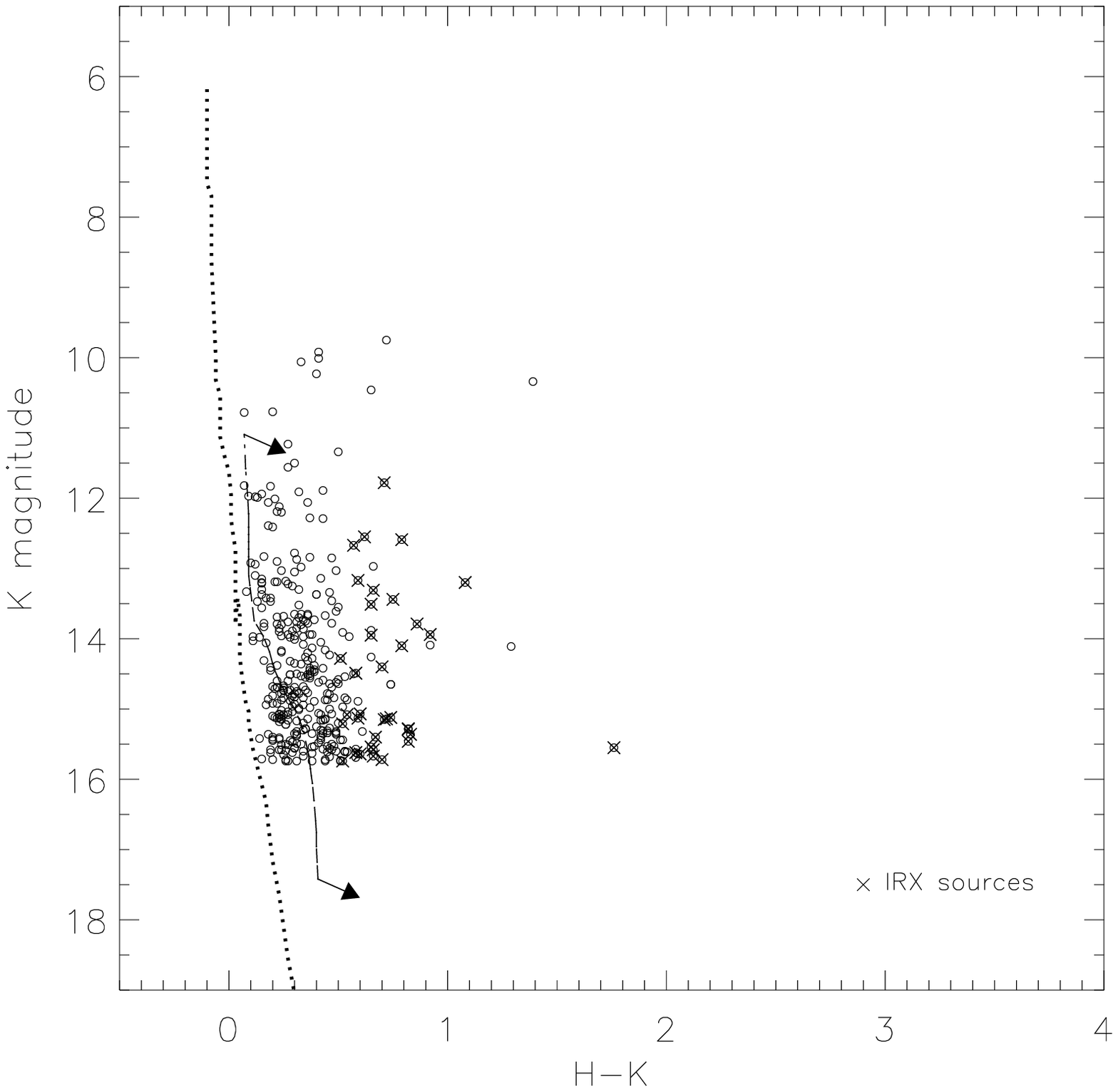}

\includegraphics[width=7cm]{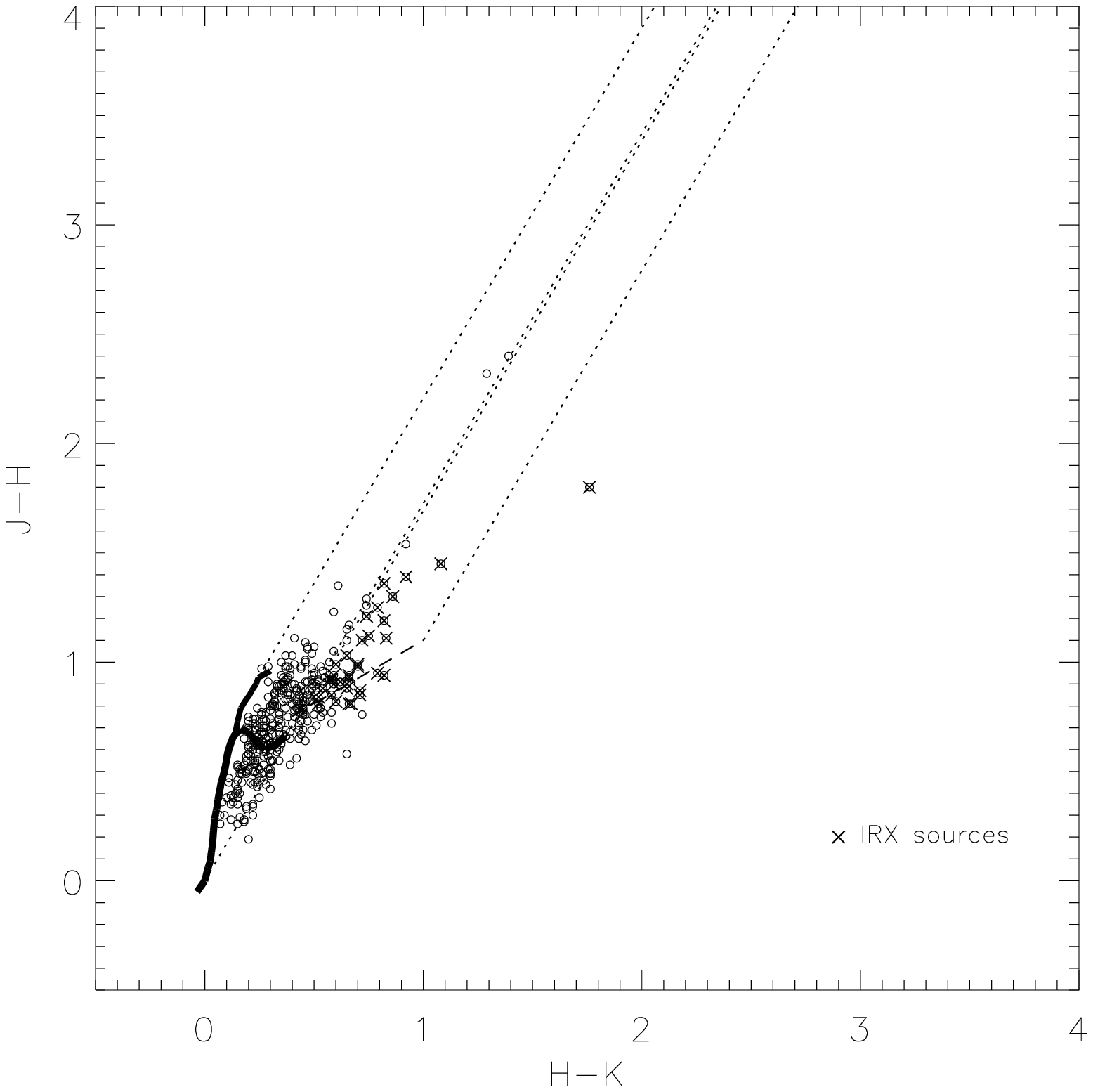}\hspace{1cm}\includegraphics[width=7cm]{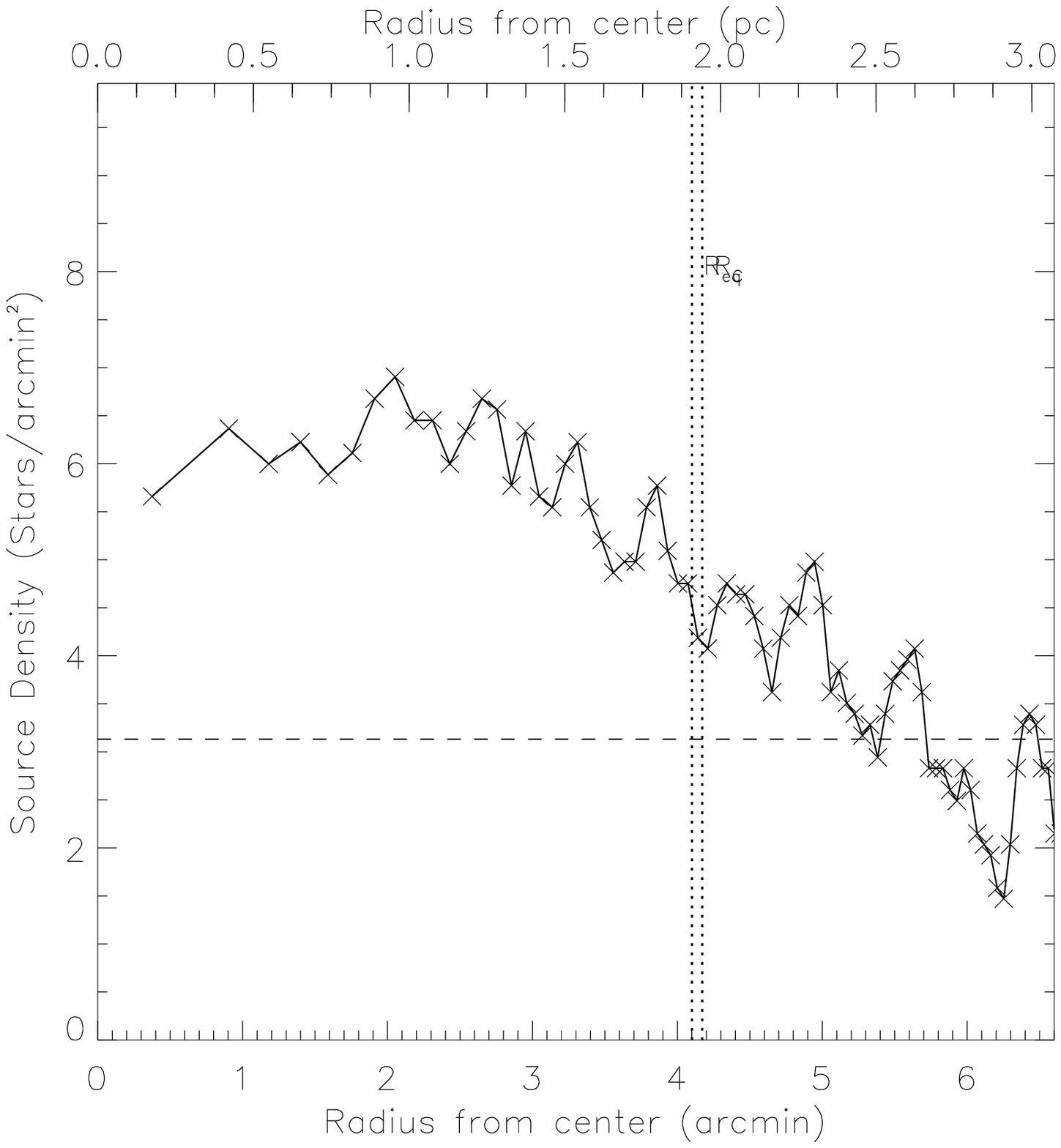}

\caption{Same as Figure \ref{fig:pl01panels}, for cluster NGC~2237.}
\label{fig:ngc2237panels} 
\end{figure}

\begin{figure}

\includegraphics[width=7cm]{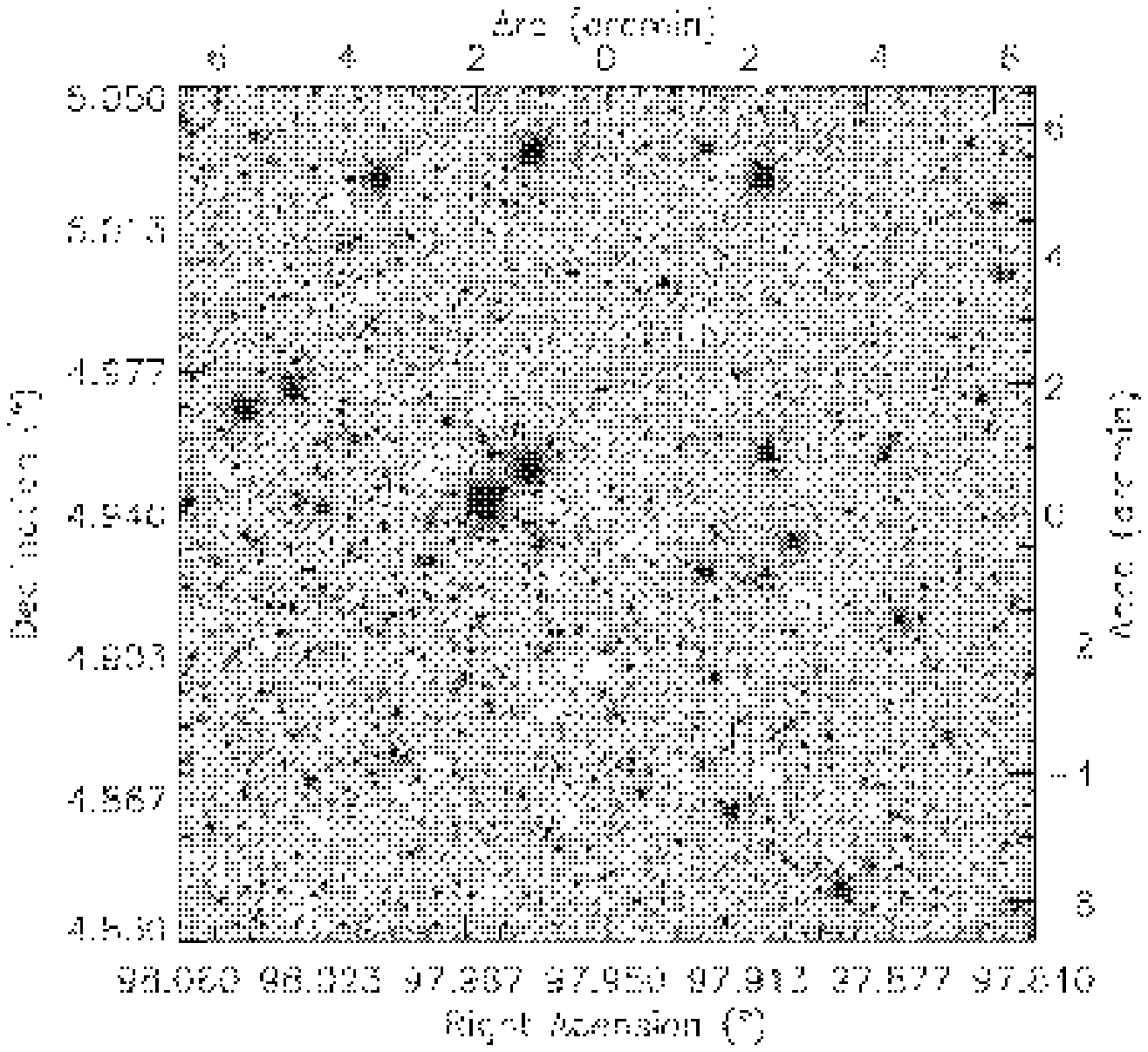}\hspace{1cm}\includegraphics[width=7cm]{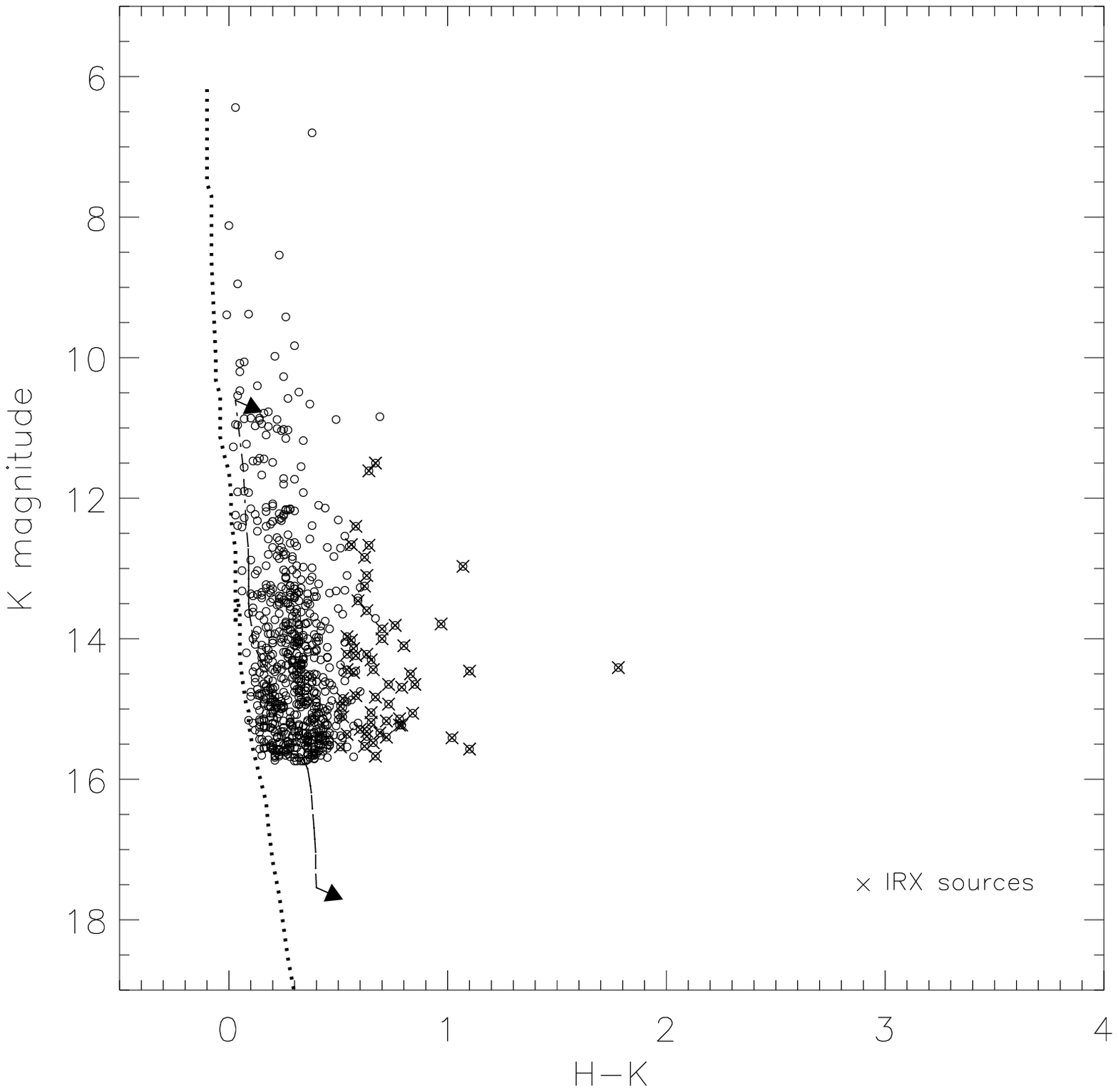}

\includegraphics[width=7cm]{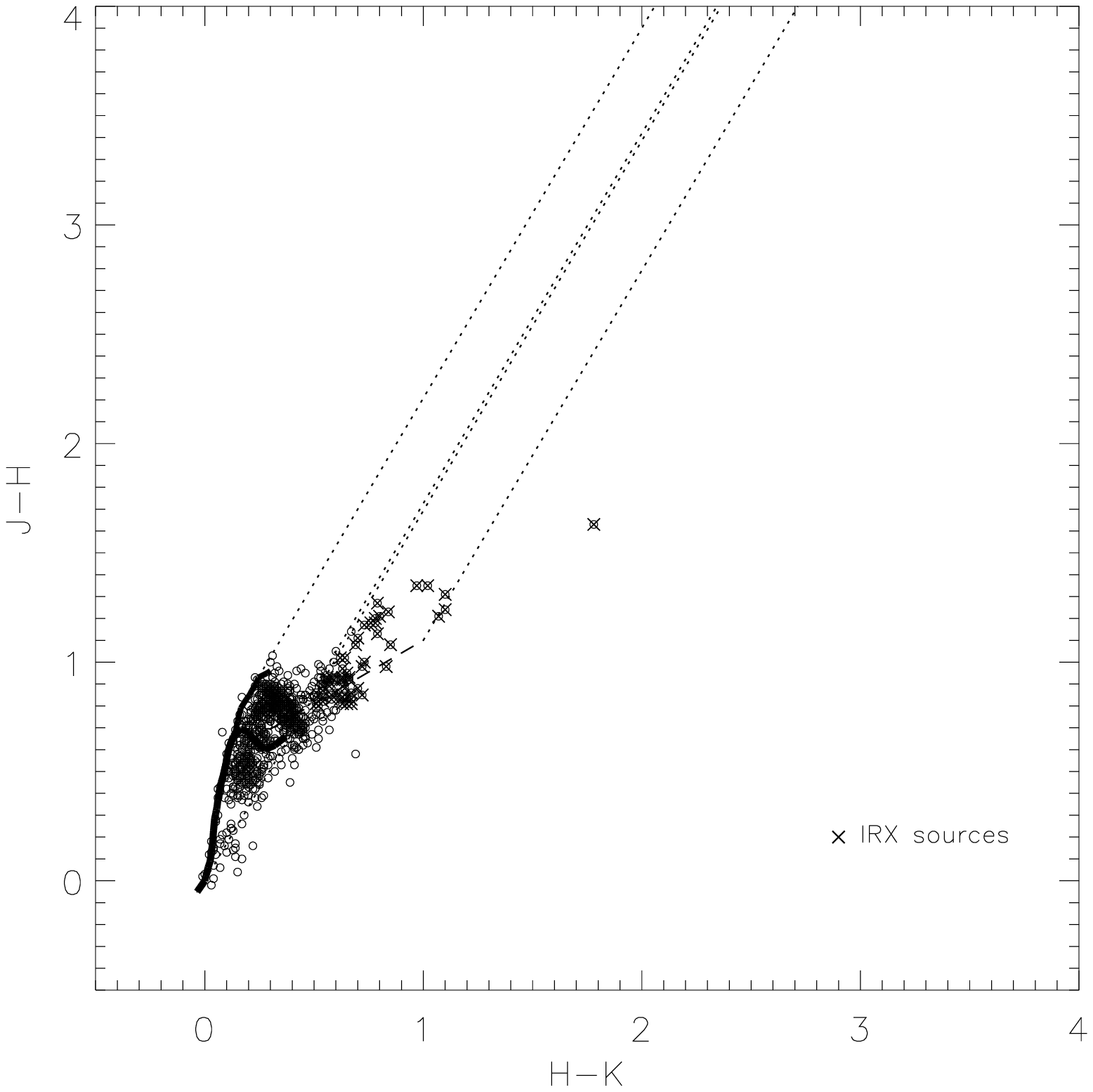}\hspace{1cm}\includegraphics[width=7cm]{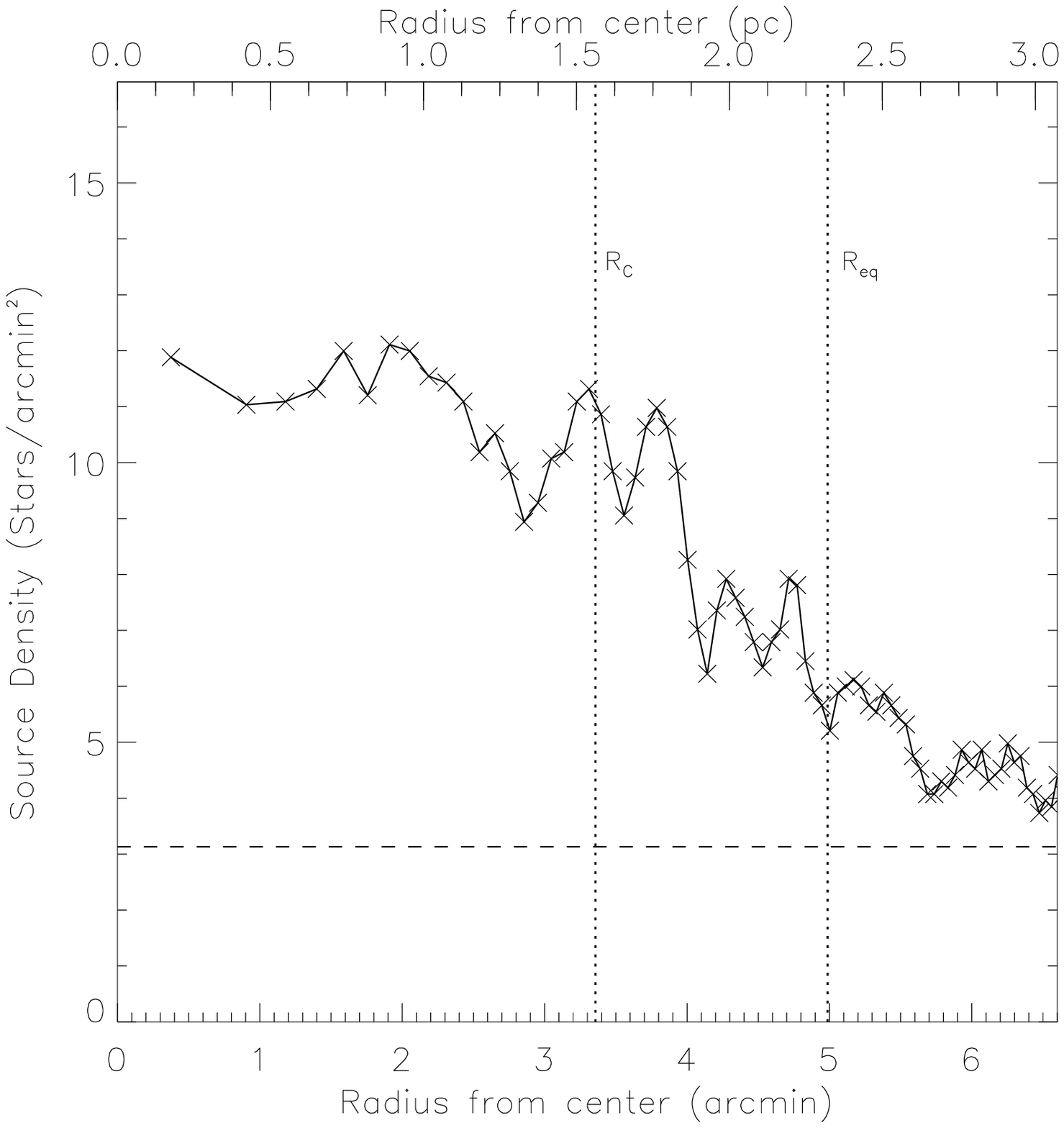}

\caption{Same as Figure \ref{fig:pl01panels}, for cluster NGC~2244.}
\label{fig:ngc2244panels} 
\end{figure}

In Figures \ref{fig:pl01panels} to \ref{fig:ngc2244panels} we show images of each of these cluster regions, as well
as the color-magnitude diagrams, color-color diagrams and stellar radial profiles for all stars within the cluster
boundaries. In the images, we marked NIRX sources brighter or equal than $K=15.75$ mag with
`X' symbols. In the $K$ vs. $H-K$ color-magnitude diagrams we plot, for each cluster, all of the stars inside the
corresponding 0.2 \samis~contour, and we marked with `X' symbols those objects with infrared excess. We also
plot the ZAMS locus and a PMS evolution isochrone of 1 Myr, both shifted to the distance of the Rosette. The
indicated extinction vectors corresponding to 3 times the mean value $\langle A_V\rangle$ in the cluster analysis
box. Stars falling to the right of the 1 Myr isochrone are clearly affected by extinction towards the line of sight
of the cluster, revealing their highly embedded nature. In the $J-H$ vs. $H-K$ color-color diagrams, the same stars
are located above the dwarf and giant star sequences along the reddening bands, with the NIRX sources located to the
right of the ZAMS reddening strip. Those objects located at or near the zero age sequences are most probably
foreground stars or evolved cloud members.

The radial density profiles in the fourth panels of each cluster figure were constructed using annuli of decreasing
width, each enclosing the same area (see e.g. \citet{mu03}) and centered on the weighted density peaks defined
above. For these radial profiles we took into account all stars down to $K=17.25$ inside the analysis boxes. In the
plots we indicate the equivalent and core radii defined above. Clusters PL02, PL06 and REFL10, have the lowest
surface densities and their profiles are poorly defined, but the rest present well defined radial profiles, which
show well extended tails. In four cases --- clusters PL01, PL04, PL07, REFL09 --- the profiles show secondary peaks
suggestive of structure. In the case of the Nebula clusters NGC~2244 and NGC~2237, their radial distribution
profiles decline slowly, which suggest a diluted core peak and a rather extended structure.

\subsubsection{Comparison of cluster properties}

\begin{figure} 
\includegraphics[clip=true]{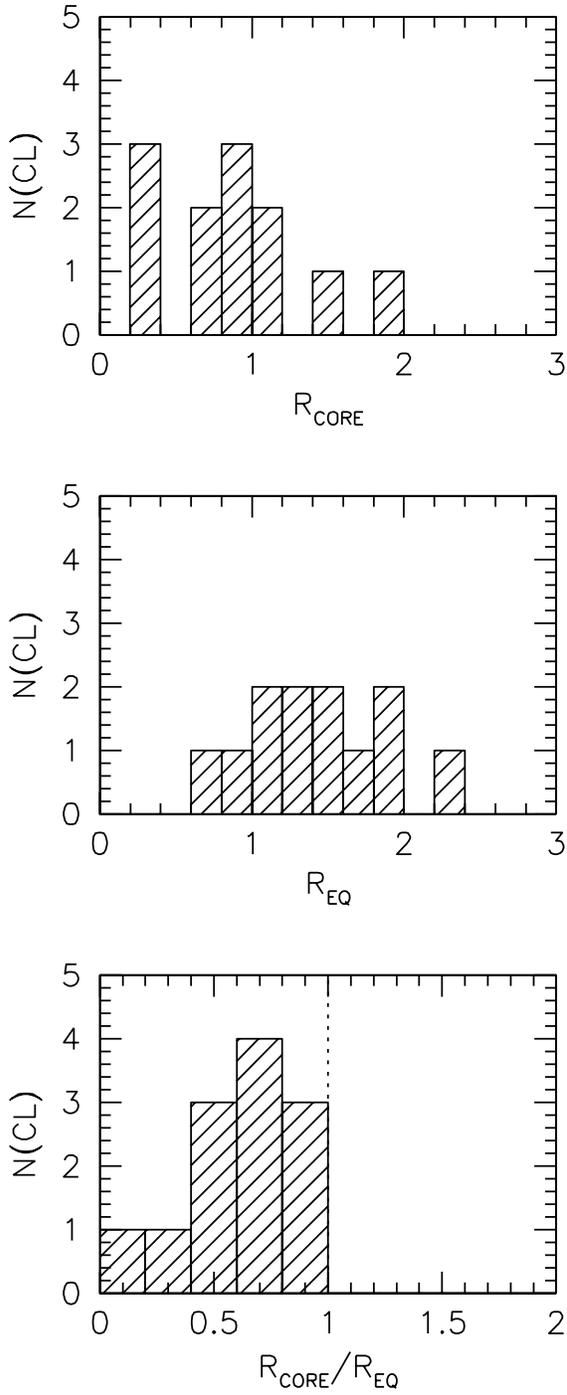} 
\caption{From top to bottom:
distribution of core radii, equivalent radii and core to equivalent radii ratios
for the Rosette clusters} \label{fig:coreq} 
\end{figure}

Under the assumption that the NIRX sources trace well the extent of the clusters, the  core and equivalent radii are
good estimations of the total sizes of the clusters.  The core radii, $R_{core}$ of the Rosette clusters  have a
range of 0.69 to 4.17 arcmin (0.32 to 19.39 pc), with an average of 2.0$\pm$1.0 arcmin (0.93$\pm$0.47 pc). Their equivalent radii, $R_{eq}$,  vary from
1.6 to 4.9 arcmin (0.74 to 2.28 pc), with an average of 3.1$\pm$1.0 arcmin (1.44$\pm$0.47 pc). The  distributions of these size estimates are shown
in the top panel of Figure \ref{fig:coreq}. 

In the bottom panel of Figure \ref{fig:coreq} we show the distribution of the  ratio $\tau=R_{core}/R_{eq}$, which
has a nominal peak at $\tau=0.7$. In the  study of FL07, where NN analysis was applied to a large sample of known
cluster  fields, they found that embedded clusters separated in two groups by their  value of $\tau$: those with
ratios below 0.5 are denominated as centrally  condensed (C-type clusters), while those with values closer to 1.0
are denominated  as having a flat profile (F-type clusters). We found that only two of the RMC clusters,  PL01 and
PL03, have C-type ratios, and for three of the clusters, PL06, REFL10 and  NGC~2237, cores could not be easily
determined and, for purposes of comparison, were given a ratio of $\tau=1.0$. 

The distribution of $\tau$ ratios is rather different from the one observed by  FL07 in their larger and more
heterogeneous sample of clusters: they found a larger fraction of compact, embedded clusters (44
percent of the sample) with  well defined cores, while in our sample, centrally condensed clusters appear to be
scarce. However, we note that the NN analysis of FL07 was not applied on color selected samples,
but on all sources detected in a given field. From the observed distribution of cluster sizes 
and core-to-total size ratios it appears that the Rosette clusters have preferentially extended profiles. 

As  listed in Table~\ref{tab:clustertable}, the observed NIRX fractions of the nebula clusters, NGC~2244 and
NGC~2237 are $12\pm 3$ and $15\pm 2$\%, respectively, and therefore they are significantly smaller than those of
the embedded clusters, which vary from $18\pm 3$\% in cluster PL05 to $76\pm 9$\% in cluster REFL09. Also, the
average NIRX fraction in the molecular cloud areas is $41\pm 6$\%, i.e almost three times as large than in NGC~2244
and NGC~2237. Curiously, the fraction observed for REFL10, which is also located at the nebula is much closer to
the average of the embedded clusters and actually slightly larger than those of clusters PL01, PL04 and PL05.

\begin{figure} 
\includegraphics[clip=true]{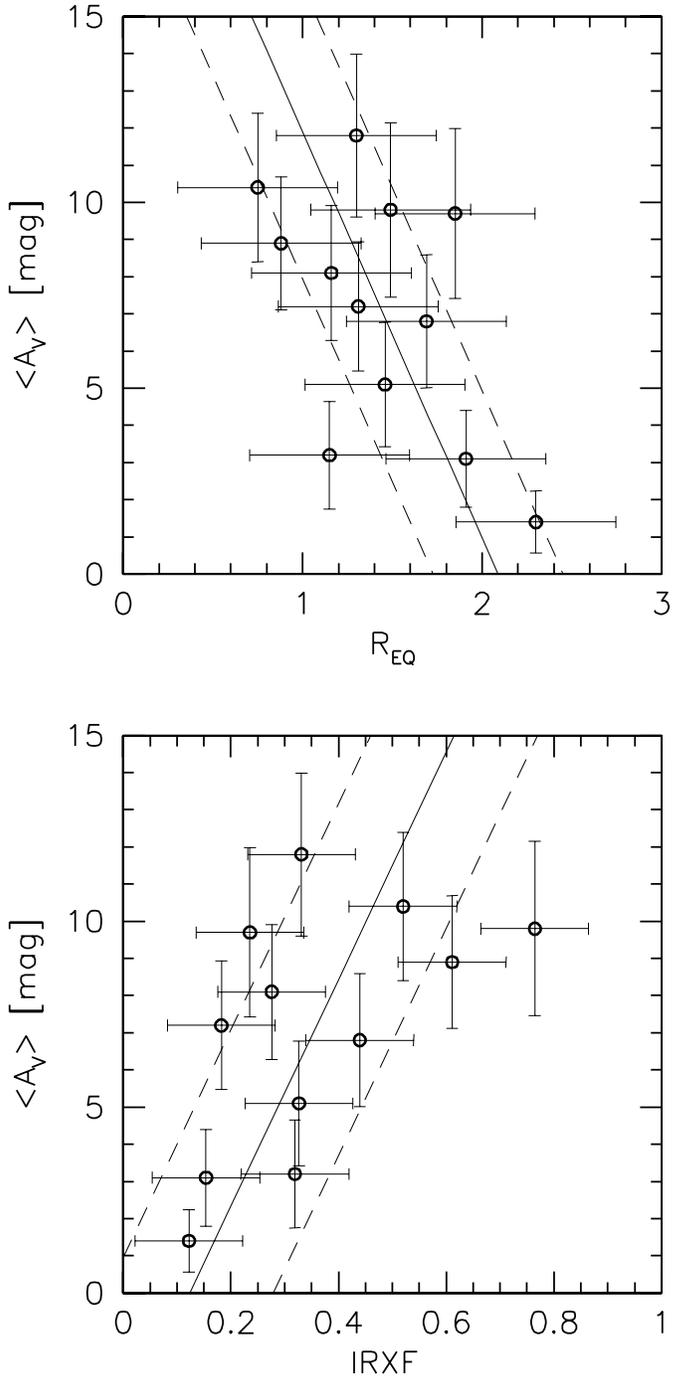} 
\caption{Top: cluster
equivalent radii vs. median extinction. The solid line and dotted lines
represent a least squares linear fitting $\pm$1 rms deviation. Bottom: NIRX
fraction ($K<15.75$~mag) vs. median extinction} \label{fig:agerel} 
\end{figure}

In Figure \ref{fig:agerel} we show the relation between the NIRX fraction, the median extinction value and the
equivalent radii of clusters. Extinction and cluster size are negatively correlated with a Pearson coefficient
$\eta=-0.55$, while  extinction and NIRX fraction are positively correlated with a Pearson  coefficient $\eta=0.57$.
Consequently, as shown in Figure \ref{fig:NIRXfsizerel}, cluster  equivalent sizes and core sizes are both
negatively correlated with NIRXF, with  Pearson coefficients $\eta=-0.61$ and $\eta=-0.54$ respectively.

We also found that the ratio $\tau$ was negatively correlated with NIRXF for clusters with a distinguishable core,
this correlation is weaker ($\eta=-0.48$) but is similar to the relation found by FL07 for a sample of nearby
clusters. The significance 
of these correlations in the context of the structure of evolution of the RMC clusters 
is discussed in section \ref{section:discussion}.

\subsection{The Fraction of Stars in Clusters}
\label{section:results:subsection:fraction}

Since NIRX sources trace the youngest population of stars within a star forming
region, we used the distribution of NIRX sources to investigate the
fraction of stars located and forming in clusters. The fraction of NIRX sources
in clusters was calculated by simply counting the number of NIRX sources
within the cluster boundaries and comparing this value with the number of
NIRX sources outside the cluster boundaries, correcting both for background
contamination. We first considered the entire area of our survey, which
provided us with an estimate of the number of stars that are currently
located within clusters across the entire complex. Since the most recent episodes of star formation
should be associated with the molecular gas, we next considered only the
area within the molecular cloud, which provided us with an estimate of the
fraction of stars currently forming in clusters.

For the first determination, we calculated that the 
total area covered by the FLAMINGOS survey is 7308 sq. arcmin (1579 sq. pc) while the 11 clusters identified with our analysis
occupy a total surface of 390 sq. arcmin (84 sq. pc), i.e. 5.3 percent of the total survey area. The total number of NIRX
sources detected  is $1169\pm 34$, out of which $549\pm 23$ are located within the boundaries of the clusters, and $620\pm25$ are outside the boundaries. In order to correct for the background contamination, we 
scaled the the number of NIRX sources observed in the control fields by a factor equal to the area
of the molecular cloud minus the areas of the clusters and then divided by the area of the control fields. This {\it expected} number of NIRX
sources due to background contamination was estimated to be $261\pm 17$. The
background correction for individual clusters is always smaller than the
Poisson uncertainties but it was also applied. Using these corrections,
the number of NIRX sources outside of clusters was $359\pm 19$, and inside clusters
$531\pm23$. Therefore, the fraction of NIRX sources located in young clusters across the Rosette
Complex is $60\pm5$\%.

The area of the molecular cloud was defined by the 0.8 \kks~integrated
intensity contours of $^{13}$CO emission \citep{HWB05}, corresponding to $A_V= 3.5\pm 0.6$ mag using a recently derived conversion ratio \citep{pineda07}.  This area
equals 2684 sq arcmin (580 sq. pc.) The 9 clusters located in the cloud occupy a total
of 242 sq. arcmin (52.3 sq. pc) or about 9\% of the RMC.
The total number of NIRX sources detected in the molecular cloud is $589\pm 24$, out
of which $436\pm 21$ are located within the cluster boundaries and $153\pm 12$ are outside
the boundaries. After correcting for field contamination, the number of NIRX sources located in clusters
is $429\pm 21$ and outside clusters is $73\pm 8$. Therefore, the fraction of NIRX 
sources currently forming in embedded clusters in the Rosette Molecular Cloud is 86$\pm$4\%.

Interestingly, a total of 208$\pm$15 NIRX sources are contained in clusters PL04, PL05, PL06 and REFL08, all located at the ``Central Core'' of the cloud. This corresponds to 48$\pm$3\% of the total number of embedded sources. Therefore, approximately half of the recent births in the RMC have occurred at the most dense region of the cloud. 

\subsection{Distribution of Sources with respect to the Rosette Nebula}
\label{section:results:subsection:clusterdistribution}

We analyzed the spatial distribution of those NIRX sources with 10th NN densities higher than the mean as a function of the
distance to the center of NGC~2244. To do this, we determined the surface density of sources
within concentric rings centered on NGC2244 having a width of 1.0 pc. For
the first 11 pc we were able to use complete rings, but further away,
we had to limit the angular extent of the rings to conform with the
irregular shape of the survey boundaries.  The surface density of sources
in each ring or segment was calculated by dividing the number of sources
in each segment by its area and then diving by the surface density inside
the first parsec circle.

This normalized radial NIRX source density distribution is shown in Figure \ref{fig:dneb} (top panel). In the plot we indicated the
approximate locations of clusters and main cores of the molecular cloud, measured in projected distance from the
center of the Nebula. We also indicate the location of the main regions of the RMC, the ``Ridge'', the
``Central Core'', and the ``Back Core'' with respect to the Rosette Nebula. 

In the bottom panel of the figure, we indicate the average fraction of NIRX sources in clusters as a
function of distance, averaged in five groups of clusters
which are roughly traced by the distribution shown in the top panel. 
The cluster groups used are as follows: first group includes NGC~2244, NGC~2237 and REFL10;
second group includes PL01 and PL02; third group includes PL04, PL05 and REFL08; fourth group includes PL03 and
PL06; finally, group 5 includes clusters PL07 and REFL09.

\subsection{Low density population}
\label{section:results:subsection:lowdensity}

In order to investigate the low density star forming component in the cloud, we closely examined the region between
the central core and the back core. This region corresponds to Field 09 of our survey. This field was chosen because the
seeing and observing conditions for it were particularly good.~Specifically, the average scatter of colors down to
$K=17.25$ remains below 0.109 mag (similar to the average for bright end bins) across the whole field (see Figure
\ref{fig:area9colorscatter}). This superior quality was achieved because the southeastern quadrant of the field,
which in other fields presents high stellar profile distortions, overlaps with the good quality northwestern
quadrant of field G1.~Selective averaging of photometry in these overlapping quadrants, permitted the reduction of
the total color scatter.

No clusters were found in this field, but a small group of NIRX sources coincide well with a roughly filamentary core in
the cloud located southwest of the PL06 cluster region (see Figure \ref{fig:NIRXpos_1575}). Our NNM analysis does
not produce a new cluster identification for this core, but we counted the number of NIRX sources in the whole field
down to brightness limits of 15.75, 16.25, 16.75 and 17.25 mag. We counted the number of NIRX in the control fields
at these same magnitude limits. In order to have make a conservative estimate, we limited the counts to a circular
area with a radius of 1500 pixels centered on pixel position [3000,2400]. This delimits the area of best
photometric quality after the polynomial correction, as described in section \ref{section:observations}.

We compared these NIRX counts in field 09 with the expected number of field NIRX sources at these same limits in an
equivalent area of high photometric confidence in the control fields, adding again, a mean extinction level of 5.0
mag to account for stars that would have excess colors after reddening. A second comparison was done by averaging
the number of observed NIRX sources in equivalent photometric confidence circular areas of survey fields 03, 13, 14
and 15. These fields are located in regions of the complex where the CO emission from the molecular cloud is lower
than average, and thus they can help to determine if the NIRX sources of field 09 could indeed be a local
enhancement near the core of the RMC.

~In figure \ref{fig:dp9hist}, we compare the counts described above in the form of cumulative histograms:
apparently, field 09 surpasses significantly the expected number of NIRX from the control field and the off-cloud
areas, with counts 4.8, 5.6, 3.5 and 2.2 times larger than those in the control field field at
$K<15.75,16.25,16.75$ and $17.25$ respectively. The counts are also higher than the averaged counts of survey
fields 3, 13, 14 and 15 by factors of 1.3, 1.5, 1.6 and 1.8 respectively.~This suggests that the region of the
cloud observed in field 09 could have a significant number of young source candidates not associated with
clusters.~Furthermore, JHK color images of selected regions of the field show a number of highly reddened
sources.~Some of these sources have thin nebulosities, and coincide or are located close to stars with near
infrared excess emission. These sources are not included in our
infrared excess catalog because they were not detected in our J band
observations and therefore we do not have three colors to determine if they
exhibit excess emission.

In addition, we counted the number of sources with no J band photometry and $H-K>1.5$ mag at the same K
brightness limits mentioned above. It is not possible to assure that these sources belong to the cloud, especially
for $K>15.75$, where the scatter increases and the contamination by background galaxies worsens. However, we
noticed that the number of these sources which locate outside the main CO emission regions of the molecular cloud
was very small; the majority of them were in fact located within the cluster areas (see Figure \ref{fig:nojotas}).
Therefore, these red sources could be tracing the location of young sources too. Also, no stars with $H-K$ colors
this large were found in the control field. In Figure \ref{fig:nojotas} we show the distribution of the sources
with $H-K>1.5$ in the survey areas down to $K<17.25$ mag respectively. In table \ref{tab:dptable} we show the
number counts of  $H-K>1.5$ sources, separating those located outside the 20.0 \kks CO contours, ($A_V\approx 3.5$
mag) and then dividing the ones located inside the contours into those within and outside the cluster areas. After
correcting by background (in this case by subtracting the scaled number of in-survey/off-cloud sources) we estimate
that the average fraction of objects with $H-K>1.5$ located inside the molecular cloud projected area but outside
the clusters is approximately 21$\pm$4\%. The fraction is slightly larger than the one obtained from counts of NIRX
sources (14$\pm$4\%), possibly due to the fact that our red source counts are less restricted, as we cannot limit
their range in $J-H$. Counts of sources with large $H-K$ values have been used before to identify highly embedded
clusters with good results \citep[e.g.][]{ha05,ojha04}; in our case, our maps show that large H-K color sources
trace well the regions we identified as clusters, but {\it also} show a population at the 10 to 20 percent level
that might be distributed along the rest of the cloud.

\section{Discussion} \label{section:discussion}

\subsection{Clusters as a Dominant Mode of Formation}
\label{section:discussion:subsection:clusterdom}

Observations of nearby molecular clouds such as Orion, Perseus and Monoceros suggest that most stars in these
clouds form in embedded clusters (LA91, CA00). In the RMC, approximately 86 percent of the present day star
formation occurs in embedded clusters, in excellent agreement with the previous results for nearby clouds. This
indicates that cluster formation is also the dominant mode of star formation in clouds beyond 1 kpc. Cluster
formation appears to be well distributed across the Rosette Complex but the clusters themselves are confined to a
rather small fraction (5 percent) of the total area of the cloud. Comparison of the distribution of embedded
clusters with the distribution of molecular gas (WBS95), reveals that the embedded clusters 
are associated with the most massive molecular clumps in the cloud, which are also the most dynamically evolved. 
This is similar to what was found in the Orion B cloud \citep{ladacs92}.

While the most massive clumps in the Rosette are responsible for the production of the embedded clusters, then some of less
massive clumps could be associated with the distributed population. The number of sources exhibiting near-IR excess in
field 09 and the number of sources with $H-K>1.5$ across the cloud could indicate that a distributed population of
young stars formed in addition to the cluster population. However, the formation of this distributed population could account
for no more than 20 percent of the total star formation within the RMC. Characterizing the distributed population
is difficult due to several observational constraints. For example, the large photometric scatter of faint sources
along with the field star contamination restrict our study to the brighter (likely higher mass) sources, which are
expected to be less numerous than the fainter (likely lower mass) ones. Spectroscopic observations and mid-infrared
imaging would help to establish the number of highly reddened sources outside the cluster regions. Also, the nature
the distributed population in molecular clouds is somewhat unclear as discussed by CA00, precisely because these
sources permeate through large areas of the molecular cloud in and between regions of cluster formation and thus do
not have a clear origin. Such a population could be the residual of a slightly older generation of star formation,
a sub-product of cluster formation resulting from mass segregation or ejection or belong to a population located in
the foreground or background of the cloud.

\subsection{Rapidly Evolving Clusters}
\label{section:discussion:subsection:rapidevol}

The mortality rate of embedded clusters in the Milky Way is quite high.
Less than 10 percent of clusters survive their emergence from molecular
clouds and live longer than 10Myr and less than 4-7 percent survive to
become bound clusters the age of the Pleiades (LL03). This early
disruption is most likely linked to low star formation efficiency and the
rapid destruction of the molecular clouds. Similar results are found in
other galaxies \citep[e.g.][]{mfall}.

The observed distribution and properties of the embedded clusters in the
Rosette Complex may be revealing some of the first signs of such early
cluster evolution. The clusters in the Rosette appear to be, on average, 
slightly larger than those located in other clouds: the cluster equivalent radii 
range from $\sim 0.75$ to $2.3$~pc, with an mean of 1.44 and a median of 1.46 pc. 
The largest cluster in the complex is NGC 2244, with an equivalent radius of 2.3~pc. 
For comparison, the sizes of clusters listed in the catalog of LL03 range from $\sim 0.3$ to $3.8$~pc, 
but have a mean of 0.8 pc and a median of 0.62 pc. Only a few clusters in this catalog, such as MonR2, 
Gem4, NGC 2282 and Trapezium/ONC, have radii larger than 1.5 pc.

\begin{figure} 
\includegraphics[clip=true]{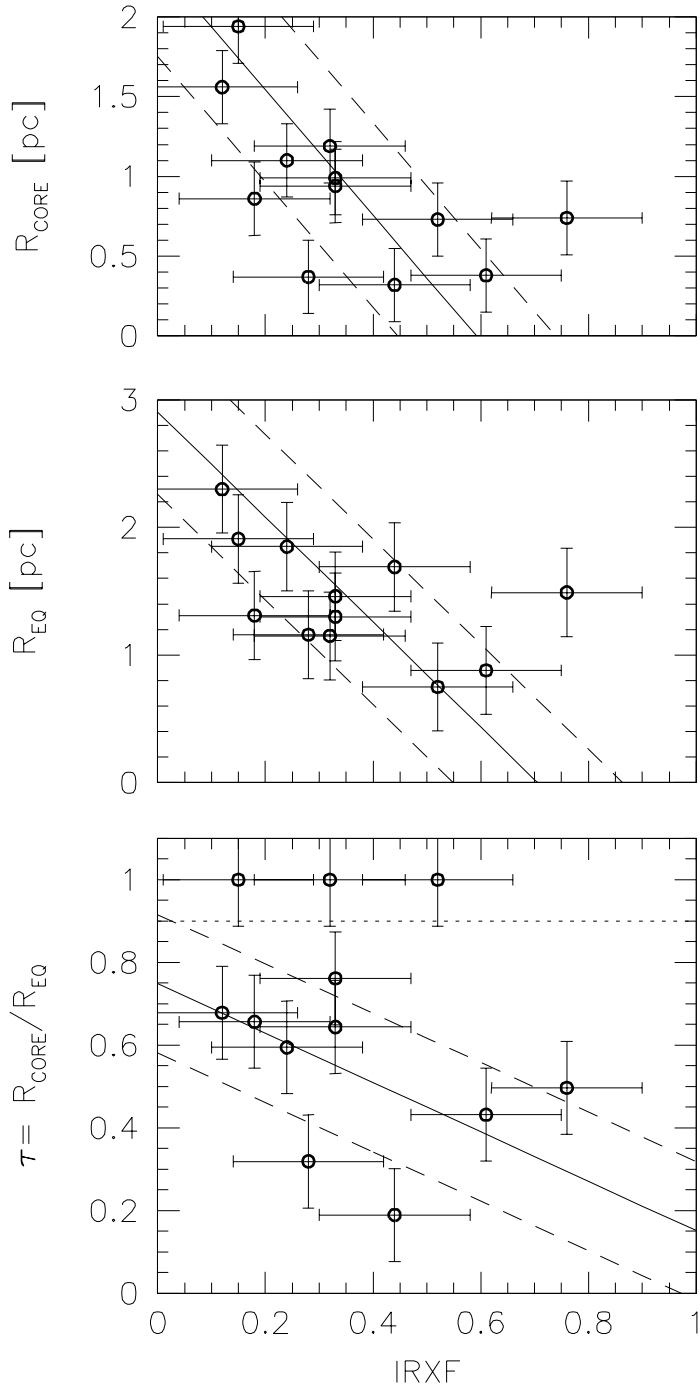} 
\caption{Clusters equivalent
sizes $R_{eq}$, cluster core sizes $R_{core}$, and core to total size ratios,
$\tau$ as a function of NIRX fraction (top, center and bottom panels
respectively). The solid line and dotted lines represent a least squares linear
fitting $\pm$1 rms deviation.} \label{fig:NIRXfsizerel} 
\end{figure}

\begin{figure} 
\epsscale{0.8} \plotone{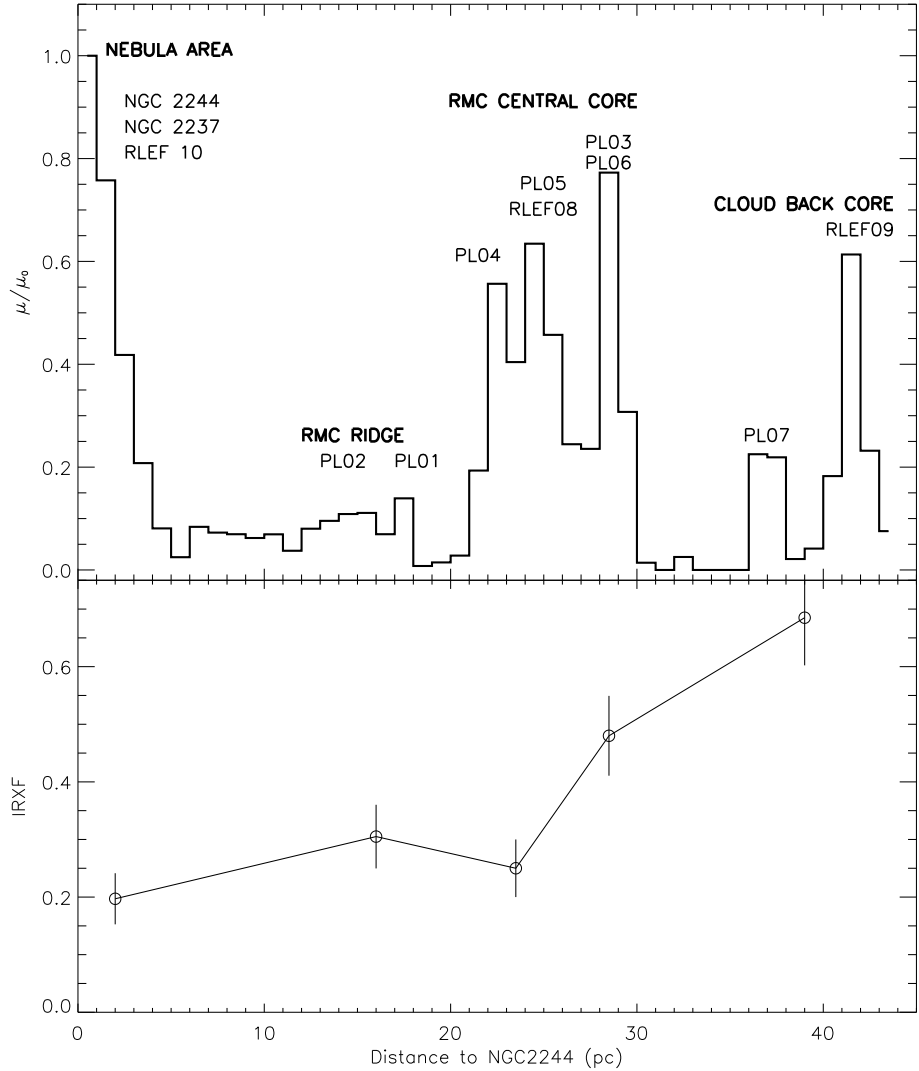} \vspace{-0.1cm} 
\caption{Top
panel: Distribution of NIRX sources density a
function of distance from the center of the Rosette Nebula (NGC~2244). The
counts are made in sectors of 1.0 pc in length and counts in each sector have
been scaled and normalized to the area and counts in the central 1.0 pc circle
in NGC~2244. Labels indicate the approximate locations of clusters described in
this paper, as well as the main 'regions' of the complex, defined by Blitz \& Thadeus (1986).
Bottom panel: averaged NIRX fractions in each of the cluster groups defined
from the top plot appear to increase with distance from the Rosette Nebula.}
\label{fig:dneb} 
\end{figure}

\begin{figure} 
\plotone{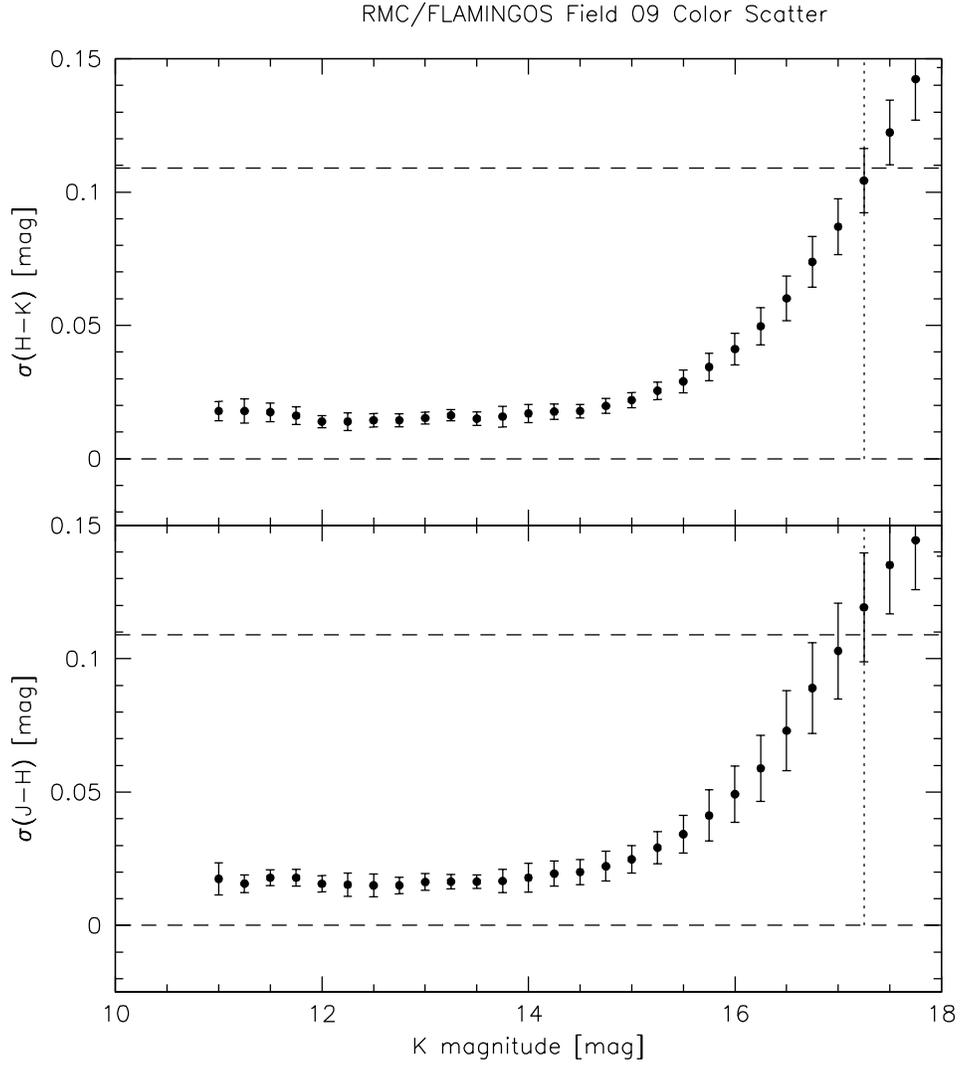} 
\caption{Distribution of color errors in Field
09 of the survey. The color scatter is low in this area due to the combination
of bad quality areas with the good quality areas of overlapping observations.
The dashed horizontal lines mark the 0.0 and 0.109 mag limits. The dashed
vertical line marks the completeness limit of the observations at $K=17.25$
mag.} \label{fig:area9colorscatter} 
\end{figure}

\begin{figure} 
\plotone{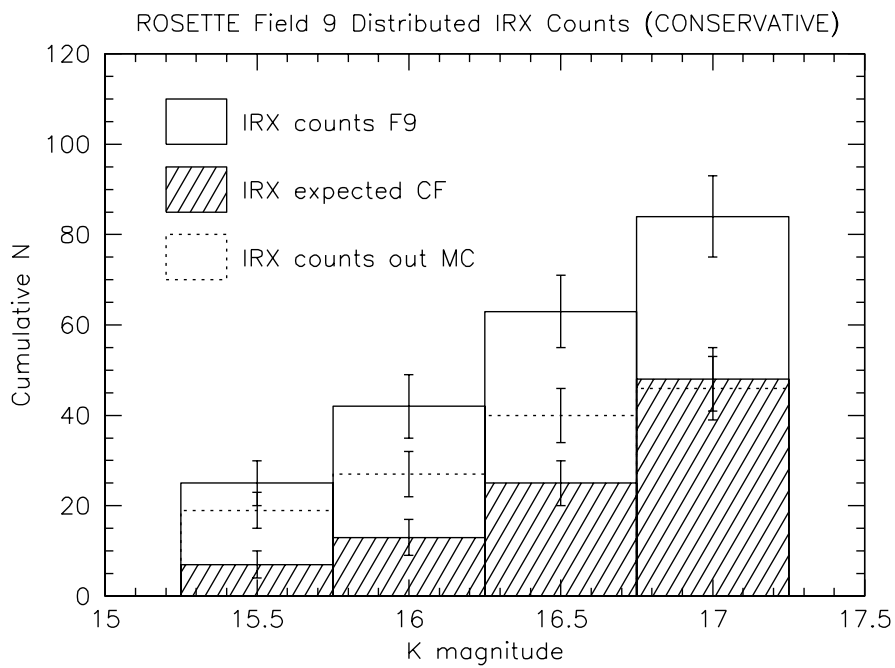} 
\caption{Cumulative Counts of NIRX sources in
field 09 of the survey.~The solid white bar histogram indicates the number of
NIRX sources in the field down to brightness limits of K=15.75,16.25,16.75 and
17.25 mag inside the area of best PSF photometry quality. The dotted line bar
histogram indicates the average of NIRX counts inside the same circular area for
fields 03, 13, 14 and 15, all located in areas mostly devoid of strong molecular
hydrogen emission.~The shaded histogram indicates the scaled number of
``expected" background NIRX sources, averaged from counts in the control fields
(also in the best PSF circular areas), and adding an uniform extinction value of
5.0 mag.} \label{fig:dp9hist} 
\end{figure}

\begin{figure} 
\plotone{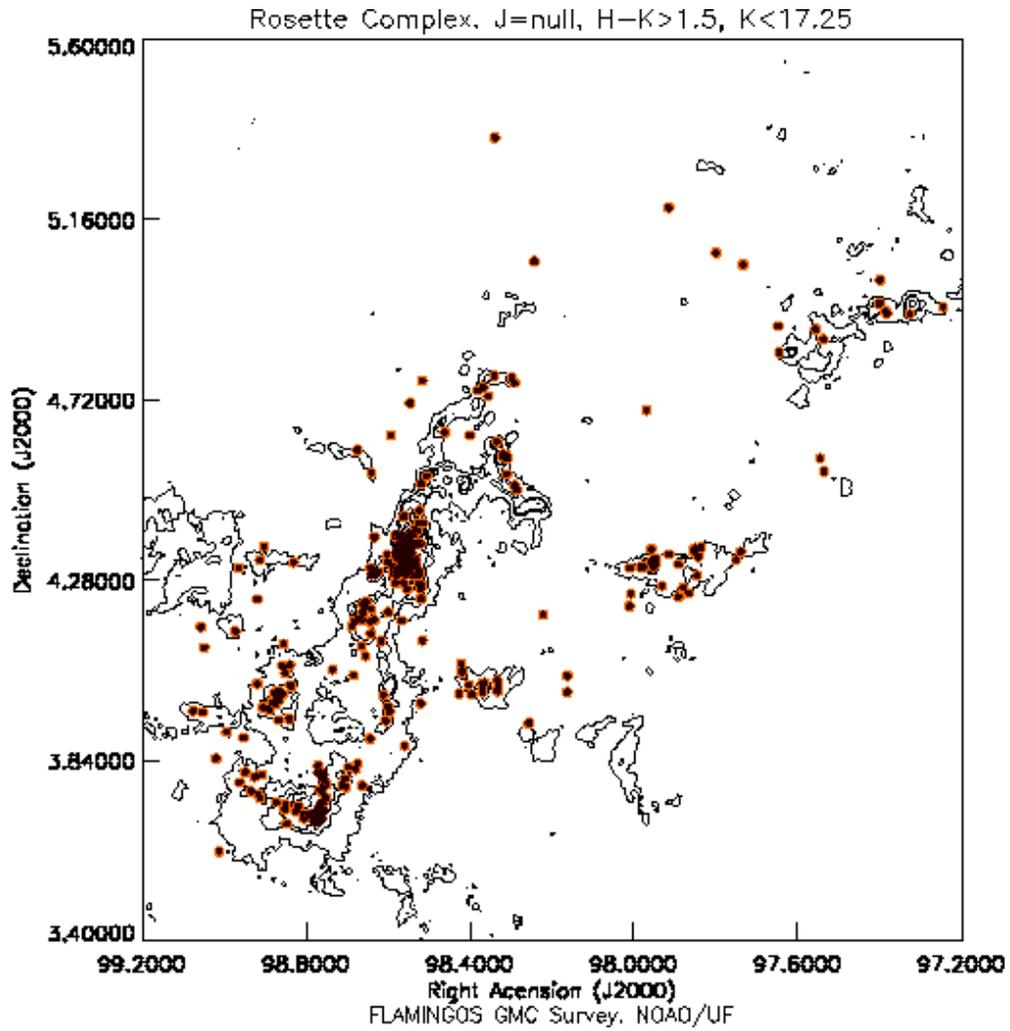} 
\caption{Distribution of sources with no detection in J and $H-K>1.5$ (circled dot symbols), limited to brightness of
$K=17.25$. The thin solid line contours indicate $^{13}$CO emission in steps of
1.0 \kks. } \label{fig:nojotas} 
\end{figure}

It is unlikely that the Rosette is systematically forming larger than average clusters. Instead, the large cluster
sizes may be due to dynamical evolution of the cloud leading to the beginning of cluster expansion and eventual
dispersal. Indeed our data suggest that cluster sizes in the Rosette may be related to cluster evolution in several
ways.  First, we found a negative correlation between extinction and cluster sizes (Fig. \ref{fig:agerel}),
indicating that the most extended clusters are the least embedded and are the ones that are emerging from the
parental molecular material. Cluster sizes are also inversely proportional to their infrared excess fraction (Fig.
\ref{fig:NIRXfsizerel}.) It has been shown that the infrared excess fraction in young clusters is initially high and decreases rapidly as a function of cluster age such that clusters with ages $>5$ Myr have no or extremely small excess fractions \citep{halala01}. In the Rosette, excess fraction decreases with increasing cluster size, implying that clusters expand relatively quickly (in a time scale comparable to the T Tauri phase). Furthermore, analysis of the density structure of the embedded clusters may also indicate the beginnings of cluster evolution. For example, even though some of the Rosette clusters exhibit a well defined core from our NN analysis, others appear to be extended well beyond this core. Indeed, the ratio of the core size to the total size, $\tau$, is 0.5 or above for 6 out of 8 young clusters with defined cores (Fig. \ref{fig:coreq}) indicating that most of the Rosette clusters do not have prominent central condensations. In addition, $\tau$  decreases with increasing infrared excess fraction (see Fig.
\ref{fig:NIRXfsizerel}). A similar correlation was found by FL07 in their analysis of local clusters. In the Rosette, clusters with $\tau\ge 0.5$ have an average infrared excess fraction of 27 percent whereas clusters with $\tau < 0.5$ have an average IRX fraction of 52\%. These results suggest that clusters form as compact units and later become less centrally condensed. This initial expansion appears to start quickly after formation and likely develops within a few Myr, favoring a scenario of rapid disruption and cluster dispersal.

\subsection{Influence of the HII Region and Sequential Star Formation}
\label{section:discussion:subsection:nebulainf}

The layout of the Rosette Complex led WBS95 and PL97 to suggest that the
formation of the clusters was triggered by the expansion of the nebula
into the cloud, resembling the picture of sequential star formation
proposed by \citet{elmelada77}. In this model, the shock front from
an expanding HII region would trigger the formation of a new OB
association in an adjacent molecular cloud which would later trigger a
third group and so on.  We can use the distribution of young stars and
clusters in our survey to investigate what effect, the HII region has had
on the star forming properties of the cloud.

Star formation in the Rosette appears to be concentrated into 4 main areas
of the complex as illustrated in Figure 24. The most prominent is the
Rosette Nebula itself, which contains the clusters NGC 2244, NGC 2237 and
REFL10.  The next concentration of star formation is located between 10 to
20 pc from the center of NGC 2244, corresponding to the RMC ridge. This
region is located the closest to and overlaps with the ionization front of
the HII region and contains clusters PL01 and PL02. It has the
lowest levels of star forming activity. Star formation may be inhibited in
this area due to the effects of the ionizing radiation. Hot ionized gas
might accelerate the evaporation of gaseous material surrounding a forming
cluster, resulting in the end of the star forming process.

The third concentration of young stars is located 20 to 30 pc from NGC
2244 and contains clusters PL04, PL05, REFL08 and PL06 at the RMC ``central
core'' and PL03 located in a gas clump (core D) outside the main area of
the cloud.  The four clusters in the central core account for 48\% of the
present day star formation in the molecular cloud  while only covering
roughly 4\% of the cloud area. These cores are associated with four of the
most massive cores in the cloud (WBS95). Similarly high concentrations of
star formation have been observed in other molecular clouds such as Orion,
W3/W4/W5 and Perseus, where approximately half of the young stars are
found in only 1 to a few embedded clusters \citep{la91,chs00,jor06}.  
The RMC central core appears to coincide with
the edge of the Rosette Nebula as shown by images of radio continuum and
25$\micron$ IRAS emission. The studies of \citet{celnik2,coxetal90}, WBS95 and
\citet{HWB05} agree that the nebula and molecular cloud
have their most intense interaction at this region and this may indicate
the recent passage of the shock front from the nebula.  The interaction
with the shock front may have triggered or stimulated the production of
stars here resulting in the second most active region of star formation in
the complex.

The last concentration of star forming activity is at the ``back core'' of
the cloud. This region contains clusters PL07 and REFL09.  These two
clusters appear to have formed in a region of the cloud located beyond the
main interaction with the nebula \citep{HWB05}. Given their
distance from the shock front, it is unlikely that the formation of these
clusters was triggered. This reinforces the suggestion by PL97
that star formation in this part of the cloud occurred
spontaneously.

Surprisingly, we do find the first evidence for a possible temporal
sequence in star forming events in the Rosette.  The average cluster
infrared excess fraction increases as a function of distance from NGC
2244, suggesting that the embedded clusters are progressively younger the
further they are from the Nebula. The existence of this apparent age
sequence indicates that star formation in the Rosette did indeed take
place sequentially in time, but not in the way proposed by 
\citeauthor{elmelada77}. Instead, it appears that the overall age sequence of cluster
formation is independent of any interaction with the expanding HII region
but rather may be primordial, possibly resulting from the formation and
evolution of the molecular cloud itself. While the existence of the HII
region cannot be responsible for the sequence of cluster ages, it does
appear to have a significant impact on the underlying sequence by either
enhancing or inhibiting the star forming process.

\subsection{Summary} \label{section:discussion:summary}

\par 1. In this paper we present results of a deep near-IR survey of the Rosette Complex. The survey was made with
the wide field imager FLAMINGOS at the Kitt Peak 2.1m telescope and covers all the main areas of the Rosette Nebula
and the Rosette Molecular Cloud

\par 2. We analyze the distribution of young stellar sources in the complex by estimating the surface densities of
objects with infrared excess, using a variation of the nearest neighbor method, which allows us to find clusters
with minimum populations of $\sim 30$ members.

\par 3. We confirmed the existence of the seven embedded clusters found by visual inspection by PL97 and found four
more clusters in the complex. Two of these clusters are deeply embedded in the molecular cloud and the other two
are located west of NGC 2244 in the Rosette Nebula.

\par 4. The young cluster population accounts for 60 percent of the stars in the complex, and approximately 86
percent of the youngest stars, located in the molecular cloud. This implies that the majority of stars in the Rosette form in embedded clusters, similar to nearby clouds.
. 
\par 5. The sizes of clusters in the Rosette Complex appear to be anti-correlated with their mean extinctions and infrared excess fractions, which suggests that clusters form as compact units then expand shortly after formation. The timescale for this process is similar or even
shorter than the T Tauri phase, as evidenced by the significant NIRX fractions and deeply embedded status.

\par 6. The distribution of young clusters suggests a division of star formation into four main regions coincident
with the largest features of the cloud: the first
is the Rosette Nebula, which contains the oldest clusters, the second is the ridge, located near to the ionization
front of the HII region, the third is the central core of the molecular cloud, where the main interaction between
the nebula and the cloud is located, and approximately 50 percent of the young stellar population was produced. The
fourth region of formation is at the back core of the cloud, located furthest from the Nebula.

\par 7. The averaged infrared excess fraction of these regions appears to
increase as a function of distance from the Rosette Nebula, which is
suggestive of a sequence of cluster ages across the cloud. Rather than
being triggered by the HII region, this sequence appears to be primordial,
possibly resulting from the formation and evolution of the molecular
cloud. The HII region appears to enhance or inhibit the underlying
pattern of star formation in the cloud.

\acknowledgments

We want to thank an anonymous referee, whose revision improved the content and quality of our manuscript. We also thank Dr. Charles J. Lada for useful discussion of this study.

Carlos Rom\'an-Z\'u\~niga wants to acknowledge CONACYT, Mexico for a fellowship that sponsored his doctoral studies at the University of Florida.

FLAMINGOS was designed and constructed by the IR instrumentation group (PI: R. Elston) at the University of Florida, Department of Astronomy with support from NSF grant (AST97-31180) and Kitt Peak National Observatory.

The data presented in this work were collected under the NOAO Survey Program {\it Towards a Complete Near-Infrared Spectroscopic Survey of Giant Molecular Clouds} (PI: E. Lada) which is supported by NSF grants AST97-3367 and AST02-02976 to the University of Florida. 

This research was also supported in part by the National Aeronautics and Space Administration under grant NNG05D66G issued through the LTSA program to the University of Florida.

The FLAMINGOS near-infrared survey of giant molecular clouds could not 
be possible without the support from the National Optical Astronomy
Observatory which is operated by the Association of Universities for
Research in Astronomy, Inc. (AURA) under cooperative agreement with
the National Science Foundation.

This publication makes use of data products from the Two Micron All Sky Survey, which is a joint project of the University of Massachusetts and the Infrared Processing and Analysis Center/California Institute of Technology, funded by the National Aeronautics and Space Administration and the National Science Foundation.

{\it Facilities:} \facility{KPNO:2.1m (FLAMINGOS)}

\clearpage

\onecolumn

\clearpage 

\clearpage 

\clearpage 

\clearpage 

\clearpage 

\clearpage 

\clearpage 

\clearpage 

\clearpage 

\clearpage 

\clearpage 

\begin{deluxetable}{lccccccc} 
\tablecolumns{8} 
\tablewidth{0pc}
\tablecaption{Young Clusters in the Rosette Complex} 
\tablehead{
\colhead{Cluster} & 
\colhead{RA} & 
\colhead{DEC} & 
\colhead{$R_{core}$} &
\colhead{$R_{equiv}$}  & 
\colhead{$N_{NIRX}\pm{\sqrt{N_{NIRX}}}$
\tablenotemark{a}} & 
\colhead{$NIRXF$ 
\tablenotemark{b}} & 
\colhead{$\langle A_v\rangle$ 
\tablenotemark{c}} \\ 
\cline{2-3} \cline{6-7} 
\colhead{ID}  &
\multicolumn{2}{c}{center, J2000} & 
\multicolumn{2}{c}{[pc]} &
\multicolumn{2}{c}{$K<15.75$} & 
\colhead{[mag]}  \\ 
\label{tab:clustertable}}

\startdata 
PL01	 & 97.96  &  4.32 &  0.37 & 1.16 &  29$\pm$5 & 0.28$\pm$0.05 & 8.1 $\pm$3.3 \\ 
PL02	 & 98.31  &  4.59 &  0.94 & 1.46 &  32$\pm$6 & 0.33$\pm$0.06 & 5.1 $\pm$2.8 \\ 
PL03	 & 98.38  &  4.00 &  0.32 & 1.69 &  80$\pm$9 & 0.44$\pm$0.05 & 6.8 $\pm$3.2 \\ 
PL04	 & 98.53  &  4.42 &  1.10 & 1.85 &  89$\pm$9 & 0.24$\pm$0.03 & 9.7 $\pm$5.2 \\ 
PL05	 & 98.63  &  4.32 & 0.86 & 1.31 &  57$\pm$8 & 0.18$\pm$0.03 & 7.2 $\pm$3.0 \\ 
PL06	 & 98.66  & 4.21 &  0.73 & 0.75 &  13$\pm$4 & 0.52$\pm$0.16 & 10.4$\pm$4.0 \\ 
PL07	 & 98.88  &  3.98 &  0.38 & 0.88 &  22$\pm$5 & 0.61$\pm$0.14 & 8.9 $\pm$3.2 \\
REFL08	 & 98.56  &  4.32 &  0.99 & 1.30 &  49$\pm$7 & 0.33$\pm$0.05 & 11.8$\pm$4.8 \\ 
REFL09	 & 98.78  &  3.69 &  0.74 & 1.49 &  65$\pm$8 & 0.76$\pm$0.09 & 9.8 $\pm$5.5 \\ 
REFL10	 & 97.78  &  5.27 &  1.19 & 1.15 & 15$\pm$4 & 0.32$\pm$0.09 & 3.2 $\pm$2.1 \\ 
NGC~2237 & 97.59  &  4.93 &  1.94 & 1.91 &  36$\pm$6 & 0.15$\pm$0.03 & 3.1 $\pm$1.7 \\ 
NGC~2244 & 97.95  &  4.94 & 1.56 & 2.30 &  62$\pm$8 & 0.12$\pm$0.02 & 1.4 $\pm$0.7 \\ 
\enddata

\tablenotetext{a}{Number of NIRX stars with 10th Nearest Neighbor densities above 0.2 \samis.} 
\tablenotetext{b}{NIRX fraction with respect to total number of stars with $K<15.75$ inside 0.2 \samis ~contour.}
\tablenotetext{c}{Average Extinction towards cluster line of sight (estimated from background source colors).}

\end{deluxetable}

\begin{deluxetable}{cccccc} 
\tablecolumns{6} 
\tablewidth{0pc}
\tablecaption{Rosette Complex: Sources with No detection in J, $H-K>1.5$} 
\tablehead{
\colhead{Brightness Limit} & 
\colhead{Total} & 
\colhead{Off Cloud} &
\multicolumn{2}{c}{In Cloud} & 
\colhead{} \\ 
\cline{4-5} 
\colhead{} & 
\colhead{} & 
\colhead{} & 
\colhead{In Clusters} & 
\colhead{Off Clusters} & 
\colhead{Loose Fraction} \\ 
\cline{2-5} 
\colhead{[mag]} & 
\multicolumn{4}{c}{[No. Sources]} &
\colhead{[\%]} \\ 
\label{tab:dptable}}

\startdata 15.75 &   112$\pm$11 & 5 $\pm$2  & 86 $\pm$9  &  21$\pm$5  &
19.3$\pm$4.6 \\ 16.25 &   173$\pm$13 & 7 $\pm$3  & 131$\pm$11 &  35$\pm$6  &
20.7$\pm$3.6 \\ 16.75 &   264$\pm$16 & 15$\pm$4  & 194$\pm$14 &  53$\pm$7  &
20.7$\pm$2.7 \\ 17.25 &   330$\pm$18 & 30$\pm$5  & 224$\pm$15 &  72$\pm$9  &
23.0$\pm$2.9 \\ \enddata

\end{deluxetable}

\end{document}